%% file: PAPER.tex
\documentclass[12pt]{article}  %fuer beidseitige Version
\usepackage[a4paper,textwidth=490.0pt,textheight=703.1pt]{geometry}
\usepackage{slashed}
\usepackage{graphicx}
\usepackage{epsfig}
\usepackage{amsmath}
\usepackage{amssymb}
\usepackage{ulem}
\usepackage{mdwlist}
\usepackage{color}                                                  
\usepackage{float}
\usepackage[rflt]{floatflt}
\usepackage{slashed}          
\usepackage{cite}
\usepackage{mathtools}
\usepackage{dsfont}
\usepackage{subfig}

\newcommand{\HA}{{\rm H}}
\usepackage{array}

\newcommand{\pqgn}{p_{qg}^{(0)}}
\newcommand{\Dpgqn}{\Delta p_{qg}^{(0)}}

\usepackage[english]{babel}
\usepackage[latin1]{inputenc}
\usepackage[T1]{fontenc}
\usepackage{ae}

\usepackage{url}

\usepackage{amsmath, amsthm, amssymb}

\newtheorem{thm}{Theorem}[section]

\usepackage{array}

\usepackage[english]{babel}
\usepackage[latin1]{inputenc}
\usepackage[T1]{fontenc}
\usepackage{ae}

\newtheorem{definition}[thm]{Definition}

\newcommand{\expm}{\raisebox{-0.07cm   } {$\, \stackrel{?}{=}\, $}}

\newcommand{\gsim}{\raisebox{-0.07cm   } {$\, \stackrel{>}{{\scriptstyle\sim}}\, $}}
\newcommand{\GeV}{\rm GeV}

\newcommand{\Ahathat}{\hat{\hspace*{0mm}\hat{A}}}
\newcommand{\Li}{{\rm Li}}
\newcommand{\Mvec}{{\rm \bf M}}

\newcommand{\ep}{\varepsilon}

\usepackage{rotating}

\usepackage{graphicx}
\newcounter{mmacnt}
\def\restartmma{\setcounter{mmacnt}{0}}
\restartmma \catcode`|=\active
\def|#1|{\mathrm{#1}}
\catcode`|=12
\newenvironment{mma}{
 \par\smallskip
 \catcode`|=\active
 \parskip=0pt\parindent=0pt % locally
 \small
 \def\In##1\\{%
   \def\linebreak{\hfill\break\null\qquad}%
   \refstepcounter{mmacnt}
   \hangindent=2.5em\hangafter=0
   \leavevmode
   \llap{\tiny\sffamily In[\arabic{mmacnt}]:=\kern.5em}%
   \mathversion{bold}\footnotesize$\displaystyle##1$\normalsize
   \mathversion{normal}\par
 }%
 \def\Print##1\\{%
   \def\linebreak{\hfill\break}%
   \hangindent=2.5em\hangafter=0
   \leavevmode ##1\par}%
 \def\Out##1\\{%
   \def\linebreak{$\hfill\break\null\hfill$}%
   \kern\abovedisplayskip\par
   \hangindent=2.5em\hangafter=0
   \leavevmode
   \llap{\tiny\sffamily Out[\arabic{mmacnt}]=\kern.5em}
   \footnotesize$\displaystyle##1$\normalsize\hfill\null\par
   \kern\belowdisplayskip
 }%
 \def\Warning##1##2\\{%
   \def\linebreak{\hfill\break}%
   \hangindent=2.5em\hangafter=0
   \leavevmode
   {\scriptsize##1 : ##2}\par}%
}{%
 \par\smallskip
}

\allowdisplaybreaks[4]

\usepackage{color}

\newenvironment{fshaded}{%
\MakeFramed {\FrameRestore}
}%
{\endMakeFramed}

\usepackage{tikz}
\usetikzlibrary{matrix}

\allowdisplaybreaks[4]

\begin{document}
\setlength{\baselineskip}{0.515cm}
\sloppy
\thispagestyle{empty}
\begin{flushleft}
DO--TH 23/12  \hfill  % {\tt arXiv:23XX.XXXXX [hep-ph]}
\\
DESY 23--142 \hfill October 2023\\
CERN-TH-2023-164 \hfill RISC Report series 23--12\\
ZU-TH 60/23 \hfill
MSUHEP-23-025 
\end{flushleft}

\setcounter{table}{0}

\begin{center}

\vspace*{2mm}
{\Large\bf \boldmath The first--order factorizable contributions to the three--loop massive
operator matrix elements $A_{Qg}^{(3)}$ and $\Delta A_{Qg}^{(3)}$}

\vspace*{1cm}
\large
J.~Ablinger$^{a,b}$, 
A.~Behring$^{c,d}$,
J.~Bl\"umlein$^d$,
A.~De Freitas$^{d,a}$, \\
A.~von Manteuffel$^{e,f}$,
C.~Schneider$^a$,
and
K.~Sch\"onwald$^{g}$

\vspace*{2mm}
{\small
{\it $^a$~Johannes Kepler University, 
Research Institute for Symbolic Computation (RISC),
Altenberger Stra\ss{}e 69,
                          A-4040, Linz, Austria}

\vspace*{2mm}
{\it $^b$~Johann Radon Institute for Computational and Applied Mathematics
(RICAM), Austrian Academy of Sciences,
Altenberger Stra\ss{}e 69, A-4040 Linz Austria}

\vspace*{2mm}
{\it $^c$~Theoretical Physics Department, CERN, 1211 Geneva 23, Switzerland}

\vspace*{2mm}
{\it $^d$~Deutsches Elektronen-Synchrotron DESY, Platanenallee 6, 15738 Zeuthen, Germany}

\vspace*{2mm}
{\it $^e$~Institut f\"ur Theoretische Physik, Universit\"at Regensburg,
93040 Regensburg, Germany}

\vspace*{2mm}
{\it $^f$~Department of Physics and Astronomy, Michigan State 
University, 
East Lansing, MI 48824, USA}

\vspace*{2mm}
{\it $^g$~Physik-Institut, Universit\"at Z\"urich, 
Winterthurerstrasse 190, CH-8057 Z\"urich, Switzerland}}
\\

\end{center}
\normalsize
\vspace{\fill}
\begin{abstract}
\noindent 
The unpolarized and polarized massive operator matrix elements $A_{Qg}^{(3)}$ and $\Delta A_{Qg}^{(3)}$
contain first--order factorizable and non--first--order factorizable contributions in the determining 
difference or differential equations of their master integrals. We compute their first--order factorizable 
contributions in the single heavy mass case for all contributing Feynman diagrams. Moreover, we present the 
complete color--$\zeta$ factors for the cases in which also non--first--order factorizable contributions 
emerge in the master integrals, but cancel in the final result as found by using the method of arbitrary high 
Mellin moments. Individual contributions depend also on generalized harmonic sums and on nested finite 
binomial and inverse binomial sums in Mellin $N$--space, and correspondingly, on Kummer--Poincar\'e and 
square--root valued alphabets in Bjorken--$x$ space. We present a complete discussion of the possibilities 
of solving the present problem in $N$--space analytically and we also discuss the limitations in the present 
case to analytically continue the given $N$--space expressions to $N \in \mathbb{C}$ by strict methods. The 
representation through generating functions allows a well synchronized representation of the first--order 
factorizable results 
over a 17--letter alphabet. We finally obtain representations in terms of iterated integrals over the 
corresponding 
alphabet in $x$--space, also containing up to weight {\sf w = 5} special constants, which can be rationalized 
to Kummer--Poincar\'e iterated integrals at special arguments. The analytic $x$--space representation requires 
separate analyses for the intervals $x \in [0,1/4], [1/4,1/2], [1/2,1]$ and $x > 1$. We also derive the small 
and large $x$ limits of the first--order factorizable contributions. Furthermore, we perform comparisons to 
a number of known Mellin moments, calculated by a different method for the corresponding subset of Feynman 
diagrams, and an independent high--precision numerical solution of the problems.
\end{abstract}

\vspace*{\fill}
\noindent
\numberwithin{equation}{section}

\newpage
%%%%%%%%%%%%%%%%%%%%%%%%%%%%%%%%%%%%%%%%%%%%%%%%%%%%%%%%%%%%%%%%%%%%%%%%%%%%%%%%%%%%%%%%%%%%%%%%%%%
\section{Introduction}
\label{sec:1}
%%%%%%%%%%%%%%%%%%%%%%%%%%%%%%%%%%%%%%%%%%%%%%%%%%%%%%%%%%%%%%%%%%%%%%%%%%%%%%%%%%%%%%%%%%%%%%%%%%%

\vspace*{1mm}
\noindent
Precision data on deep--inelastic scattering structure functions allow precision measurements
of the strong coupling constant $a_s(Q^2) = \alpha_s(Q^2)/(4\pi)$ \cite{Bethke:2011tr,Moch:2014tta,
Alekhin:2016evh,dEnterria:2022hzv}, the extraction of the parton distribution functions (PDFs), 
cf. e.g.~\cite{Accardi:2016ndt,Alekhin:2017kpj}, and the measurement of the charm quark mass $m_c$ 
\cite{Alekhin:2012vu}. To suppress higher twist effects \cite{Blumlein:2008kz,Blumlein:2012se,Alekhin:2012ig} 
one chooses $Q^2 \gsim 25 \GeV^2$, which is also the asymptotic region for the charm contribution to the 
structure function $F_2(x,Q^2)$ \cite{Buza:1995ie}. Besides the detailed knowledge of the 
evolution of the PDFs in Quantum Chromodynamics (QCD) \cite{Moch:2004pa,Vogt:2004mw,Vermaseren:2005qc,
Ablinger:2010ty,Blumlein:2012vq,Ablinger:2014lka,Ablinger:2014vwa,
Ablinger:2014nga,Moch:2014sna,Anastasiou:2015vya,Ablinger:2017tan,   
Mistlberger:2018etf,Behring:2019tus,Luo:2019szz, Duhr:2020seh, Ebert:2020yqt,Ebert:2020unb,
Luo:2020epw,
Blumlein:2021enk, 
Blumlein:2021ryt,
Blumlein:2022gpp,  
Baranowski:2022vcn,
Gehrmann:2023ksf}
one needs the massless \cite{Vermaseren:2005qc,Blumlein:2022gpp} and massive Wilson coefficients in the 
single--mass 
\cite{
Behring:2014eya,Ablinger:2014vwa,Ablinger:2014nga,
Ablinger:2014lka,Ablinger:2019etw,
Blumlein:2021xlc,Behring:2021asx,Ablinger:2022wbb
} and two--mass cases  
\cite{
Ablinger:2017err,Ablinger:2017xml,Ablinger:2018brx,Ablinger:2019gpu,
Ablinger:2020snj
} to three--loop order for neutral current interactions. 

While many of the contributing massive operator 
matrix elements (OMEs) have been calculated both in the unpolarized and polarized case
\cite{Ablinger:2010ty,
Blumlein:2012vq,
Behring:2014eya,
Ablinger:2014lka,
Ablinger:2014vwa,
Ablinger:2014nga,
Ablinger:2017err,
Ablinger:2017xml,
Blumlein:2017wxd,
Ablinger:2018brx,
Ablinger:2019etw,
Ablinger:2019gpu,
Ablinger:2020snj,
Behring:2021asx,
Blumlein:2021enk, 
Blumlein:2021xlc,
Blumlein:2021ryt,
Blumlein:2022gpp,  
Ablinger:2022wbb}, the constant part 
to the unrenormalized OMEs $A_{Qg}^{(3)}$ and $\Delta A_{Qg}^{(3)}$, denoted by $a_{Qg}^{(3)}$ and $\Delta a_{Qg}^{(3)}$, 
are still missing. All logarithmic contributions are known, however, \cite{Behring:2014eya,Blumlein:2021xlc}.
Furthermore, the massive OMEs provide the transition matrix elements in the variable--flavor number scheme 
in the single--mass  \cite{Buza:1996wv} and the two--mass case \cite{Ablinger:2017err}. The transitions of 
heavy flavors becoming light to two--loop order were studied in 
Refs.~\cite{Blumlein:2018jfm,Bierenbaum:2022biv}, also including two--mass effects.

The OMEs can be expressed in terms of a basis of Feynman integrals
called master integrals. These master integrals fulfill systems of
first--order differential equations. Equivalently, one can uncouple
these to scalar linear higher--order differential operators. One
important question, for example to classify which function spaces occur
in the solutions, is whether the differential operators can be
factorized into first--order factors. In the following we call those
parts of the final result which are determined by master integrals that
fulfill first--order factorizable differential equations
{\it first--order factorizable contributions} and the remaining part 
{\it non--first--order factorizable contributions}.

In this paper we present all first--order factorizable contributions (also called 
d'Alembertian solutions) for the Feynman diagrams contributing  
to $a_{Qg}^{(3)}$ and $\Delta a_{Qg}^{(3)}$. 
We compute the polarized OMEs in the Larin scheme \cite{Larin:1993tq}.\footnote{To describe the scale 
dependence of the polarized structure function $g_1(x,Q^2)$, the Wilson coefficients have to be computed in the 
same 
scheme \cite{Blumlein:2022gpp} and one needs to refer to parton distributions in this scheme, the evolution of which
is ruled by the anomalous dimensions in the Larin scheme \cite{Moch:2014sna,Blumlein:2021ryt}. For the non--singlet 
case see Ref.~\cite{Blumlein:2021lmf}.}
The results for the complete OMEs in Mellin $N$--space can also be
subdivided according to color factors and $\zeta$ values, which are
multiplied by functions that evaluate to rational numbers for fixed
values of $N$. These rational numbers fulfill recurrence relations
that can be determined by using the method of arbitrarily high Mellin
moments \cite{Blumlein:2017dxp}. In this way, we find solutions to complete color--$\zeta$
factors, even in some cases where non--first--order factorizable master
integrals emerge,  since their contributions cancel in the final result.
Out of 25 color--$\zeta$ values, 10 remain to be 
computed and we will deal with their first--order factorizable terms here for all contributing 
diagrams. The remaining terms, containing also non--first--order factorizable contributions,
are the subject of a forthcoming paper \cite{AQGFIN}, since the algorithms to compute them are rather different 
from the ones of the present paper. 

Furthermore, we also present the solutions in Mellin $N$--space we have 
obtained for the complete project. In a series of color--$\zeta$ terms with non--first--order 
factorizable 
contributions we computed closed form difference equations at very high degree and order. These define 
recurrent functions which may be used for shift relations of the analytic continuations from $N \in \mathbb{N}$
to $N \in \mathbb{C}$
within the analyticity region of the problems. Moreover, one may calculate the asymptotic solutions
of difference equations of this kind and of individual building blocks also in the first--order 
factorizable case, such as generalized harmonic sums \cite{Ablinger:2013cf} and nested finite binomial sums 
\cite{Ablinger:2014bra}. These serve as numeric initial conditions for the shift relations in $N \in 
\mathbb{C}$. The asymptotic expansions for harmonic sums \cite{Vermaseren:1998uu, Blumlein:1998if} were 
derived in Refs.~\cite{Blumlein:2009ta,Blumlein:2009fz}. Besides the results in the first--order factorizable
case, we will also present details of the technologies used in the present calculation. 

The paper is organized as follows. In Section~\ref{sec:2} we present the basic computation steps to obtain
the results in Mellin $N$--space. A subset of Feynman diagrams turns out to be given in terms of master integrals 
which are first--order factorizable and therefore lead to product--sum representations
using the algorithms of Refs.~\cite{Karr:1981,Bron:00,Schneider:01,Schneider:04a,Schneider:05a,Schneider:05b,
Schneider:07d,Schneider:2009rcr,
Schneider:10c,Schneider:15a,Schneider:08d,Schneider:08e,Schneider:2017,ABPET1} 
implemented in the package {\tt Sigma} \cite{SIG1,SIG2}.  
As we mentioned above, it is also possible to compute a 
series of complete 
color--$\zeta$ values, 
even though 
their master integrals 
may contain non--first--order factorizable contributions, as long as
these contributions cancel in the determining recurrence, 
which then turns out to be first--order factorizable.
In Section~\ref{sec:3} we present these contributions to 
$a_{Qg}^{(3)}(N)$ and $\Delta a_{Qg}^{(3)}(N)$, except for irreducible diagrams resulting in
purely rational and $\zeta_3$ terms. The structure 
of the first--order factorizable contributions to Feynman diagrams contributing to the purely 
rational and $\zeta_3$ terms 
in $N$--space is discussed in Section~\ref{sec:4}. Here we also consider the principal structure of high moments to
all contributions. In Section~\ref{sec:5} we compute the $t$--space representation of the 
first--order factorizable 
contributions of the OMEs from the associated set of differential equations to the required depth 
in the dimensional parameter $\ep = D - 4$. Here $t \in \mathbb{R}$ denotes a resummation variable, 
cf.~Eq.~(\ref{eq:FT}). From this representation we perform the analytic continuation to 
$x$--space. The final expressions are given by G--functions, see Eq.~(\ref{eq:GL1}), over root--valued 
alphabets and 
corresponding 
G--constants at special values $x = x_0$. The contributing G--functions of $x$ over the root--valued 
alphabet can be rationalized 
and mapped to Kummer--Poincar\'e iterated integrals 
\cite{KUMMER1,KUMMER2,KUMMER3,POINCARE,LANDAN,CHEN,GONCHAROV}, 
at the expense of a  root--valued main argument. We expand 
the  G--functions of $x$ into series  around $x = 0, 
1/2$ and 1 and give precise numerical representations for the G--functions at special values 
of $x_0$. The letters of the
latter quantities can be rationalized.\footnote{We will also say that the corresponding G--functions are 
rationalized.} In Section~\ref{sec:6} we perform the expansions around $x=0$ and $x=1$ for the first--order 
factorizable terms to determine their contributions to the most singular terms of $a_{Qg}^{(3)}$ and 
$\Delta a_{Qg}^{(3)}$. This requires the 
calculation of a series of G--functions, Eq.~(\ref{eq:GL1}), with root valued letters at $x=1$. We also 
investigate color rescaling relations. In Section~\ref{sec:7} we present numerical results, and
Section~\ref{sec:8} contains the conclusions. In the Appendices~\ref{sec:A}--\ref{sec:D} we summarize 
a series of technical 
aspects, such as the asymptotic expansions of contributing generalized harmonic sums,
of characteristic aspects of nested (inverse) binomial  sums and their asymptotic expansion,
the calculation of special G--constants, and the analytic continuation to $N$--space.
%%%%%%%%%%%%%%%%%%%%%%%%%%%%%%%%%%%%%%%%%%%%%%%%%%%%%%%%%%%%%%%%%%%%%%%%%%%%%%%%%%%%%%%%%%%%%%%%%%%
\section{The main steps of the calculation}
\label{sec:2}
%%%%%%%%%%%%%%%%%%%%%%%%%%%%%%%%%%%%%%%%%%%%%%%%%%%%%%%%%%%%%%%%%%%%%%%%%%%%%%%%%%%%%%%%%%%%%%%%%%%

\vspace*{1mm}
\noindent
The Feynman diagrams of the OMEs $(\Delta) A_{Qg}^{(3)}$ are generated by {\tt QGRAF} \cite{Nogueira:1991ex} 
using the Feynman rules of Refs.~\cite{YND,Bierenbaum:2009mv}. The Lorentz- and Dirac algebra has been 
performed with {\tt Form} \cite{Vermaseren:2000nd,Tentyukov:2007mu}, the color algebra by using 
{\tt color} \cite{vanRitbergen:1998pn}, and the integration by parts reduction 
\cite{IBP1,IBP2,IBP3,IBP4,Chetyrkin:1981qh,Laporta:2001dd} by using the package {\tt Reduze~2} 
\cite{Studerus:2009ye,vonManteuffel:2012np}. The diagrams have been calculated in Mellin $N$--space using 
different techniques which are described in Refs.~\cite{Blumlein:2018cms,Blumlein:2022qci}
for the first--order--factorizable contributions. 
These included summation 
technologies based on difference ring theory 
\cite{Karr:1981,Bron:00,Schneider:01,Schneider:04a,Schneider:05a,Schneider:05b,Schneider:07d,Schneider:2009rcr,
Schneider:10c,Schneider:15a,Schneider:08d,Schneider:08e,Schneider:2017,ABPET1}, encoded in the package 
{\tt Sigma} \cite{SIG1,SIG2}, the solution of 
differential equations \cite{Ablinger:2015tua,Ablinger:2018zwz} 
and using {\tt SolveDE} of the package 
{\tt 
HarmonicSums}
\cite{Vermaseren:1998uu,
Blumlein:1998if,
Ablinger:2013cf,
Ablinger:2011te,
Ablinger:2014bra,
Remiddi:1999ew,
Blumlein:2003gb,
Blumlein:2009ta,
Blumlein:2009cf,
Ablinger:2010kw,
Ablinger:2013hcp,
Ablinger:2014rba,
Ablinger:2015gdg,
ALL2016,
ALL2018,
Ablinger:2018cja,
Ablinger:2019mkx,
Ablinger:2021fnc} 
as well as the differential equation solver for first--order factorizable systems
of Ref.~\cite{Ablinger:2018zwz}.
Differential equations are decoupled using the package {\tt OreSys} 
\cite{ORESYS1,ORESYS2,ORESYS3}. Finally, we applied also the multiple 
Almkvist--Zeilberger algorithm \cite{AZ,AZ1}  as implemented in the package 
{\tt MultiIntegrate} \cite{Ablinger:2021dfb}. We thus obtain first a representation in $N$--space for
all first--order factorizable contributions.

One may even envisage the complete solution of the problem in $N$--space. Here the first step is to obtain 
closed form difference equations for all color--$\zeta$ values by using the method of arbitrary high Mellin 
moments \cite{Blumlein:2017dxp} and guessing algorithms \cite{GUESS,Blumlein:2009tj} implemented in {\tt Sage} \cite{SAGE,
GSAGE}. For the color factors $C_{F,A}^2 T_F$ and $C_F C_A T_F$ this task is very demanding 
in terms of computer time and requires an amount of moments far beyond 15000, 
which is the level currently obtained for other color factors in the unpolarized case. In the polarized 
case we computed 11000 moments. 
At present it is only possible to solve those color--$\zeta$
contributions which are related to first--order factorizable difference
equations. Solving them leads to nested product--sum representations.
Even though individual Feynman diagrams may contain master integrals
that fulfill non--first--order factorizable differential equations, in
some cases it is still possible to solve the corresponding
color--$\zeta$ contribution if the non--first--order factorizable
terms cancel in the sum over all Feynman diagrams.

Because of the fact that we cannot easily solve some of the large difference equations, which will also apply to
the yet missing ones, we have chosen a different strategy in the cases of non--first--order 
factorizable difference
equations by referring to $x$--space directly. The $N$--space expressions are first resummed into the 
$t$--space representations \cite{Ablinger:2012qm,Ablinger:2014yaa} by
%----------------------------------------------------------------------------------------------
\begin{eqnarray}
F(t) = \sum_{N=1}^\infty t^N F(N),~\text{with}~t \in \mathbb{R}.
\label{eq:FT}
\end{eqnarray}
%----------------------------------------------------------------------------------------------
Likewise, we can do this also for the master integrals directly and express the first--order factorizable 
master integrals as functions of $t$. This covers 1009 of the total 1233 Feynman
diagrams which contain only first--order factorizable master integrals.
We obtain the corresponding $x$--space representation $\tilde{F}(x)$ from $F(t)$ using the method described in 
Ref.~\cite{Behring:2023rlq} by  computing the
discontinuity of $F(1/x)$,
%----------------------------------------------------------------------------------------------
\begin{align}
  \tilde{F}(x)
    &= -\frac{1}{2\pi i} \mathrm{Disc}_x F\left(\frac{1}{x}\right)
  \,.
\end{align}
%----------------------------------------------------------------------------------------------
This leads to G--functions, which are defined by 
%----------------------------------------------------------------------------------------------
\begin{eqnarray}
\label{eq:GL1}
G(\{f_1(\tau), \vec{f}(\tau)\},x) = \int_0^x dy f_1(y) G(\{\vec{f}(\tau)\},y).
\end{eqnarray}
%----------------------------------------------------------------------------------------------
Here the letters $f_i$ belong to an associated alphabet $\mathfrak{A}$ of length $m$,
%----------------------------------------------------------------------------------------------
\begin{eqnarray}
\mathfrak{A} = \left. \{f_i(x)\}\right|_{i=1}^m.
\end{eqnarray}
%----------------------------------------------------------------------------------------------
In general some of the letters are given by higher transcendental functions.
Both the G--functions in $t$--and $x$--space have to have representations as Riemann integrals
individually, which requires to remove singularities if  they are present in some letters.
In the course of the calculation
different constants will emerge as G--functions, (\ref{eq:GL1}), evaluated at a series of special values
of $x \in [0,1]$. All G--functions of $x$ and the constants shall be further reduced algebraically 
at the end of the calculation and, if possible, simplified to known special functions,
using algorithms of the package {\tt HarmonicSums}.
As it turns out later, it will also be useful to apply the $t$--space representation to the nested binomial 
sum 
contributions.
 
The results in $N$-- and in $x$--space are related by a Mellin 
transform \cite{RIEMANN,CAHEN,MELLIN1,MELLIN2,TITCHMARSH}
%------------------------------------------------------------------------
\begin{eqnarray}
\Mvec[f(x)](N) = \int_0^1 dx x^{N-1} f(x).
\end{eqnarray}
%------------------------------------------------------------------------
In the following we will present the results in Mellin $N$--space. 
The unrenormalized OME~$\Ahathat_{Qg}$ has the following structure \cite{Bierenbaum:2009mv}
both in the unpolarized and polarized cases\footnote{In the polarized case the symbol $\Delta$
is put in front of the respective coefficient. Structurally the relations are the same
as those of (\ref{OME:1}--\ref{OME:3}).}
%---------------------------------------------------------------------------------
\begin{eqnarray}
    \Ahathat_{Qg}\left(\frac{\hat{m}^2}{\mu^2},\hat{a}_s,\ep,N\right) 
&=& \sum_{l=1}^\infty \hat{a}_s^l \Ahathat_{Qg}^{(l)}\left(\frac{\hat{m}^2}{\mu^2},\ep,N\right), 
\\
\label{OME:1}
    \Ahathat_{Qg}^{(1)}&=&
             \Bigl(\frac{\hat{m}^2}{\mu^2}\Bigr)^{\ep/2}\Biggl[
                          \frac{\hat{\gamma}_{qg}^{(0)}}{\ep}
                         +a_{qg,Q}^{(1)}
                         +\ep\overline{a}_{qg,Q}^{(1)}
                         +\ep^2\overline{\overline{a}}_{qg,Q}^{(1)}
                        \Biggr] + O(\ep^3),
\\
%---------------------------------------------------------------------------------
    \Ahathat_{Qg}^{(2)}&=&
             \Bigl(\frac{\hat{m}^2}{\mu^2}\Bigr)^{\ep}
                  \Biggl[
                          \frac{1}{\ep^2} c_{Qg,(2)}^{(-2)}
                         + \frac{1}{\ep} c_{Qg,(2)}^{(-1)}
                         +  c_{Qg,(2)}^{(0)}
                         + \ep c_{Qg,(2)}^{(1)} \Biggr]+ O(\ep^2),
\\
%---------------------------------------------------------------------------------
\label{OME:3}
    \Ahathat_{Qg}^{(3)}&=&
             \Bigl(\frac{\hat{m}^2}{\mu^2}\Bigr)^{3\ep/2}
                  \Biggl[
                          \frac{1}{\ep^3} c_{Qg,(3)}^{(-3)}
                         + \frac{1}{\ep^2} c_{Qg,(3)}^{(-2)}
                         + \frac{1}{\ep} c_{Qg,(3)}^{(-1)}
                         + a_{Qg}^{(3)}\Biggr] + O(\ep)~,
 \end{eqnarray}
%---------------------------------------------------------------------------------
with $\hat{m}$ the unrenormalized mass, $\hat{a}_s = \hat{g}^2_s/(16\pi^2)$
the unrenormalized strong coupling constant and $\mu$ the renormalization and factorization scale.
The renormalization of the OMEs proceeds in four steps, cf.~\cite{Bierenbaum:2009mv}: the renormalization 
of the heavy quark mass, of the strong 
coupling~\cite{Tarasov:1980au,Larin:1993tp,vanRitbergen:1997va,Czakon:2004bu,
Chetyrkin:2004mf,Baikov:2016tgj,Herzog:2017ohr,Chetyrkin:2017bjc,Luthe:2017ttg,Luthe:2017ttc,Chetyrkin:1999ys,
Chetyrkin:1999qi,Melnikov:2000qh,Broadhurst:1991fy,Marquard:2018rwx,Marquard:2016dcn,Marquard:2015qpa}, of 
the local composite 
operators, and the subtraction of the collinear singularities due to massless 
sub--graphs \cite{Bierenbaum:2009mv}.
%%%%%%%%%%%%%%%%%%%%%%%%%%%%%%%%%%%%%%%%%%%%%%%%%%%%%%%%%%%%%%%%%%%%%%%%%%%%%%%%%%%%%%%%%%%%%%%%%%%
\section{\boldmath First-order factorizable recurrences for complete color-$\zeta$ contributions}
\label{sec:3}
%%%%%%%%%%%%%%%%%%%%%%%%%%%%%%%%%%%%%%%%%%%%%%%%%%%%%%%%%%%%%%%%%%%%%%%%%%%%%%%%%%%%%%%%%%%%%%%%%%%

\vspace*{1mm}
\noindent
In the following we will consider the constant parts of the unrenormalized unpolarized and polarized OMEs,
$a_{Qg}^{(3)}(N)$ and $\Delta a_{Qg}^{(3)}(N)$. The additional contributions resulting from lower--order 
terms due to renormalization were given in Refs.~\cite{Behring:2021asx,Blumlein:2021xlc}.
 
The calculation of 2000 Mellin moments has been sufficient to determine all color--$\zeta$ contributions
to $a_{Qg}^{(3)}(N)$ and $\Delta a_{Qg}^{(3)}(N)$ obeying first--order--factorizable recurrences by using 
the methods of  Refs.~\cite{Blumlein:2017dxp,GUESS,Blumlein:2009tj} for generating the
recurrences and the methods of Refs.~\cite{Karr:1981,Bron:00,Schneider:01,Schneider:04a,Schneider:05a,
Schneider:05b,Schneider:07d,Schneider:2009rcr,Schneider:10c,Schneider:15a,Schneider:08d,Schneider:08e,
Schneider:2017,ABPET1} 
for computing the closed form solutions. These are all contributions $\propto 
\textcolor{blue}{N_F}$, with $\textcolor{blue}{N_F}$ the number of massless flavors,\footnote{A direct 
computation of the \textcolor{blue}{$N_F$} terms in the unpolarized case using summation methods has been 
performed before in \cite{Ablinger:2010ty}.} and all 
other terms except the purely rational ones and those $\propto \zeta_3$ of $O(\textcolor{blue}{T_F})$ and 
$O(\textcolor{blue}{T_F^2})$ of the irreducible Feynman diagrams, which contain $_2F_1$  terms 
\cite{Ablinger:2017bjx,Behring:2023rlq} 
in $x$--space. This implies that their recurrences are not first--order factorizable. Here we include 
also the reducible contributions\footnote{These are self--energy insertions on external lines of the Feynman diagrams.}, 
which leads to a part of the 
rational and $\zeta_3$ terms. In the unpolarized case also the Feynman diagrams with external Faddeev--Popov ghosts 
\cite{Faddeev:1967fc} contribute to the 
first--order factorizable terms. The yet missing part concerns only  irreducible Feynman diagrams.
%------------------------------------------------------------------------

For the color-$\zeta$ structures which can be obtained fully in closed form,  we obtain
%------------------------------------------------------------------------
\begin{eqnarray}
\label{eq:aunpol}
\input{a3unp}
\end{eqnarray}
%----------------------------------------------------------------------
and 
%------------------------------------------------------------------------
\begin{eqnarray}
p_{qg}^{(0)} = \frac{N^2+N+2}{N(N+1)(N+2)}.
\end{eqnarray}
%----------------------------------------------------------------------
%------------------------------------------------------------------------
The above expressions can all be represented in terms of nested harmonic sums \cite{Vermaseren:1998uu,
Blumlein:1998if}
%---------------------------------------------------------------------------------
\begin{eqnarray}
S_{b,\vec{a}}(N) = \sum_{k=1}^N \frac{({\rm sign}(b))^k}{k^{|b|}} S_{\vec{a}}(k),~~~S_\emptyset = 1,~~~~b, a_i \in 
\mathbb{Z} \backslash \{0\},
\end{eqnarray}
%---------------------------------------------------------------------------------
for which we use the shorthand notation $S_{\vec{a}}(N) \equiv S_{\vec{a}}$.
The constant {\sf B}$_4$ is given by
%------------------------------------------------------------------------
\begin{eqnarray}
{\sf B}_4 &=& -4 \zeta_2 \ln^2(2) + \frac{2}{3} \ln^4(2) - \frac{13}{2} \zeta_4 + 16 
\Li_4\left(\frac{1}{2}\right),
\end{eqnarray}
%------------------------------------------------------------------------
and
%------------------------------------------------------------------------
\begin{eqnarray}
\Li_n(x) = \sum_{k=1}^\infty \frac{x^k}{k^n},~~~|x| \leq 1.
\end{eqnarray}
%------------------------------------------------------------------------
The constants $\zeta_n,~~n \geq 2$, denote the Riemann $\zeta$ 
\cite{RIEMANN,TITCH1} function evaluated at integer argument $n$,
%------------------------------------------------------------------------
\begin{eqnarray}
\zeta_n =  \sum_{k=1}^\infty \frac{1}{k^n}.
\end{eqnarray}
%------------------------------------------------------------------------
The $SU(N_c)$ color factors are given by $\textcolor{blue}{C_A} = N_c, \textcolor{blue}{C_F} = (N_c^2-1)/(2 N_c), 
\textcolor{blue}{T_F} = 1/2$, with $N_c = 3$ for QCD. 

The contribution to the constant part\footnote{Here the same conditions as for 
Eq.~(\ref{eq:aunpol}) apply.} $\Delta a_{Qg}^{(3)}(N)$ of the polarized three--loop OME 
$\Delta A_{Qg}^{(3)}$ in the Larin scheme is given by
%------------------------------------------------------------------------
\input{a3pol}
%----------------------------------------------------------------------
and 
%------------------------------------------------------------------------
\begin{eqnarray}
\Delta p_{qg}^{(0)} = \frac{N-1}{N(N+1)}.
\end{eqnarray}
%----------------------------------------------------------------------
The first moment of $\Delta a_{Qg}^{(3)}$ vanishes for the computed color--$\zeta$ contributions,
as it does at first \cite{Watson:1981ce} and second order \cite{Buza:1996xr,Bierenbaum:2022biv}.
At first order, this is even true for general values of $m^2/Q^2$, 
cf.~\cite{Buza:1996xr,Bierenbaum:2022biv}.\footnote{For other OMEs and Wilson coefficients, as e.g.
in the polarized non--singlet case, the first moment (corresponding to the polarized Bjorken sum rule) is 
not vanishing \cite{Baikov:2010je,Mason:2023mud} and also 
obtains power corrections of $O((m^2/Q^2)^k)$ \cite{Blumlein:2016xcy}.}
The available parts of
$a_{Qg}^{(3)}$ and $\Delta a_{Qg}^{(3)}$ are expressed in terms of  the following set of 14 harmonic sums 
%---------------------------------------------------------------------------------
\begin{eqnarray}
\big\{S_{-4}, S_{-3}, S_{-2}, S_1, S_2, S_3,
S_4, S_{-3,1}, S_{-2,1}, S_{-2,2}, S_{2,1},
S_{3,1}, S_{-2,1,1}, S_{2,1,1}\big\}.
\end{eqnarray}
%---------------------------------------------------------------------------------
Correspondingly, the following set of 21 harmonic polylogarithms \cite{Remiddi:1999ew}
spans the expressions in $x$--space
%---------------------------------------------------------------------------------
\begin{eqnarray}
&& \big\{\HA_{-1}, 
\HA_0, 
\HA_1, 
\HA_{0,-1}, 
\HA_{0,1}, 
\HA_{0,-1,-1}, 
\HA_{0,-1,1},
\HA_{0,0,-1},
\HA_{0,0,1},
\HA_{0,1,-1},
\HA_{0,1,1},
\HA_{0,-1,-1,-1},
\HA_{0,-1,0,1},
\nonumber\\ &&
\HA_{0,0,-1,-1},
\HA_{0,0,-1,1},
\HA_{0,0,0,-1},
\HA_{0,0,0,1},
\HA_{0,0,1,-1}, \HA_{0,0,1,1}, \HA_{0,1,1,1}, \HA_{0,0,0,0,1}\big\},
\end{eqnarray}
%---------------------------------------------------------------------------------
after algebraic reduction \cite{Blumlein:2003gb}. The harmonic polylogarithms are defined by
%---------------------------------------------------------------------------------
\begin{eqnarray}
\HA_{b,\vec{a}}(x) = \int_0^x dy f_b(y) \HA_{\vec{a}}(y),~~f_b(y) \in \left\{\frac{1}{y}, \frac{1}{1-y}, 
\frac{1}{1+y} \right\}.
\end{eqnarray}
%---------------------------------------------------------------------------------
The corresponding expressions in $N$-- and $x$--space are given in ancillary files in 
computer readable form. 
%%%%%%%%%%%%%%%%%%%%%%%%%%%%%%%%%%%%%%%%%%%%%%%%%%%%%%%%%%%%%%%%%%%%%%%%%%%%%%%%%%%%%%%%%%%%%%%%%%%
\section{\boldmath The $N$-space structure of the remaining diagrams}
\label{sec:4}
%%%%%%%%%%%%%%%%%%%%%%%%%%%%%%%%%%%%%%%%%%%%%%%%%%%%%%%%%%%%%%%%%%%%%%%%%%%%%%%%%%%%%%%%%%%%%%%%%%%

\vspace*{1mm}
\noindent
In the following we summarize the results we have obtained in Mellin $N$--space, both for the 
cases of first--order factorizable and non--first--order factorizable recurrences. One possible 
strategy
to follow is to obtain closed form recursion relations for all color--$\zeta$ contributions
and to perform an analytic continuation to $N \in \mathbb{C}$. One has to derive the asymptotic expansion 
of these recurrences and  use their shift properties
%---------------------------------------------------------------------------------
\begin{eqnarray}
N + 1 \rightarrow N
\end{eqnarray}
%---------------------------------------------------------------------------------
to reach any point in the analyticity region for $N \in \mathbb{C}$. 
It finally turns out that it is very time--consuming to obtain the recurrences for the purely rational terms of 
$O(\textcolor{blue}{T_F})$, 
while those of $O(\textcolor{blue}{T_F} \zeta_3)$ still can be obtained on the basis of up to 15000 Mellin 
moments, see Table~\ref{tab:nfrec} below. In the polarized case a maximal number of 11000 Mellin moments has been 
computed.

Analyzing the sequences of Mellin moments for the purely rational terms
of the color factors $\textcolor{blue}{T_F C_F^2}$,
$\textcolor{blue}{T_F C_A^2}$, $\textcolor{blue}{T_F C_F C_A}$,
$\textcolor{blue}{T_F^2 C_F}$ and $\textcolor{blue}{T_F^2 C_A}$ one
observes that these contributions to $a_{Qg}^{(3)}$ and
$\Delta a_{Qg}^{(3)}$ 
%---------------------------------------------------------------------
\begin{center}
\begin{table}[H]\centering
\renewcommand*{\arraystretch}{1.4}
\begin{tabular}{|r|r|r|r|}
\hline
$c_i$  & 200 & 1000 &  2000\\
\hline
\textcolor{blue}{$C_F T_F^2$}   & $-5.89 \cdot 10^{-5}$  & $- 2.14 \cdot 10^{-43}$ & $-1.65 \cdot 10^{-93}$
\\ 
\textcolor{blue}{$C_A T_F^2$}   & $-1.38 \cdot 10^{-6}$  & $- 8.12 \cdot 10^{-45}$ & $-6.89 \cdot 10^{-95}$ 
\\ 
\textcolor{blue}{$C_F^2 T_F$}   & $1.96 \cdot 10^{-58}$  & $9.83 \cdot 10^{-299}$  & $1.41 \cdot 10^{-599}$
\\ 
\textcolor{blue}{$C_F C_A T_F$} & $9.84 \cdot 10^{-59}$ & $4.29 \cdot 10^{-299}$ & $5.93 \cdot 10^{-600}$ 
\\ 
\textcolor{blue}{$C_A^2 T_F$}   & $2.91 \cdot 10^{-58}$ & $1.16 \cdot 10^{-298}$ & $1.59 \cdot 10^{-599}$ 
\\
\hline
\hline
$\Delta c_i$  & 200 & 1000 &  2000\\
\hline
\textcolor{blue}{$C_F T_F^2$}   & $-5.53 \cdot 10^{-5}$  & $- 2.82 \cdot 10^{-43}$ & $-3.68 \cdot 10^{-93}$
\\ 
\textcolor{blue}{$C_A T_F^2$}   & $-1.06 \cdot 10^{-6}$  & $- 4.48 \cdot 10^{-43}$ & $-5.57 \cdot 10^{-93}$ 
\\ 
\textcolor{blue}{$C_F^2 T_F$}   & $-1.52 \cdot 10^{-57}$  & $-3.51 \cdot 10^{-297}$  & $9.90 \cdot 10^{-598}$
\\ 
\textcolor{blue}{$C_F C_A T_F$} & $-8.21 \cdot 10^{-57}$ & $-1.68 \cdot 10^{-296}$ & $-1.19 \cdot 10^{-597}$ 
\\ 
\textcolor{blue}{$C_A^2 T_F$}   & $2.71 \cdot 10^{-58}$ & $2.61 \cdot 10^{-294}$ & $1.14 \cdot 10^{-597}$ 
\\
\hline
\end{tabular}
\caption[]{\sf Relative approximation of the ratio of color factors, cf.~Eq.~(\ref{eq:RAT}),
as a function of $N = 200, 1000, 2000$ for $a_{Qg}^{(3)}$ and  $\Delta a_{Qg}^{(3)}$. 
The corresponding coefficients are $c_i$ and $\Delta c_i$.
\label{TAB1}}
\renewcommand*{\arraystretch}{1}
\end{table}
\end{center}
%---------------------------------------------------------------------
%---------------------------------------------------------------------
\begin{center}
\begin{table}[H]\centering
\renewcommand*{\arraystretch}{1.4}
\begin{tabular}{|r|r|r|r|}
\hline
$N$  & $a_{Qg}^{(3),\rm sol,irr}$ & $N$ & $\Delta a_{Qg}^{(3),\rm sol, irr}$ 
\\
\hline
2      & --201.6595414             &    &
\\ 
4      &  --1525.640364            &  3  &   --847.6187716   
\\ 
6      &  --1715.840721            &  5  &   --1460.511965      
\\ 
10     &  --1741.066914            &  9  &   --1687.025772           
\\ 
100    &  --966.5291789            &  99 &   --969.8344024
\\ 
200    &  --737.1136471            &  199 &  --738.5607476
\\ 
1000    &  --358.5858699           &  999 &  --358.7549068
\\ 
2000    &  --254.2324483           & 1999 &  --254.2957895             
\\
5000    &  --156.9872766           & 4999 &  --157.0039294    
\\
\hline
\end{tabular}
\caption[]{\sf Values of some moments of the irreducible 
contributions of the first--order factorizable diagrams $(\Delta) a_{Qg}^{(3)}(N)$ 
in QCD for $\textcolor{blue}{N_F} = 0$.\label{tab:over}}
\renewcommand*{\arraystretch}{1}
\end{table}
\end{center}
%---------------------------------------------------------------------
\noindent
individually diverge 
strongly for large values of
$N$. The coefficients of the same color factors with an additional factor of
$\zeta_3$ show the same behavior. The sum over the purely rational
terms and the $\zeta_3$ terms for each color factor separately, however, tends to
zero, i.e.
%---------------------------------------------------------------------
\begin{eqnarray}
\lim_{N \rightarrow \infty} \frac{r[c_i](N)}{r[c_i \zeta_3](N)} + \zeta_3 = 0. 
\label{eq:RAT}
\end{eqnarray}
%---------------------------------------------------------------------
Here $r[c_i]$ denotes the corresponding rational pre--factor of the color factors, which we illustrate in 
Table~\ref{TAB1}. 
Therefore, 
the respective recurrences are not independent. On the other hand, they cannot be easily joined in an exact manner,
but only approximately by rationalizing $\zeta_3$ with a high number of digits in the numerator and denominator.
One therefore would have to deal with diverging asymptotic representations, which have to be handled analytically.

In $N$--space individual terms rise with factors $2^N$ or $4^N$ and it is hard to see how these contributions cancel 
analytically. We therefore list a 
series of moments for the sum of the first--order factorizable diagrams to the irreducible 
contribution to $(\Delta) a_{Qg}^{(3)}(N)$ in the unpolarized and polarized cases, setting the 
known \textcolor{blue}{$N_F$}--terms to zero and the color factors to those of QCD in 
Table~\ref{tab:over}. 
They are first rising and then slowly falling towards $N \rightarrow \infty$, which 
suggests that intermediate contributions rising  $\propto 2^N$ or larger do finally cancel
in the set of the first--order factorizable terms.
Similar to what has been observed in Ref.~\cite{Ablinger:2022wbb} for $A_{gg}^{(3)}$, even the
values of the irreducible contributions of the first--order factorizable diagrams in the unpolarized and 
polarized cases  approach each other for large values of $N$. 
As will be shown in Section~\ref{sec:6}, both $a_{Qg}^{(3)}$ and $\Delta a_{Qg}^{(3)}$ tend to zero
as $N \rightarrow \infty$ for the first--order--factorizable contributions. In $x$--space the most singular 
contributions are $\propto \ln^k(1-x),~~k > 0, 
k \in \mathbb{N}$.
Because the $Qg$--channel is off--diagonal, no $\delta(1-x)$ and 
$\left[\ln^k(1-x)/(1-x)\right]_+$--distributions, with $k \geq 0$, 
will be present in $x$--space.

%---------------------------------------------------------------------
\begin{center}
\begin{table}[H]\centering
\renewcommand*{\arraystretch}{1.4}
\begin{tabular}{|r|l|r|r|r|r|r|}
\hline
\multicolumn{1}{|c}{Unpolarized} & 
\multicolumn{1}{|c}{Color/$\zeta$} & 
\multicolumn{1}{|c}{Moments} & 
\multicolumn{1}{|c}{Order} & 
\multicolumn{1}{|c}{Degree} & 
\multicolumn{1}{|c}{First order} &
\multicolumn{1}{|c|}{Size of rec.} \\
\multicolumn{1}{|c}{} & 
\multicolumn{1}{|c}{} & 
\multicolumn{1}{|c}{} & 
\multicolumn{1}{|c}{} & 
\multicolumn{1}{|c}{} & 
\multicolumn{1}{|c}{factors} & 
\multicolumn{1}{|c|}{[Mbyte]} \\
\hline
            &$\textcolor{blue}{C_F T_F^2}$& 3150& 27&  654& 15&  11.75                                 \\  
            &$\textcolor{blue}{C_A T_F^2}$& 9858& 46& 1407&   30 &  105.08                  \\  
            &$\textcolor{blue}{C_F T_F^2}   \zeta_3$& 1092& 15& 238&  7&  0.89           \\  
            &$\textcolor{blue}{C_A T_F^2}   \zeta_3$& 2156& 24& 447& 14& 5.52             \\  
            &$\textcolor{blue}{C_F^2 T_F}   \zeta_3$&  9858&  58& 2024&     
& 304.79                                   \\  
            &$\textcolor{blue}{C_F C_A T_F} \zeta_3$& 12826& 65& 2602& &  563.50                             
\\  
            &$\textcolor{blue}{C_A^2 T_F}   \zeta_3$& 14036& 68 & 2848&  & 709.63                            
\\  
\hline
\multicolumn{1}{|c}{Polarized} & 
\multicolumn{1}{|c}{} & 
\multicolumn{1}{|c}{} & 
\multicolumn{1}{|c}{} & 
\multicolumn{1}{|c}{} & 
\multicolumn{1}{|c}{} & 
\multicolumn{1}{|c|}{} \\
\hline
            &$\textcolor{blue}{C_F T_F^2}$& 1395&  18& 279&           9  &   1.69  \\  
%            &$\textcolor{blue}{C_A T_F^2}$&   ?&  ?& ?&                  &  \\  
            &$\textcolor{blue}{C_F T_F^2}   \zeta_3$& 480& 10& 104&   4  &   0.15  \\  
            &$\textcolor{blue}{C_A T_F^2}   \zeta_3$& 1702& 21& 352&  11 &   3.46   \\  
            &$\textcolor{blue}{C_F^2 T_F}   \zeta_3$& 8787&  55& 1803&   & 233.36 \\  
            &$\textcolor{blue}{C_F C_A T_F} \zeta_3$& 10340& 60& 2146&   & 363.35             \\  
%            &$\textcolor{blue}{C_A^2 T_F}   \zeta_3$& & & &              &  \\  
\hline
\end{tabular}
\caption[]{\sf Characteristics of non--first--order factorizable recurrences in the unpolarized 
and
polarized cases, by the required number of moments, their order and degree, their first--order 
factors, and their size.
\label{tab:nfrec}}
\renewcommand*{\arraystretch}{1}
\end{table}
\end{center}
%---------------------------------------------------------------------
A further problem is given by the fact that the asymptotic representation is not easily
obtainable to a sufficient number of terms for non-first-order factorizable recurrences if the recurrences are
very large.
On the other hand, in the first--order factorizable cases, analytic techniques are available to compute the 
asymptotic representations as will be outlined in Appendices~\ref{sec:A} and \ref{sec:B}. While for 
non--first--order factorizable recurrences their first--order factors can all be split off, the 
respective factors of higher than first--order cannot 
be algorithmically 
determined yet. Therefore one is left with  some recurrences of a larger order, the final solution 
of which is not given 
by product--sum representations, but by higher transcendental functions to be determined.\footnote{There 
exist only very few studies for solutions of this kind, cf.~\cite{MP}.} In Table~\ref{tab:nfrec} we 
summarize the characteristics of the cases of non--first--order factorizable recurrences we have 
computed.

To illustrate the complexity of the problem, the largest rational number in the input for the 
determination of the recurrences has a size of 31k digits in the numerator and of 26.6k in 
the denominator. The largest recursion obtained has a size of $\sim 0.7$~GB.

The first--order factors are split off by using algorithms of {\tt Sigma} \cite{SIG1,SIG2} 
leaving
a non--first--order factorizable remainder. For very large recurrences this process can take
several months of computation time, and we did not perform this computation in these cases, since
a very large remainder recurrence is obtained for which the analytic solution cannot be given
at present.

One may consider asymptotic solutions of the non--first--order factorizable difference equations 
following 
Refs.~\cite{BIRK1,BIRK2,WIMP85,Kauers}. For the color factors $\textcolor{blue}{C_F T_F^2}\zeta_3$ related 
to the smallest corresponding recurrences we calculated the fundamental system of order {\sf o = 15} 
and {\sf o = 10} in the unpolarized and polarized case, respectively. These systems are computed by the 
{\tt HarmonicSums} command  {\tt REAsymptotics[rec,f[n],7]},\footnote{Commands described here and in the 
following refer to the package {\tt HarmonicSums}, unless denoted otherwise.} 
where {\tt f[n]} is the 
function obeying the homogeneous recurrence {\tt rec}
and 7 is the desired expansion depth. The systems of the asymptotic solutions are given by
%---------------------------------------------------------------------
\begin{eqnarray}
\lefteqn{\left\{\left. T_k^{C_F T_F^2 
\zeta_3}\right|_{k=1}^{15}\right\} = 
} \nonumber\\ && 
\Biggl\{\Biggl[
 \frac{1}{N^4}
+\frac{174}{197 N^5}
-\frac{115915}{197 N^6}
-\frac{10928670}{197 N^7}
\Biggr]
\Biggl(-\frac{9}{8}\Biggr)^N, 
\frac{(-1)^N}{N},
\nonumber\\ &&
\Biggl[
-\frac{158191326}{19272263 N^5}
+\frac{3264014438}{57816789 N^6}
        -\frac{449608338428}{1561053303 N^7}
+\frac{1}{N^2}\Biggr] 
(-1)^N,
\nonumber\\ &&
\Biggl[
 \frac{1}{N^3}
-\frac{321703313}{19272263 N^5}
+\frac{7167720182}{57816789 N^6}
        -\frac{1124854321331}{1561053303 N^7}
\Biggr](-1)^N,
\nonumber\\ &&
\Biggl[
 \frac{1}{N^4}
-\frac{156571794}{19272263 N^5}
+\frac{2861638081}{57816789 N^6}
        -\frac{431900140522}{1561053303 N^7}
\Biggr] 
(-1)^N,
\nonumber\\ &&
\Biggl[
        -\frac{3486911695}{77089052 N^5}
        +\frac{3980355789289}{6938014680 N^6}
        -\frac{859592977355719}{187326396360 N^7}
\nonumber\\ && 
        +\Biggl(
 \frac{1}{N^2}
-\frac{3}{N^3}
+\frac{7}{N^4}
-\frac{15}{N^5}
+\frac{31}{N^6}
                -\frac{63}{N^7}
\Biggr) \ln(N)
\Biggr] (-1)^N,
\nonumber\\ && 
\Biggl[
        -\frac{1}{10}
        +\frac{1260823915355237}{1046073508500 N^6}
        -\frac{24450999262196856571}{296561839659750 N^7}
        + \Biggl(
 \frac{1}{N}
-\frac{370565089177}{34869116950 N^2}
\nonumber\\ &&  
+\frac{1286040852281}{34869116950 N^3}
-\frac{7299389618221}{52303675425 N^4}
+\frac{1795444741338}{3486911695 N^5}
-\frac{124663841290259}{69738233900 N^6}
\nonumber\\ && 
             +   \frac{412778573182657}{69738233900 N^7}
\Biggr) \ln(N)
        +\Biggl(
 \frac{1}{N^2}
-\frac{3}{N^3}
+\frac{7}{N^4}
-\frac{15}{N^5}
+\frac{31}{N^6}
                -\frac{63}{N^7}
\Biggr) \ln^2(N)
\Biggr] (-1)^N,
\nonumber\\ && 
\Biggl[
 \frac{1}{N^3}
+\frac{162553}{18054 N^4}
+\frac{388414}{9027 N^5}
+\frac{29532113}{162486 N^6}
        + \frac{571451720}{731187 N^7}
\Biggr]
\Biggl(-\frac{1}{8}\Biggr)^N,
\nonumber\\ && 
\Biggl[
 \frac{1}{N^3}
+\frac{191225}{26118 N^4}
+\frac{147338}{4353 N^5}
+\frac{31187665}{235062 N^6}
       + \frac{591255400}{1057779 N^7}
\Biggr] \left(\frac{1}{8}\right)^{N},
\nonumber\\ && 
 \frac{1}{N}
+\frac{10178885}{237654 N^4}
-\frac{42829585}{118827 N^5}
+\frac{1605361621}{712962 N^6}
-\frac{122618276090}{9624987 N^7},
\nonumber\\ && 
 \frac{1}{N^2}
+\frac{46443703}{475308 N^4}
-\frac{197209085}{237654 N^5}
+\frac{7403011091}{1425924 N^6}
-\frac{282535272647}{9624987 N^7},
\nonumber\\ && 
 \frac{1}{N^3}
+\frac{2528753}{158436 N^4}
-\frac{11998345}{79218 N^5}
+\frac{460898245}{475308 N^6}
-\frac{17701180264}{3208329 N^7}
,
\nonumber\\ && 
 \frac{76310107}{633744 N^4}
-\frac{845948561}{792180 N^5}
+\frac{66371498737}{9506160 N^6}
-\frac{36692791628777}{898332120 N^7}
\nonumber\\ && 
+  \Biggl(
 \frac{1}{N}
-\frac{1}{N^2}
+\frac{3}{N^3}
-\frac{7}{N^4}
+\frac{15}{N^5}
-\frac{31}{N^6}
        +\frac{63}{N^7}
\Biggr) \ln(N),
\nonumber\\ && 
-\frac{1}{10}
-\frac{551687998205741}{824149155600 N^5}
+\frac{7787774926751531}{824149155600 N^6}
-\frac{11810146290512240171}{77882095204200 N^7}
\nonumber\\ && 
+ \Biggl( 
 \frac{2317251527}{3433954815 N}
-\frac{19487025602}{3433954815 N^2}
+\frac{29435287829}{2289303210 N^3}
-\frac{546390092023}{6867909630 N^4}
\nonumber\\ && 
+\frac{1751952232097}{4578606420 N^5}
-\frac{20751649513859}{13735819260 N^6}
   +     \frac{514700153390909}{96150734820 N^7}
\Biggr) \ln(N)
+ \Biggl(
 \frac{1}{N}
-\frac{1}{N^2}
\nonumber\\ && 
-\frac{7}{N^4}
+\frac{15}{N^5}
+\frac{3}{N^3}
-\frac{31}{N^6}
      +  \frac{63}{N^7}
\Biggr) \ln^2(N),
\nonumber\\ && 
\Biggl[
\frac{1}{N^4}
+\frac{11886}{325 N^5}
+\frac{111473}{65 N^6}
        +\frac{27764898}{325 N^7}
\Biggr] \Biggl(\frac{9}{8}\Biggr)^N 
\Biggr\}  
+ O\left(\left(\frac{9}{8}\right)^N \frac{1}{N^8}\right),
\\
%-------------------------------------------------------------------
\lefteqn{\left\{\left.\Delta T_k^{C_F T_F^2 
\zeta_3}\right|_{k=1}^{10}\right\} = } \nonumber\\ && 
\Biggl\{
\left(-\frac{9}{8}\right)^N \Biggl[
 \frac{1}{N^4} 
+ \frac{183}{4 N^5} 
+ \frac{2123}{2 N^6} 
+ \frac{369303}{4 N^7} 
\Biggr],
\nonumber\\ &&
\Biggl[
 \frac{1}{N}
-\frac{6167375}{2538 N^4}
+\frac{251053021}{2538 N^5}
-\frac{6845892761}{2538 N^6}
        +\frac{269421699821}{4374 N^7}
\Biggr] (-1)^N,
\nonumber\\ &&
\Biggl[
 \frac{1}{N^2}
-\frac{1377791}{2538 N^4}
+\frac{49777453}{2538 N^5}
-\frac{1270724105}{2538 N^6}
        + \frac{47946548525}{4374 N^7}
\Biggr] (-1)^N,
\nonumber\\ && 
\Biggl[
 \frac{1}{N^3}
-\frac{205031}{5076 N^4}
+\frac{5548237}{5076 N^5}
-\frac{125746313}{5076 N^6}
        +\frac{4443239285}{8748 N^7}
\Biggr] 
(-1)^N,
\nonumber\\ && 
\Biggl[
        \frac{44549521}{60912 N^4}
        -\frac{9803624797}{304560 N^5}
        +\frac{281386223183}{304560 N^6}
        -\frac{80403763171913}{3674160 N^7}
\nonumber\\ && 
        + \Biggl(
\frac{1}{N}
-\frac{14}{N^2}
+\frac{194}{N^3}
-\frac{2666}{N^4}
+\frac{36386}{N^5}
-\frac{493754}{N^6}
+ \frac{6667634}{N^7}
\Biggr) \ln(N)
\Biggr] (-1)^N,
\nonumber\\ && 
\Biggl[
         \frac{176076128920087}{32075655120 N^5}
        -\frac{2170077025499773}{8018913780 N^6}
        +\frac{153855500975294933063}{18186896453040 N^7}
\nonumber\\ &&       
  + \Biggl(
-\frac{96551923}{44549521 N}
+\frac{2198167821}{44549521 N^2}
-\frac{205984981993}{267297126 N^3}
+\frac{1479535025071}{133648563 N^4}
\nonumber\\ && 
-\frac{409866696159859}{2672971260 N^5}
+\frac{926453767962681}{445495210 N^6}
                -\frac{390121843311426469}{14033099115 N^7}
\Biggr) \ln(N)
\nonumber\\ && 
        + \Biggl(
 \frac{1}{N}
-\frac{14}{N^2}
+\frac{194}{N^3}
-\frac{2666}{N^4}
+\frac{36386}{N^5}
-\frac{493754}{N^6}
+\frac{6667634}{N^7}
\Biggr) \ln^2(N)
\Biggr] (-1)^N,
\nonumber\\ && 
\Biggl[
        \frac{263556904}{8019 N^7}
-\frac{5053141}{891 N^6}
+\frac{47194}{99 N^5}
-\frac{2765}{99 N^4}
+\frac{1}{N^3}
\Biggr] \left(-\frac{1}{8}\right)^N,
\nonumber\\ && 
\Biggl[
        -\frac{299001079}{729 N^7}
+\frac{1211485}{81 N^6}
-\frac{2521}{9 N^5}
-\frac{97}{9 N^4}
+\frac{1}{N^3}\Biggr] \left(\frac{1}{8}\right)^N,
\nonumber\\ &&
 \frac{1}{N}
-\frac{403557}{1346 N^3}
+\frac{1066131187}{109026 N^4}
-\frac{25497634607}{109026 N^5}
+\frac{1600616428213}{327078 N^6}
\nonumber\\ && 
-\frac{838217215706339}{8831106 N^7},
\nonumber\\ &&
 \frac{1}{N^2}
-\frac{664689}{18844 N^3}
+\frac{1356819775}{1526364 N^4}
-\frac{29465309123}{1526364 N^5}
+\frac{1762154712985}{4579092 N^6}
\nonumber\\ && 
-\frac{897121773060599}{123635484 N^7}
\Biggr\} 
+ O\left(\left(\frac{9}{8}\right)^N \frac{1}{N^8}\right).
\end{eqnarray} 
%---------------------------------------------------------------------
More efforts are needed to compute the respective systems for the larger recurrences. In a final numerical step
one has to combine these solutions. This combination is not unique, as the result is necessarily 
approximate.
Details on this will be given in a later publication.
Here the problem is also that a series of particular solutions strongly diverges as 
$N \rightarrow \infty$. These contributions cancel against contributions in other color--$\zeta$ 
factors as outlined above.

Let us now turn back to the first--order factorizable contributions and consider
the Feynman diagrams which are solely determined by these. 
Still one may use the techniques available for first--order factorizable problems for individual 
Feynman diagrams 
contributing to the sets of color--$\zeta$ factors not yet being covered by the results in 
Section~\ref{sec:3}.
The final strategy is then to transform these results to $x$--space where also the non--first--order factorizable 
contributions will be solved, cf.~\cite{AQGFIN}. Working on a diagram--by--diagram basis we have obtained 
the $N$--space 
solutions for the first--order factorizable cases, which we will discuss now.

The results are given in terms of  generalized harmonic sums \cite{Borwein:1999js,Moch:2001zr,Ablinger:2013cf}, cyclotomic sums 
\cite{Ablinger:2011te}, and finite (inverse) binomial sums \cite{Ablinger:2014bra} in form of polynomials over 
$\mathbb{Q}(N)$ in Mellin $N$--space, beyond the harmonic sums. The cyclotomic sums can be shown to be reducible to 
harmonic sums in all contributing Feynman diagrams. 
The generalized harmonic sums are defined by 
%---------------------------------------------------------------------
\begin{eqnarray}
S_{b,\vec{a}}(\{c,\vec{d}\},N) = \sum_{k=1}^N 
\frac{c^k}{k^b}  S_{\vec{a}}(\{\vec{d}\},k),~~~b, a_i \in \mathbb{N} \backslash \{0\},
~~~c, d_i \in \mathbb{Z} \backslash \{0\}.
\label{eq:gens}
\end{eqnarray}
%---------------------------------------------------------------------
All these quantities obey first--order shift 
relations in a hierarchy of terms, such as
%---------------------------------------------------------------------
\begin{eqnarray}
S_{b,\vec{a}}(\{c,\vec{d}\},N+1) - S_{b,\vec{a}}(\{c,\vec{d}\},N) = \frac{c^N}{N^b} 
S_{\vec{a}}(\{\vec{d}\},N)
\label{eq:rec1}
\end{eqnarray}
%---------------------------------------------------------------------
for the harmonic and generalized harmonic sums and synonymous relations for nested (inverse) binomial sums. 
Their recurrences are of the type
%--------------------------------------------------------------------------------------------------
\begin{eqnarray}
{\sf BS}_1(N) &=& \sum_{k=1}^N f(k),
\\
{\sf BS}_2(N) &=& \sum_{k=1}^N g(k) {\sf BS_3}(k),
\end{eqnarray}
%--------------------------------------------------------------------------------------------------
and one obtains
%--------------------------------------------------------------------------------------------------
\begin{eqnarray}
{\sf BS}_1(N) - {\sf BS}_1(N-1) &=& f(N),
\label{eq:rec2}
\\
{\sf BS}_2(N) - {\sf BS}_2(N-1) &=& g(N) {\sf BS_3}(N).
\label{eq:rec3}
\end{eqnarray}
%--------------------------------------------------------------------------------------------------
Using Eqs.~(\ref{eq:rec1}, \ref{eq:rec2}, \ref{eq:rec3}) the 
respective outermost sum is removed. What remains is to provide these sums in the asymptotic region $N \in 
\mathbb{C},~~|N| \rightarrow \infty$.

The following sums contribute to the first--order factorizable contributions. In the linear 
representation 
these are 72 harmonic sums up to weight {\sf w = 5}, while the following 33 harmonic sums remain after 
algebraic reduction \cite{Blumlein:2003gb},
%-------------------------------------------------------------------------------------------------
\begin{eqnarray}
&& \Biggl\{
S_{-1},
S_1,
S_{-2},
S_2,
S_{-3},
S_3,
S_{-4},
S_4,
S_{-5},
S_5,
S_{-2,-1},
S_{-2,1},
S_{2,-1},
S_{2,1},
S_{-2,2},
S_{-2,-3},
S_{-2,3},
S_{2,-3},
\nonumber\\ &&
S_{2,3},
S_{-3,1},
S_{3,1},
S_{-4,1},
S_{4,1},
S_{2,1,1},
S_{-2,1,-2},
S_{-2,1,1},
S_{-2,2,1},
S_{2,1,-2},
S_{2,2,1},
S_{-3,1,1},
S_{3,1,1},
S_{-2,1,1,1},
\nonumber\\ &&
S_{2,1,1,1}
\Biggr\}.
\end{eqnarray}
%-------------------------------------------------------------------------------------------------
Their asymptotic representations have been given in Ref.~\cite{Blumlein:2009ta}. 

Furthermore, the following 45 generalized harmonic sums 
%-------------------------------------------------------------------------------------------------
\begin{eqnarray}
&&\Biggl\{
S_1(\{-2\}),
S_1\left(\left\{\frac{1}{2}\right\}\right),
S_1(\{2\}),
S_2(\{-2\}),
S_2\left(\left\{\frac{1}{2}\right\}\right),
S_3(\{2\}),
S_4(\{2\}),
S_{1,1}\left(\left\{\frac{1}{2},2\right\}\right),
\nonumber\\ &&
S_{1,1}\left(\left\{1,\frac{1}{2}\right\}\right),
S_{1,1}\left(\{1,2\}\right),
S_{1,2}\left(\left\{\frac{1}{2},-1\right\}\right),
S_{1,2}\left(\left\{\frac{1}{2},1\right\}\right), 
S_{1,2}(\{2,1\}),
S_{1,3}\left(\left\{\frac{1}{2},2\right\}\right),
\nonumber\\ &&
S_{1,3}(\{2,1\}),
S_{1,4}\left(\left\{\frac{1}{2},2\right\}\right),
S_{2,1}(\{2,1\}),
S_{2,3}\left(\left\{\frac{1}{2},2\right\}\right),
S_{1,1,1}\left(\left\{\frac{1}{2},1,1\right\}\right),
S_{1,1,1}(\{2,1,1\}),
\nonumber\\ &&
S_{1,1,2}\left(\left\{-2,\frac{1}{2},-1\right\}\right),
S_{1,1,2}\left(\left\{-2,\frac{1}{2},1\right\}\right),
S_{1,1,2}\left(\left\{\frac{1}{2},2,1\right\}\right),
S_{1,1,2}\left(\left\{2,\frac{1}{2},-1\right\}\right),
\nonumber\\ && 
S_{1,1,2}\left(\left\{2,\frac{1}{2},1\right\}\right),
S_{1,1,3}\left(\left\{\frac{1}{2},2,1\right\}\right),
S_{1,1,3}\left(\left\{1,\frac{1}{2},2\right\}\right),
S_{1,2,1}\left(\left\{\frac{1}{2},2,1\right\}\right),
\nonumber\\ &&
S_{2,1,1}(\{2,1,1\}),
S_{2,1,2}\left(\left\{-2,\frac{1}{2},-1\right\}\right),
S_{2,1,2}\left(\left\{-2,\frac{1}{2},1\right\}\right),
S_{2,1,2}\left(\left\{\frac{1}{2},2,1\right\}\right),
\nonumber\\ &&
S_{2,2,1}\left(\left\{\frac{1}{2},2,1\right\}\right),
S_{1,1,1,1}\left(\left\{\frac{1}{2},2,1,1\right\}\right),
S_{1,1,1,1}\left(\left\{2,\frac{1}{2},1,1\right\}\right),
S_{1,1,1,1}(\{2,1,1,1\}),
\nonumber\\ &&
S_{1,1,1,2}\left(\left\{1,\frac{1}{2},2,1\right\}\right),
S_{1,1,1,2}\left(\left\{1,2,\frac{1}{2},-1\right\}\right),
S_{1,1,1,2}\left(\left\{1,2,\frac{1}{2},1\right\}\right),
\nonumber\\ &&
S_{1,1,2,1}\left(\left\{1,\frac{1}{2},2,1\right\}\right),
S_{1,2,1,1}\left(\left\{\frac{1}{2},2,1,1\right\}\right),
S_{2,1,1,1}\left(\left\{\frac{1}{2},2,1,1\right\}\right),
\nonumber\\ &&
S_{1,1,1,1,1}\left(\left\{\frac{1}{2},2,1,1,1\right\}\right),
S_{1,1,1,1,1}\left(\left\{1,\frac{1}{2},2,1,1\right\}\right),
S_{1,1,1,1,1}\left(\left\{1,2,\frac{1}{2},1,1\right\}\right)
\Biggr\}
\end{eqnarray}
%-------------------------------------------------------------------------------------------------
contribute. 

The asymptotic expansion of these sums can be performed by using the {\tt HarmonicSums}
commands
{\tt SExpansion} and {\tt BSExpansion}, respectively. In course of this the following additional 
17 sums emerge, which have to be dealt with in the same way
%-------------------------------------------------------------------------------------------------
\begin{eqnarray}
&& \Biggl\{
S_2(\{2\}),
S_{1,1}(\{2,1\}),
S_{1,1,3}\left(\left\{\frac{1}{2},1,2\right\}\right),
S_{1,2,2}\left(\left\{\frac{1}{2},2,1\right\}\right),
S_{1,3,1}\left(\left\{\frac{1}{2},2,1\right\}\right),
\nonumber\\ && 
S_{1,1,1,2}\left(\left\{\frac{1}{2},2,1,1\right\}\right),
S_{1,1,1,2}\left(\left\{2,1,1,\frac{1}{2}\right\}\right),
S_{1,1,2,1}\left(\left\{\frac{1}{2},2,1,1\right\}\right),
\nonumber\\ && 
S_{1,1,1,1,1}\left(\left\{\frac{1}{2},1,2,1,1\right\}\right),
S_{2,2}(\{2,1\}),
S_{1,3}(\{1,2\}),
S_{1,1,2}(\{1,2,1\}),
S_{1,1,1,1}(\{2,1,1,1\}),
\nonumber\\ && 
S_{2,1}(\{2,1\}),
S_{1,1,1,2}\left(\left\{1,\frac{1}{2},2,1\right\}\right),
S_{1,1,1,2}\left(\left\{\frac{1}{2},1,2,1\right\}\right),
S_{1,1,1,2}\left(\left\{1,\frac{1}{2},2,1\right\}\right)
\Biggr\}.
\nonumber\\
\end{eqnarray}
%-------------------------------------------------------------------------------------------------
In some cases a certain generalized harmonic sum has to be re--shuffled before by using 
{\tt SRemoveLeadingIndex},
{\tt SRemoveTrailingIndex}, or by using more general shuffling relations.
The set of constants, which are multiple zeta values in the case of harmonic sums \cite{Blumlein:2009cf},
is now extended to those of generalized harmonic sums at infinity with the additional numerator weights
%-------------------------------------------------------------------------------------------------
\begin{eqnarray}
\Biggl\{-\frac{1}{2}, -2, \frac{1}{2}, 2\Biggr\}.
\end{eqnarray}
%-------------------------------------------------------------------------------------------------
One may map these constants to G--functions at argument $x=1$ or the associated generalized harmonic
polylogarithmic constants \cite{Ablinger:2013cf}.  Here one applies first 
the command {\tt GLRemovePole[fct,a]} with $a \in [0,1]$ the pole positions, to obtain 
the Cauchy principal value of the respective integrals.
%-------------------------------------------------------------------------------------------------
Examples for constants even reducing to multiple zeta values \cite{Blumlein:2009cf} are
%-------------------------------------------------------------------------------------------------
\begin{eqnarray}
\HA_{0, 0, 0, 1/2}(1) &=& 
- \Li_4\left(\frac{1}{2}\right) - \frac{1}{24} \ln^4(2) + \ln^2(2) \zeta_2 + \frac{4}{5} \zeta_2^2,
\\
\HA_{1/2, 0, 1, 1, 1}(1) &=& \Li_5\left(\frac{1}{2}\right)
- \frac{1}{120} \ln^5(2) + \frac{1}{3} \ln^3(2) \zeta_2 + \frac{4}{5} \ln(2) \zeta^2_2
+ \frac{21}{16} \zeta_2 \zeta_3 
\nonumber\\ &&
- \frac{155}{32} \zeta_5.
\end{eqnarray}
%-------------------------------------------------------------------------------------------------
The generalized harmonic polylogarithms are defined by
%-------------------------------------------------------------------------------------------------
\begin{eqnarray}
\HA_{a,\vec{b}}(x) = \int_0^x dy f_a(y) \HA_{\vec{b}}(y),~~~\text{with}~~~f_a(y) = \frac{1}{y-a}.
\end{eqnarray}
%-------------------------------------------------------------------------------------------------

One  example of an asymptotic expansion of a contributing generalized harmonic 
sum occurring in the calculation of $a_{Qg}^{(3)}$ is given by 
%-------------------------------------------------------------------------------------------------
\begin{eqnarray}
T_1 &=& 
-\frac{32 \textcolor{blue}{C_A} \textcolor{blue}{T_F} (\textcolor{blue}{C_A}
-2 \textcolor{blue}{C_F}) (N-4)}{N (1+N) (2+N)}
S_{2,1,1,1}\left(\left\{\frac{1}{2},2,1,1\right\},N\right)
\nonumber\\ &=&
\textcolor{blue}{C_A T_F} (\textcolor{blue}{C_A} - 2 \textcolor{blue}{C_F})
\Biggl\{
        \Li_5\left(\frac{1}{2}\right) \Biggl[
 \frac{64}{N^2}
-\frac{448}{N^3}
+\frac{1216}{N^4}
-\frac{2752}{N^5}
+\frac{5824}{N^6}
-\frac{11968}{N^7}
\Biggr]
\nonumber\\ &&
        +\frac{8}{N^4}
        -\frac{1832}{27 N^5}
        +\frac{6632}{27 N^6}
        -\frac{2135236}{3375 N^7}
        +\ln^5(2) \Biggl[
-\frac{8}{15 N^2}
+\frac{56}{15 N^3}
-\frac{152}{15 N^4}
+\frac{344}{15 N^5}
\nonumber\\ &&
-\frac{728}{15 N^6}
                +\frac{1496}{15 N^7}
\Biggr]
        +\ln^3(2) \Biggl[
\frac{88}{3 N^2}
-\frac{616}{3 N^3}
+\frac{1672}{3 N^4}
-\frac{3784}{3 N^5}
+\frac{8008}{3 N^6}
                -\frac{16456}{3 N^7}
\Biggr] \zeta_2
\nonumber\\ &&
        +\Biggl[
                 \frac{16}{N^4}
                -\frac{352}{3 N^5}
                +\frac{1072}{3 N^6}
                -\frac{12832}{15 N^7}
                +\Biggl[
-\frac{14}{N^2}
+\frac{98}{N^3}
-\frac{266}{N^4}
+\frac{602}{N^5}
-\frac{1274}{N^6}
+\frac{2618}{N^7}
\Biggr]
\zeta_3
        \Biggr] \zeta_2
\nonumber\\ &&
        +\ln(2) \Biggl[
 \frac{196}{5 N^2}
-\frac{1372}{5 N^3}
+\frac{3724}{5 N^4}
-\frac{8428}{5 N^5}
+\frac{17836}{5 N^6}
                -\frac{36652}{5 N^7}
\Biggr] \zeta_2^2
        +\ln^2(2) \Biggl[
-\frac{14}{N^2}
+\frac{98}{N^3}
\nonumber\\ &&
-\frac{266}{N^4}
+\frac{602}{N^5}
-\frac{1274}{N^6}
+\frac{2618}{N^7}
\Biggr] \zeta_3
        +\Biggl[
-\frac{279}{2 N^2}
+\frac{1953}{2 N^3}
-\frac{5301}{2 N^4}
+\frac{11997}{2 N^5}
-\frac{25389}{2 N^6}
\nonumber\\ &&
                +\frac{52173}{2 N^7}
\Biggr] \zeta_5
        +\Biggl[
 \frac{16}{N^4}
-\frac{1040}{9 N^5}
+\frac{2648}{9 N^6}
                -\frac{142112}{225 N^7}
\Biggr] 
L
        +\Biggl[
\frac{16}{N^4}
-\frac{352}{3 N^5}
+\frac{1072}{3 N^6}
\nonumber\\ &&          
      -\frac{12832}{15 N^7}
\Biggr] L^2
\Biggr\} + O\left(\frac{1}{N^8}\right).
\end{eqnarray}
%-------------------------------------------------------------------------------------------------
with
%------------------------------------------------------------------------------------------------- 
\begin{eqnarray}
L = \ln(N) + \gamma_E,
\end{eqnarray}
%------------------------------------------------------------------------------------------------- 
and $\gamma_E$ the Euler--Mascheroni constant.
The asymptotic expansions of the contributing generalized harmonic sums are discussed in
Appendix~\ref{sec:A} and are given in an ancillary file in computer--readable form. Here also 
generalized harmonic polylogarithms 
beyond multiple zeta values contribute.

Now we turn to the remaining sums, which are nested binomial or inverse binomial sums. We derive linear 
representations and eliminate algebraic relations between the binomial sums. Moreover, they are reduced 
to a standard form removing summation index shifts. By these operations also sums of lower kind are 
generated. The following 58 sums contribute
%-------------------------------------------------------------------------------------------------
\begin{eqnarray}
&& \Biggl\{
%--1
\sum_{\tau_1=1}^N 
\frac{\big(
        \tau_1!\big)^2}{\big(
        2 \tau_1\big)!},
%--2
\sum_{\tau_1=1}^N \frac{\big(
        2 \tau_1\big)!}{\big(
        \tau_1!\big)^2},
%--3
\sum_{\tau_1=1}^N \frac{\big(
        \tau_1!\big)^2 
\sum_{\tau_2=1}^{\tau_1} \frac{\big(
        2 \tau_2\big)!}{\big(
        \tau_2!\big)^2}}{\big(
        2 \tau_1\big)!},
%--4
\sum_{\tau_1=1}^N \frac{\big(
        \tau_1!\big)^2 
\sum_{\tau_2=1}^{\tau_1} \frac{(-1)^{\tau_2} \big(
        2 \tau_2\big)!}{\big(
        \tau_2!\big)^2 \tau_2^3}}{\big(
        2 \tau_1\big)!},
\nonumber\\ &&
%--5
\sum_{\tau_1=1}^N \frac{\big(
        \tau_1!\big)^2 
\sum_{\tau_2=1}^{\tau_1} \frac{\big(
        2 \tau_2\big)! 
\sum_{\tau_3=1}^{\tau_2} \frac{1}{\tau_3}}{\big(
        \tau_2!\big)^2 \tau_2^2}}{\big(
        2 \tau_1\big)!},
%--6
\sum_{\tau_1=1}^N \frac{\big(
        \tau_1!\big)^2 
\sum_{\tau_2=1}^{\tau_1} \frac{\big(
        2 \tau_2\big)!}{\big(
        \tau_2!\big)^2 \tau_2}}{\big(
        2 \tau_1\big)!},
%--7
\sum_{\tau_1=1}^N \frac{\big(
        \tau_1!\big)^2 
\sum_{\tau_2=1}^{\tau_1} \frac{(-1)^{\tau_2} \big(
        2 \tau_2\big)!}{\big(
        \tau_2!\big)^2 \tau_2}}{\big(
        2 \tau_1\big)!},
\nonumber\\ &&
%--8
\sum_{\tau_1=1}^N \frac{\big(
        \tau_1!\big)^2 
\sum_{\tau_2=1}^{\tau_1} \frac{\big(
        2 \tau_2\big)! 
\sum_{\tau_3=1}
^{\tau_2} 
\frac{1}{\tau_3^2}}{\big(
        \tau_2!\big)^2 \tau_2}}{\big(
        2 \tau_1\big)!},
%--9
\sum_{\tau_1=1}^N \frac{\big(
        \tau_1!\big)^2 
\sum_{\tau_2=1}^{\tau_1} \frac{(-1)^{\tau_2} \big(
        2 \tau_2\big)! 
\sum_{\tau_3=1}^{\tau_2} \frac{1}{\tau_3^2}}{\big(
        \tau_2!\big)^2 \tau_2}}{\big(
        2 \tau_1\big)!},
\nonumber\\ &&
%--10
\sum_{\tau_1=1}^N \frac{\big(
        \tau_1!\big)^2 
\sum_{\tau_2=1}^{\tau_1} \frac{\big(
        2 \tau_2\big)! 
\sum_{\tau_3=1}^{\tau_2} \frac{(-1)^{\tau_3}}{\tau_3^2}}{\big(
        \tau_2!\big)^2 \tau_2}}{\big(
        2 \tau_1\big)!},
%--11
\sum_{\tau_1=1}^N \frac{\big(
        \tau_1!\big)^2 
\sum_{\tau_2=1}^{\tau_1} \frac{(-1)^{\tau_2} \big(
        2 \tau_2\big)! 
\sum_{\tau_3=1}^{\tau_2} \frac{(-1)^{\tau_3}}{\tau_3^2}}{\big(
        \tau_2!\big)^2 \tau_2}}{\big(
        2 \tau_1\big)!},
\nonumber\\ &&
%--12
\sum_{\tau_1=1}^N \frac{\big(
        \tau_1!\big)^2 
\sum_{\tau_2=1}^{\tau_1} \frac{\big(
        2 \tau_2\big)! 
\sum_{\tau_3=1}^{\tau_2} \frac{
\sum_{\tau_4=1}^{\tau_3} \frac{1}{\tau_4}}{\tau_3}}{\big(
        \tau_2!\big)^2 \tau_2}}{\big(
        2 \tau_1\big)!},
%--13
\sum_{\tau_1=1}^N \frac{\big(
        \tau_1!\big)^2 
\sum_{\tau_2=1}^{\tau_1} \frac{(-2)^{\tau_2} \big(
        2 \tau_2\big)! \tau_2^2}{\big(
        \tau_2!\big)^2}}{\big(
        2 \tau_1\big)!},
\nonumber\\ && 
%--14
\sum_{\tau_1=1}^N 
\frac{\big(
        \tau_1!\big)^2 
\sum_{\tau_2=1}^{\tau_1} \frac{(-1)^{\tau_2} \big(
        2 \tau_2\big)! \tau_2^2}{\big(
        \tau_2!\big)^2}}{\big(
        2 \tau_1\big)!},
%--15
\sum_{\tau_1=1}^N \frac{\big(
        \tau_1!\big)^2 
\sum_{\tau_2=1}^{\tau_1} \frac{(-1)^{\tau_2} \big(
        2 \tau_2\big)! \big(
        \sum_{\tau_3=1}^{\tau_2} \frac{1}{\tau_3^2}\big) 
\tau_2^2}{\big(
        \tau_2!\big)^2}}{\big(
        2 \tau_1\big)!},
\nonumber\\ &&
%--16
\sum_{\tau_1=1}^N \frac{\big(
        \tau_1!\big)^2 
\sum_{\tau_2=1}^{\tau_1} \frac{(-1)^{\tau_2} \big(
        2 \tau_2\big)! S_{-2}(\tau_3)
\tau_2^2}{\big(
        \tau_2!\big)^2}}{\big(
        2 \tau_1\big)!},
%--17
\sum_{\tau_1=1}^N \frac{\big(
        \tau_1!\big)^2 
\sum_{\tau_2=1}^{\tau_1} \frac{(-2)^{\tau_2} \big(
        2 \tau_2\big)! \big(
        \sum_{\tau_3=1}^{\tau_2} \frac{2^{-\tau_3} S_{2}(\tau_3)
}{\tau_3}\big) 
\tau_2^2}{\big(
        \tau_2!\big)^2}}{\big(
        2 \tau_1\big)!},
\nonumber\\ &&
%--18
\sum_{\tau_1=1}^N \frac{\big(
        \tau_1!\big)^2 
\sum_{\tau_2=1}^{\tau_1} \frac{(-2)^{\tau_2} \big(
        2 \tau_2\big)! 
\big(
        \sum_{\tau_3=1}^{\tau_2} \frac{2^{-\tau_3}
\sum_{\tau_4=1}^{\tau_3} 
        \frac{(-1)^{\tau_4}}{\tau_4^2}}{\tau_3}\big) 
\tau_2^2}{\big(
        \tau_2!\big)^2}}{\big(
        2 \tau_1\big)!},
%--19
\sum_{\tau_1=1}^N \frac{(-1)^{\tau_1} \big(
        2 \tau_1\big)!}{\big(
        \tau_1!\big)^2 \tau_1^4},
%--20
\sum_{\tau_1=1}^N \frac{(-1)^{\tau_1} \big(
        2 \tau_1\big)!}{\big(
        \tau_1!\big)^2 \tau_1^3},
\nonumber\\ &&
%--21
\sum_{\tau_1=1}^N \frac{\big(
        2 \tau_1\big)! 
\sum_{\tau_2=1}^{\tau_1} \frac{1}{\tau_2}}{\big(
        \tau_1!\big)^2 \tau_1^3},
%--22
\sum_{\tau_1=1}^N \frac{\big(
        \tau_1!\big)^2}{\big(
        2 \tau_1\big)! \tau_1^2},
%--23
\sum_{\tau_1=1}^N \frac{\big(
        \tau_1!\big)^2 
\sum_{\tau_2=1}^{\tau_1} \frac{\big(
        2 \tau_2\big)!}{\big(
        \tau_2!\big)^2}}{\big(
        2 \tau_1\big)! \tau_1^2},
%--24
\sum_{\tau_1=1}^N \frac{\big(
        \tau_1!\big)^2 
\sum_{\tau_2=1}^{\tau_1} \frac{(-1)^{\tau_2} \big(
        2 \tau_2\big)!}{\big(
        \tau_2!\big)^2 \tau_2^3}}{\big(
        2 \tau_1\big)! \tau_1^2},
\nonumber\\ &&
%--25
\sum_{\tau_1=1}^N \frac{\big(
        2 \tau_1\big)! 
\sum_{\tau_2=1}^{\tau_1} \frac{(-1)^{\tau_2}}{\tau_2^2}}{\big(
        \tau_1!\big)^2 \tau_1^2},
%--26
\sum_{\tau_1=1}^N \frac{(-1)^{\tau_1} \big(
        2 \tau_1\big)! 
\sum_{\tau_2=1}^{\tau_1} \frac{(-1)^{\tau_2}
}{\tau_2^2}}{\big(
        \tau_1!\big)^2 \tau_1^2},
%--27
\sum_{\tau_1=1}^N 
\frac{\big(
        \tau_1!\big)^2 
\sum_{\tau_2=1}^{\tau_1} \frac{\big(
        2 \tau_2\big)! 
\sum_{\tau_3=1}^{\tau_2} \frac{1}{\tau_3}}{\big(
        \tau_2!\big)^2 \tau_2^2}}{\big(
        2 \tau_1\big)! \tau_1^2},
\nonumber\\ &&
%--28
\sum_{\tau_1=1}^N \frac{\big(
        2 \tau_1\big)! 
\sum_{\tau_2=1}^{\tau_1} \frac{1}{\tau_2}}{\big(
        \tau_1!\big)^2 \tau_1^2},
%--29
\sum_{\tau_1=1}^N \frac{\big(
        \tau_1!\big)^2 
\sum_{\tau_2=1}^{\tau_1} \frac{\big(
        2 \tau_2\big)!}{\big(
        \tau_2!\big)^2 \tau_2}}{\big(
        2 \tau_1\big)! \tau_1^2},
%--30
\sum_{\tau_1=1}^N \frac{\big(
        \tau_1!\big)^2 
\sum_{\tau_2=1}^{\tau_1} \frac{(-1)^{\tau_2} \big(
        2 \tau_2\big)!}{\big(
        \tau_2!\big)^2 \tau_2}}{\big(
        2 \tau_1\big)! \tau_1^2},
\nonumber\\ &&
%--31
\sum_{\tau_1=1}^N \frac{\big(
        \tau_1!\big)^2 
\sum_{\tau_2=1}^{\tau_1} \frac{\big(
        2 \tau_2\big)! 
\sum_{\tau_3=1}^{\tau_2} \frac{1}{\tau_3^2}}{\big(
        \tau_2!\big)^2 \tau_2}}{\big(
        2 \tau_1\big)! \tau_1^2},
%--32
\sum_{\tau_1=1}^N \frac{\big(
        \tau_1!\big)^2 
\sum_{\tau_2=1}^{\tau_1} \frac{(-1)^{\tau_2} \big(
        2 \tau_2\big)! 
\sum_{\tau_3=1}^{\tau_2} \frac{1}{\tau_3^2}}{\big(
        \tau_2!\big)^2 \tau_2}}{\big(
        2 \tau_1\big)! \tau_1^2},
\nonumber\\ &&
%--33
\sum_{\tau_1=1}^N 
\frac{\big(
        \tau_1!\big)^2 
\sum_{\tau_2=1}^{\tau_1} \frac{\big(
        2 \tau_2\big)! 
\sum_{\tau_3=1}^{\tau_2} \frac{(-1)^{\tau_3}}{\tau_3^2}}{\big(
        \tau_2!\big)^2 \tau_2}}{\big(
        2 \tau_1\big)! \tau_1^2},
%--34
\sum_{\tau_1=1}^N \frac{\big(
        \tau_1!\big)^2 
\sum_{\tau_2=1}^{\tau_1} \frac{(-1)^{\tau_2} \big(
        2 \tau_2\big)! 
\sum_{\tau_3=1}^{\tau_2} \frac{(-1)^{\tau_3}}{\tau_3^2}}{\big(
        \tau_2!\big)^2 \tau_2}}{\big(
        2 \tau_1\big)! \tau_1^2},
\nonumber\\ && 
%--35
\sum_{\tau_1=1}^N \frac{\big(
        \tau_1!\big)^2 
\sum_{\tau_2=1}^{\tau_1} \frac{\big(
        2 \tau_2\big)! 
\sum_{\tau_3=1}^{\tau_2} \frac{
\sum_{\tau_4=1}^{\tau_3} \frac{1}{\tau_4}}{\tau_3}}{\big(
        \tau_2!\big)^2 \tau_2}}{\big(
        2 \tau_1\big)! \tau_1^2},
%--36
\sum_{\tau_1=1}^N \frac{\big(
        \tau_1!\big)^2 
\sum_{\tau_2=1}^{\tau_1} \frac{(-2)^{\tau_2} \big(
        2 \tau_2\big)! \tau_2^2}{\big(
        \tau_2!\big)^2}}{\big(
        2 \tau_1\big)! \tau_1^2},
\nonumber\\ &&
%--37
\sum_{\tau_1=1}^N \frac{\big(
        \tau_1!\big)^2 
\sum_{\tau_2=1}^{\tau_1} \frac{(-1)^{\tau_2} \big(
        2 \tau_2\big)! \tau_2^2}{\big(
        \tau_2!\big)^2}}{\big(
        2 \tau_1\big)! \tau_1^2},
%--38
\sum_{\tau_1=1}^N 
\frac{\big(
        \tau_1!\big)^2 
\sum_{\tau_2=1}^{\tau_1} \frac{(-1)^{\tau_2} \big(
        2 \tau_2\big)! S_2(\tau_2)
\tau_2^2}{\big(
        \tau_2!\big)^2}}{\big(
        2 \tau_1\big)! \tau_1^2},
\nonumber\\ &&
%--39
\sum_{\tau_1=1}^N \frac{\big(
        \tau_1!\big)^2 
\sum_{\tau_2=1}^{\tau_1} \frac{(-1)^{\tau_2} \big(
        2 \tau_2\big)! S_{-2}(\tau_2)
\tau_2^2}{\big(
        \tau_2!\big)^2}}{\big(
        2 \tau_1\big)! \tau_1^2},
%--40
\sum_{\tau_1=1}^N \frac{\big(
        \tau_1!\big)^2 
\sum_{\tau_2=1}^{\tau_1} \frac{(-2)^{\tau_2} \big(
        2 \tau_2\big)! \big(
        \sum_{\tau_3=1}^{\tau_2} \frac{2^{-\tau_3} 
        \sum_{\tau_4=1}^{\tau_3} \frac{1}{\tau_4^2}}{\tau_3}\big) 
\tau_2^2}{\big(
        \tau_2!\big)^2}}{\big(
        2 \tau_1\big)! \tau_1^2},
\nonumber\\ &&
%--41
\sum_{\tau_1=1}^N \frac{\big(
        \tau_1!\big)^2 
\sum_{\tau_2=1}^{\tau_1} \frac{(-2)^{\tau_2} \big(
        2 \tau_2\big)! \big(
        \sum_{\tau_3=1}^{\tau_2} \frac{2^{-\tau_3} 
        \sum_{\tau_4=1}^{\tau_3} 
\frac{(-1)^{\tau_4}}{\tau_4^2}}{\tau_3}\big) \tau_2^2}{\big(
        \tau_2!\big)^2}}{\big(
        2 \tau_1\big)! \tau_1^2},
%--42
\sum_{\tau_1=1}
^N 
\frac{\big(
        2 \tau_1\big)!}{\big(
        \tau_1!\big)^2 \tau_1},
%--43
\sum_{\tau_1=1}^N \frac{(-1)^{\tau_1} \big(
        2 \tau_1\big)!}{\big(
        \tau_1!\big)^2 \tau_1},
\nonumber\\ &&
%--44
\sum_{\tau_1=1}^N \frac{\big(
        2 \tau_1\big)! 
\sum_{\tau_2=1}^{\tau_1} \frac{1}{\tau_2^3}}{\big(
        \tau_1!\big)^2 \tau_1},
%--45
\sum_{\tau_1=1}^N \frac{\big(
        2 \tau_1\big)! 
\sum_{\tau_2=1}^{\tau_1} \frac{(-1)^{\tau_2}}{\tau_2^3}}{\big(
        \tau_1!\big)^2 \tau_1},
%--46
\sum_{\tau_1=1}^N \frac{\big(
        2 \tau_1\big)! 
\sum_{\tau_2=1}^{\tau_1} \frac{1}{\tau_2^2}}{\big(
        \tau_1!\big)^2 \tau_1},
%--47
\sum_{\tau_1=1}^N \frac{(-1)^{\tau_1} \big(
        2 \tau_1\big)! 
\sum_{\tau_2=1}^{\tau_1} \frac{1}{\tau_2^2}}{\big(
        \tau_1!\big)^2 \tau_1},
\nonumber\\ &&
%--48
\sum_{\tau_1=1}^N \frac{\big(
        2 \tau_1\big)! 
\sum_{\tau_2=1}^{\tau_1} \frac{(-1)^{\tau_2}}{\tau_2^2}}{\big(
        \tau_1!\big)^2 \tau_1},
%--49
\sum_{\tau_1=1}^N \frac{(-1)^{\tau_1} \big(
        2 \tau_1\big)! 
\sum_{\tau_2=1}^{\tau_1} \frac{(-1)^{\tau_2}}{\tau_2^2}}{\big(
        \tau_1!\big)^2 \tau_1},
%--50
\sum_{\tau_1=1}^N \frac{\big(
        2 \tau_1\big)! 
\sum_{\tau_2=1}^{\tau_1} \frac{
\sum_{\tau_3=1}^{\tau_2} \frac{1}{\tau_3}}{\tau_2^2}}{\big(
        \tau_1!\big)^2 \tau_1},
\nonumber\\ &&
%--51
\sum_{\tau_1=1}^N 
\frac{\big(
        2 \tau_1\big)! 
\sum_{\tau_2=1}^{\tau_1} \frac{
\sum_{\tau_3=1}^{\tau_2} 
\frac{(-1)^{\tau_3}}{\tau_3^2}}{\tau_2}}{\big(
        \tau_1!\big)^2 \tau_1},
%--52
\sum_{\tau_1=1}^N \frac{\big(
        2 \tau_1\big)! 
\sum_{\tau_2=1}^{\tau_1} \frac{
\sum_{\tau_3=1}^{\tau_2} \frac{1}{\tau_3}}{\tau_2}}{\big(
        \tau_1!\big)^2 \tau_1},
%--53
\sum_{\tau_1=1}^N \frac{(-2)^{\tau_1} \big(
        2 \tau_1\big)! \tau_1^2}{\big(
        \tau_1!\big)^2},
\nonumber\\ &&
%--54
\sum_{\tau_1=1}^N \frac{(-1)^{\tau_1} \big(
        2 \tau_1\big)! \tau_1^2}{\big(
        \tau_1!\big)^2},
%--55
\sum_{\tau_1=1}^N \frac{(-1)^{\tau_1} \big(
        2 \tau_1\big)! \big(
        \sum_{\tau_2=1}^{\tau_1} \frac{1}{\tau_2^2}\big) 
\tau_1^2}{\big(
        \tau_1!\big)^2},
%--56
\sum_{\tau_1=1}^N \frac{(-1)^{\tau_1} \big(
        2 \tau_1\big)! \big(
        \sum_{\tau_2=1}^{\tau_1} \frac{(-1)^{\tau_2}}{\tau_2^2}\big) 
\tau_1^2}{\big(
        \tau_1!\big)^2},
\nonumber\\ &&
%--57
\sum_{\tau_1=1}^N \frac{(-2)^{\tau_1} \big(
        2 \tau_1\big)! \big(
        \sum_{\tau_2=1}^{\tau_1} \frac{2^{-\tau_2} 
        \sum_{\tau_3=1}^{\tau_2} \frac{1}{\tau_3^2}}{\tau_2}\big) 
\tau_1^2}{\big(
        \tau_1!\big)^2},
%--58
\sum_{\tau_1=1}^N \frac{(-2)^{\tau_1}
 \big(
        2 \tau_1\big)! \big(
        \sum_{\tau_2=1}^{\tau_1} \frac{2^{-\tau_2} 
        \sum_{\tau_3=1}^{\tau_2} 
\frac{(-1)^{\tau_3}}{\tau_3^2}}{\tau_2}\big) \tau_1^2}{\big(
        \tau_1!\big)^2}\Biggr\}.
\nonumber\\
\end{eqnarray}
%--------------------------------------------------------------------------------------------------

The nested binomial sums do not obviously have such a systematic representation like the case for 
harmonic, generalized harmonic and cyclotomic sums. This is implied by the different building blocks
entering the different summands, which are central binomials in the numerators and denominators, 
lower sums of different kind, rational expressions and powers of 
$m^k, m \in \mathbb{N}, k \in \mathbb{Z}$. Yet one may find a basis set by summation technologies of {\tt 
Sigma} \cite{SIG1,SIG2} using the underlying difference ring theory 
\cite{Schneider:08d,Schneider:08e} and additional term synchronization with algorithms in {\tt HarmonicSums}.

A more synchronized picture emerges in $t$--space, as is outlined in Section~\ref{sec:5}.
The corresponding generating functions will then have a representation in terms of letters
smoothing out the more involved structures in the sum representations.

One may compute the asymptotic expansion of most of these binomial sums using the command  {\tt BSExpansion}, 
which relies on the inverse Mellin transform of the respective binomial sum. This may not be 
easily derived 
because regularizations beyond the one given by the $+$-operation are necessary. 
In these cases one needs to reformulate 
the corresponding expressions by individual partial integrations first. By considering the Mellin inversion 
for the nested binomial sums only, one needs to account for Mellin convolutions with the corresponding 
pre--factors, cf.~\cite{Ablinger:2014bra}. In Appendix~\ref{sec:B} we will list a series of expansions of 
nested binomial sums in the asymptotic region. More expansions are given in an ancillary file. 

Finally, we are going to use a different strategy to deal with the nested binomial sum contributions 
to the unpolarized and polarized 
amplitudes in Section~\ref{sec:5}. In the end our goal is to obtain first the $x$--space representation for
all first--order factorizable contributions.\footnote{Let us note that also in the two--mass case 
the
unpolarized and polarized pure singlet OMEs could not be computed in terms of first--order factorizable structures 
in $N$--space, but it has been possible in $x$--space, cf.~Refs.~\cite{Ablinger:2017xml,Ablinger:2019gpu}.} At 
the end of the calculation analytic structures are obtained in $x$--space which can be finally Mellin--transformed 
and allow for  representations in Mellin $N$--space evolution programs \cite{BS}, see~Appendix~\ref{sec:D}.
%%%%%%%%%%%%%%%%%%%%%%%%%%%%%%%%%%%%%%%%%%%%%%%%%%%%%%%%%%%%%%%%%%%%%%%%%%%%%%%%%%%%%%%%%%%%%%%%%%%
\section{\boldmath From $t$-space to $x$-space}
\label{sec:5}
%%%%%%%%%%%%%%%%%%%%%%%%%%%%%%%%%%%%%%%%%%%%%%%%%%%%%%%%%%%%%%%%%%%%%%%%%%%%%%%%%%%%%%%%%%%%%%%%%%%

\vspace*{1mm}
\noindent
To obtain an even more uniform approach to the present problem, in particular for the nested binomial sum 
contributions, we went back to the amplitude representation in $t$--space, the resummation of the $N$--space 
representation into a generating function, cf.~Eq.~(\ref{eq:FT}), in the unpolarized and polarized case. We
solved the first--order factorizable contributions in terms of G--functions in the region around 
$t=0$. The
$t$--representation in the unpolarized case is even in $t$ and in the polarized case odd in $t$,
\cite{Politzer:1974fr,Blumlein:1996vs}.
I.e. it is sufficient to consider one of the regions
%-------------------------------------------------------------------------------------------------
\begin{eqnarray}
(-\infty, 0]~~~~~~~~\text{or}~~~~~~~~[0, \infty).
\end{eqnarray}
%-------------------------------------------------------------------------------------------------
The following 
alphabet of 17 letters spans the contributing master integrals
to the required order in the dimensional parameter $\ep$ in the basis originally obtained by the 
integration--by--parts reduction
\cite{Studerus:2009ye,vonManteuffel:2012np}
%-------------------------------------------------------------------------------------------------
\begin{eqnarray}
\mathfrak{A} &=& \Biggl\{
\frac{1}{t}, 
\frac{1}{1-t}, 
\frac{1}{1+t},
\frac{1}{2-t},
\frac{1}{2+t},
\frac{1}{4-t},
\frac{1}{4+t},
\frac{1}{1 - 2t}, 
\frac{1}{1 + 2t}, 
\sqrt{t(4-t)}, 
\sqrt{t(4+t)}, 
\nonumber\\ &&
\frac{\sqrt{t(4-t)}}{1-t}, 
\frac{\sqrt{t(4+t)}}{1+t}, 
\frac{\sqrt{t(4-t)}}{1+t}, 
\frac{\sqrt{t(4+t)}}{1-t}, 
\frac{\sqrt{t(4-t)}}{1 + 2t}, 
\frac{\sqrt{t(4+t)}}{1 - 2t}\Biggr\}. 
\label{eq:ALPH1}
\end{eqnarray}
%-------------------------------------------------------------------------------------------------
The alphabet exhibits the symmetry 
%-------------------------------------------------------------------------------------------------
\begin{eqnarray}
t \leftrightarrow -t, 
\end{eqnarray}
%-------------------------------------------------------------------------------------------------
which is essential for the occurrence of either only even or only odd moments. With the exception of  
$1/t$, there are 
therefore only eight essential letters. The $t$--space representation is still very close to the $N$--space
representation, since the latter is obtained by performing a formal Taylor expansion of the former one.
In this way now also the finite nested (inverse) binomial sums received a more systematic representation.
As can be seen in the alphabet $\mathfrak{A}$, Eq.~(\ref{eq:ALPH1}), there are letters for harmonic 
polylogarithms, Kummer--Poincar\'e integrals 
and root--valued letters, as well as products of those
with the former ones. Furthermore, the corresponding G--functions in $t$ contain respective combinations
of subsets of all these letters.

We will consider the region $t \in [0, \infty)$, containing the following (pseudo)thresholds 
%-------------------------------------------------------------------------------------------------
\begin{eqnarray}
t_0 \in \Biggl\{\frac{1}{2}, 1, 2, 4 \Biggr\}.
\end{eqnarray}
%-------------------------------------------------------------------------------------------------
The differential equations have to be solved in the regions $t \in [0,1/2], [1/2,1], [1,2], [2,4]$, and 
$[4,\infty)$,
corresponding in $x$ to the potential ranges for $x \in [2,\infty), [1,2], [1/2,1], [1/4,1/2], [0,1/4]$.
Finally, it will turn out that $x \in [0,1]$, because the amplitude will exhibit an imaginary part
after the transformation 
%-------------------------------------------------------------------------------------------------
\begin{eqnarray}
t \rightarrow \frac{1}{x}
\label{eq:TRAF}
\end{eqnarray}
%-------------------------------------------------------------------------------------------------
in the physical region only.

We begin by solving the amplitude in the region $t \in [0,1/2]$. Here 10, 20, 44, and up to 1046
G--functions contribute for the terms $O(1/\ep^k),~~k = -3, ... ,0$, referring to the G--basis representation 
after algebraic reduction. On the other hand, there are only 730 contributing G--functions at $O(\ep^0)$ in the 
original unpolarized and polarized amplitudes, where products of iterated integrals are expanded into their 
linear representations. 
This is typical for large alphabets, cf. also~ Ref.~\cite{Ablinger:2014yaa}. Therefore we did not 
perform the algebraic reduction in the present case.  

The formal Taylor series expansion in $t$ reproduces the moments computed by {\tt MATAD} 
\cite{Steinhauser:2000ry,Bierenbaum:2009mv} for $N = 2,4,6,8,10$
in the unpolarized case and for $N = 3,5,7,9$ in the polarized case.\footnote{We did not compute 
the moment for $N=1$.} The solution in terms of G--functions shows that after the transformation   
(\ref{eq:TRAF}) 
the amplitude both in the unpolarized and the polarized case has no imaginary parts and, furthermore, no 
singularity at $t = 1/2$.
The solution of the differential equations in the present case and for the subsequent regions are obtained as 
follows. 
As we have the solutions in terms of G--functions, we can establish a hirachical system of coupled
     differential equations by differentiating with respect to $t$.
     To base the iterated integrals at a new point $t_0$, we can now transform the system to a new 
variable
     $t'=t_0+t$ and integrate the differential equation again.
     The boundary values at the point $t' = 0$ can be obtained by evaluating the previous representation at
     $t = t_0$.
     If the leftmost letter is singular at $t=t_0$ we can shuffle this letter to the right and obtain 
logarithmic
     singularities at $t=t_0$ which have to match the ones generated by the expansion around $t' = 0$ of the 
new
     representation.
     The hierarchical structure of the system allows to proceed from iterated integrals of weight 1 up to the
     ones with weight 6.
     One then inserts the G--functions into the amplitudes and checks
     whether an imaginary part remains.
     For t = 1/2 this is not the case and the real part turns out to be non--singular.
     This implies that there are no contributions to the amplitude for $x > 1$.

In the next step we use the representation at $t = 1/2$ and solve the corresponding differential equations in the 
region $t \in [1/2,1]$ repeating the above steps, with initial values at $t=1/2$ and correspondingly for the
thresholds $t = 1, 2, 4$. 
One may test the result in $x$--space by computing Mellin moments and by comparing with the results above by 
{\tt MATAD}. 

The analytic continuation at $t = 2$ and at $t = 4$ does not generate additional imaginary parts, which we
     established by flagging the respective imaginary parts occurring at each new threshold for individual
     integrals and found that these contributions vanish in the amplitudes.
     This requires the precise calculation of the contributing constants.
     The analytic proof of the vanishing of the combination of the G--functions with different main arguments
     would be much more difficult. This means that the result in $x$--space is continuous in $x \in  (0, 1]$.

The results in $x$--space are given by G--functions over the 14 letter alphabet $\mathfrak{A}_x$, 
%-------------------------------------------------------------------------------------------------
\begin{eqnarray}
\mathfrak{A}_x &=&  
\Biggl\{
\frac{1}{x},
\frac{1}{1-x},
\frac{1}{1+x},
\frac{1}{1-2 x},
\frac{1}{2+x},
\frac{1}{2-x},
\frac{\sqrt{1-4 x}}{x},
\frac{\sqrt{1-4 x}}{1-x},
\frac{\sqrt{1-4 x}}{1+x},
\frac{\sqrt{1-4 x}}{2+x},
\nonumber\\ &&
\frac{\sqrt{1+4 x}}{x},
\frac{\sqrt{1+4 x}}{1-x}
\frac{\sqrt{1+4 x}}{1+x},
\frac{\sqrt{1+4 x}}{2-x}
\Biggr\}.
\end{eqnarray}
%-------------------------------------------------------------------------------------------------
For the G--functions at argument $x$ we perform algebraic reductions. This reduces 386 original G--functions 
in the unpolarized case to 322, and 360 G--functions in the polarized case to 315. The number of letters of the 
contributing G--constants is larger than the one in $\mathfrak{A}_x$. 
There are 697 constants in the unpolarized case and 659 in the polarized case given by
root--valued G--functions at main argument $x = 1/4, 1/2, 3/4$ and 1. The algebraic reduction to a basis
would enlarge the number of contributing constants. 

Since in all cases only one type of square--root factor appears in the letters, the letters for all 
contributing G--functions can be rationalized.
In a series of cases one has to remove poles in the integration domain at $x = 1/2$. 
The letter $1/(1-2 x)$ emerges for {\tt GL[\{...\},1/2]} only, i.e. at the integration boundary,
which can be removed by {\tt GLRemovePole}.\footnote{Here {\tt GL[...]} denotes the respective G--function 
in the notation of  {\tt HarmonicSums}.}

In the unpolarized case, 155 G--constants at $x=1/4$ emerge, 161 at $x=1/2$, 112 at $x=3/4$ and
269 at $x=1$,  while in the polarized case 155 G--constants at $x=1/4$, 127 at $x=1/2$, 112 at $x=3/4$ and
265 at $x=1$ contribute. The constants with main argument $x=1/2$ are all generalized harmonic polylogarithms.
These constants reduce to
%----------------------------------------------------------------------------------------------------------
\begin{eqnarray}
&& \Biggl\{\ln(2) 
\zeta_2,
\zeta_3,
\Li_4\left(\frac{1}{2}\right),
\zeta_5,
\Li_5\left(\frac{1}{2}\right),
\HA_{0,0,-2,1}(1),
\HA_{0,0,1,-2}(1),
\HA_{0,1,0,-2}(1),
\HA_{0,3,1,-1}(1),
\nonumber\\ &&
\HA_{3,0,1,-1}(1),
\HA_{3,1,0,-1}(1),\HA_{0,1,0,0,-2}(1),\HA_{3,1,-1,0,-1}(1),\HA_{3,1,-1,0,1}(1),\HA_{3,1,1,0,-1}(1)\Biggr\}.
\end{eqnarray}
%----------------------------------------------------------------------------------------------------------
For the generalized harmonic polylogarithms which do not reduce to multiple zeta values or 
logarithms and polylogarithms of different argument, we present numerical values in an ancillary 
file. 

For the G--constants of main argument $x = 1/4, 3/4$ one has first to rescale the letters such that 
the main  argument is $x = 1$. An example is
%----------------------------------------------------------------------------------------------------------
\begin{eqnarray}
T_2 =
\text{G} \Biggl[\Biggl\{\frac{1}{1-\tau},\frac{\sqrt{1-4\tau}}{\tau}, \frac{\sqrt{1-4\tau}}{1-\tau} \Biggr\},
\frac{1}{4}\Biggr] = 
\text{G}\Biggl[\Biggl\{
        \frac{1}{4-\tau},\frac{\sqrt{1-\tau}}{\tau},\frac{\sqrt{1-\tau}}{4-\tau }\Biggr\},1\Biggr].
\end{eqnarray}
%----------------------------------------------------------------------------------------------------------
Then the command {\tt SpecialGLToH} rationalizes the G-function. Furthermore, 
denominators containing quadratic forms need to be decomposed by the command {\tt LToGL[GLToL[GL[\{....\},c]]} 
which yields a proper 
input form for a numerical precision calculation. One obtains
%----------------------------------------------------------------------------------------------------------
\begin{eqnarray}
\lefteqn{T_2 = 6 + \sqrt{3} i
} \nonumber\\ && 
\times \Biggl[
        \text{G}\Biggl[\Biggl\{\frac{1}{i \sqrt{3}+\tau }\Biggr\},1\Biggr]
\Biggl[-4
              - \text{G}\Biggl[\Biggl\{\frac{1}{-i \sqrt{3}+\tau },\frac{1}{-1+\tau }\Biggr\},1\Biggr]
              + \text{G}\Biggl[\Biggl\{\frac{1}{-i \sqrt{3}+\tau },\frac{1}{1+\tau }\Biggr\},1\Biggr]
\nonumber\\ && 
              + \text{G}\Biggl[\Biggl\{\frac{1}{i \sqrt{3}+\tau },\frac{1}{-1+\tau }\Biggr\},1\Biggr]
  - \text{G}\Biggl[\Biggl\{\frac{1}{i \sqrt{3}+\tau },\frac{1}{1+\tau }\Biggr\},1\Biggr]
        \Biggr]
        + \sqrt{3} i \text{G}\Biggl[\Biggl\{\frac{1}{-i \sqrt{3}+\tau }\Biggr\},1
        \Biggr]
\Biggl[4 
\nonumber\\ && 
         - \text{G}\Biggl[\Biggl\{ \frac{1}{-i \sqrt{3}+\tau },\frac{1}{-1+\tau }\Biggr\},1\Biggr]
         + \text{G}\Biggl[\Biggl\{\frac{1}{-i \sqrt{3}+\tau },\frac{1}{1+\tau }\Biggr\},1\Biggr]
       + \text{G}\Biggl[\Biggl\{\frac{1}{i \sqrt{3}+\tau },\frac{1}{-1+\tau }\Biggr\},1\Biggr]
\nonumber\\ &&         
 - \text{G}\Biggl[\Biggl\{\frac{1}{i \sqrt{3}+\tau },\frac{1}{1+\tau }\Biggr\},1\Biggr]
        \Biggr]
+ 2 i \sqrt{3} \Biggl\{    
                \text{G}\Biggl[\Biggl\{
                        \frac{1}{-i \sqrt{3}+\tau },\frac{1}{-i \sqrt{3}+\tau },\frac{1}{-1+\tau 
}\Biggr\},1\Biggr]
\nonumber\\ && 
               - \text{G}\Biggl[\Biggl\{
                        \frac{1}{-i \sqrt{3}+\tau },\frac{1}{-i \sqrt{3}+\tau },\frac{1}{1+\tau 
}\Biggr\},1\Biggr]
        -
                \text{G}\Biggl[\Biggl\{
                        \frac{1}{i \sqrt{3}+\tau },\frac{1}{i \sqrt{3}+\tau },\frac{1}{-1+\tau 
}\Biggr\},1\Biggr]
\nonumber\\ && 
        +  
                \text{G} \Biggl[\Biggl\{
                \frac{1}{i \sqrt{3}+\tau },\frac{1}{i \sqrt{3}+\tau },\frac{1}{1+\tau 
}\Biggr\},1\Biggr]\Biggr\}
+\Biggl(
        -8
        +4 \ln(2)
        +12 \text{G}\Biggl[\Biggl\{
                \frac{1}{i \sqrt{3}+\tau }\Biggr\},1\Biggr]
\Biggr) 
\nonumber\\ && \times 
\text{G}\Biggl[\Biggl\{
        \frac{1}{-i \sqrt{3}+\tau }\Biggr\},1\Biggr]
+4 (-2+\ln(2)) \text{G}\Biggl[\Biggl\{
        \frac{1}{i \sqrt{3}+\tau }\Biggr\},1\Biggr]
-4 \text{G}\Biggl[\Biggl\{
        \frac{1}{-i \sqrt{3}+\tau },\frac{1}{1+\tau }\Biggr\},1\Biggr]
\nonumber\\ && 
-4 \text{G}\Biggl[\Biggl\{
        \frac{1}{i \sqrt{3}+\tau },\frac{1}{1+\tau }\Biggr\},1\Biggr],
\end{eqnarray}
%----------------------------------------------------------------------------------------------------------
with 
%----------------------------------------------------------------------------------------------------------
\begin{eqnarray}
\text{G}\left[\left\{\frac{1}{\pm i \sqrt{3} + \tau}\right\},1\right] = \ln(2) - \frac{1}{2} \ln(3)
\mp \frac{i}{6} \pi.
\end{eqnarray}
%----------------------------------------------------------------------------------------------------------
In this example no poles in $x \in [0,1]$ are present.
The G--constants are given in terms of Kummer--Poincar\'e integrals. Their amount is more 
than an order of magnitude larger than the number of original constants. Kummer--Poincar\'e 
integrals can be evaluated by using
methods of {\tt Pari GP} \cite{PARI} or by \cite{Vollinga:2004sn}, using
the method of H\"older convolution from Ref.~\cite{Borwein:1999js}.

Finally, we performed formal Taylor series expansions of the results in $x$ around $x = 0, 1/2$ and 1 with
100 terms. The number of G--constants which emerge in the expansions around $x = 0$ and are not multiple 
zeta values \cite{Blumlein:2009cf} is 604, corresponding to 12719 terms after rationalization and decomposition 
into Kummer--Poincar\'e integrals. For $x = 
1/2$, 502 
constants contribute, and for $x = 1$ the number of constants is 213, see Appendix~\ref{sec:C}. 
The set of the necessary constants in the polarized case are a subset of those in the unpolarized case.
The representations of the expansions are given by
%-------------------------------------------------------------------------------------------------
\begin{eqnarray}
\label{eq:F0}
\mathbb{F}_0(x)     &=&  \sum_{k=-1}^{100} \sum_{l=0}^5 a_{k,l}^{(0)} \ln^l(x) x^k,
\\
\mathbb{F}_{1/2}(x) &=&  \sum_{k=0}^{100}  a_{k}^{(1/2)}  \left(x-\frac{1}{2}\right)^k,
\\
\label{eq:F1}
\mathbb{F}_1(x)     &=&  \sum_{k=0}^{100} \sum_{l=0}^5 a_{k,l}^{(1)} \ln^l(1-x) (1-x)^k.
\end{eqnarray}
%-------------------------------------------------------------------------------------------------
These representations can be matched at $x = 2/10$ and $x = 7/10$ and 
one may compute a series of  lower Mellin 
moments numerically at high precision, and also compare with the direct numerical solution, 
see also Refs.~\cite{Maier:2017ypu,Fael:2021kyg,Fael:2022miw}. The representations  
(\ref{eq:F0}--\ref{eq:F1}) can also be 
Mellin transformed to construct a $N$--space representation for $N \in \mathbb{C}$, see~Appendix~\ref{sec:D}.
The method of iterated integrals in
G--space shows the cancellation of diverging terms in the large $N$ limit as $a^N, a > 1$. 
The largest number of constants contributes for the representation around $x=0$, and those for the expansion 
around $x=1$ are, apart from very few new constants, a subset of those around $x=0$. The contributing constants 
are the same in the unpolarized and polarized cases. 
%%%%%%%%%%%%%%%%%%%%%%%%%%%%%%%%%%%%%%%%%%%%%%%%%%%%%%%%%%%%%%%%%%%%%%%%%%%%%%%%%%%%%%%%%%%%%%%%%%%
\section{\boldmath The small and large $x$ expansions}
\label{sec:6}
%%%%%%%%%%%%%%%%%%%%%%%%%%%%%%%%%%%%%%%%%%%%%%%%%%%%%%%%%%%%%%%%%%%%%%%%%%%%%%%%%%%%%%%%%%%%%%%%%%%

\vspace*{1mm}
\noindent
We can now perform the expansion of the results for the already solved parts of $a_{Qg}^{(3)}$ and 
$\Delta a_{Qg}^{(3)}$ and for the first--order factorizable contributions both for $x \rightarrow 
0$ 
and $x \rightarrow 1$. The final result is not obtained directly, but will require a series of technical 
steps to be carried out because of the emergence of quite a series of Kummer--Poincar\'e integrals 
at special numbers, cf.~Appendix~\ref{sec:C}. In the polarized case, even terms that cannot contribute
seem to be present. It will, however, turn out that the corresponding coefficients in front of the 
respective structures will arrange to zero, requiring to solve three--fold iterated integrals with 
letters in root--valued alphabets.

In the small $x$ limit there is a prediction in the unpolarized case from  
Ref.~\cite{Catani:1990eg}, given by
%-------------------------------------------------------------------------------------------------
\begin{eqnarray}
\label{eq:SX1}
a_{Qg}^{(3), x \rightarrow 0}(x) &=& \frac{64}{243} \textcolor{blue}{C_A^2 T_F} \left[
1312 + 135 \zeta_2 - 189 \zeta_3\right] \frac{\ln(x)}{x}.
\end{eqnarray}
%-------------------------------------------------------------------------------------------------
This expression rescales with $\textcolor{blue}{C_F/C_A}$ in the pure singlet case, first calculated 
in Ref.~\cite{Ablinger:2014nga}. The contribution $\propto \zeta_2$ has been computed 
in Eq.~(\ref{eq:aunpol}) and agrees with the corresponding term in Eq.~(\ref{eq:SX1}). 

In the polarized case the leading singularity is located at $N=0$ and does not derive from the same 
dynamics as in the unpolarized case. Instead, one uses so--called infrared evolution equations or 
similar techniques \cite{Bartels:1996wc,Adamiak:2023okq} in the massless case.\footnote{The 
corresponding massive calculation has not been performed.} Whether in this case also color rescaling 
works is not clear a priori. From  Eq.~(\ref{eq:apol}) we obtain the small $x$ contribution
%-------------------------------------------------------------------------------------------------
\begin{eqnarray}
\frac{4}{3} \textcolor{blue}{C_F T_F^2 N_F} \ln^5(x).
\end{eqnarray}
%-------------------------------------------------------------------------------------------------
This term cannot be color rescaled by $\textcolor{blue}{C_F/C_A}$ to the pure singlet contribution, 
cf.~Ref.~\cite{Ablinger:2019etw}, which after color rescaling would yield
%------------------------------------------------------------------------------------------------- 
\begin{eqnarray} 
\Delta a_{Qg}^{(3), x \rightarrow 0, \rm  resc.}(x) &=& \frac{2}{15} \textcolor{blue}{C_A 
T_F} \left[ 8 
\textcolor{blue}{C_A} + 9 \textcolor{blue}{C_F} \right] \ln^5(x). 
\label{eq:sxpol2}
\end{eqnarray} 
%------------------------------------------------------------------------------------------------- 
but no term $\propto \textcolor{blue}{N_F}$.
This situation is similar to the case of $\Delta a_{qg,Q}^{(3)}$ and $\Delta a_{qq,Q}^{{\rm PS},(3)}$,
cf.~\cite{Blumlein:2021xlc},
where the leading singularity at small $x$ of $\Delta a_{qq,Q}^{{\rm PS},(3)}$ is $\propto \ln^4(x)$,
while one obtains $\Delta a_{qg,Q}^{(3)} \propto (8/3) \textcolor{blue}{C_F T_F^2 N_F} \ln^5(x)$,
due to which there is no color rescaling in this case either. 
The corresponding color factors in Eq.~(\ref{eq:sxpol2}) can still receive contributions from the 
non--first--order factorizable contributions.

In deriving the small $x$ limit of the irreducible contributions to the first order factorizable
terms in the polarized case, also potential contributions of $O(\ln(x)/x)$ and $O(1/x)$ emerge. Their 
pre--factor has to be proven to vanish, which requires the calculation of special constants over 
root--valued alphabets. This is shown in Appendix~\ref{sec:C}. While the vanishing of the coefficient 
in front of the $O(\ln(x)/x)$ term can be shown analytically, we decided to show the cancellation of 
the pre--factor of the $O(1/x)$ term numerically to a precision of $\sim 1000$ digits, which is 
equivalent to methods used in `experimental mathematics', cf. e.g. \cite{Bailey:2010aj}.
There are terms $\propto \ln^5(x)$ in the irreducible first--order factorizing terms to color factors
which also contribute to the non--first--order factorizing contributions.
Since it has been known for longer that it is  very difficult to determine the small $x$ behavior of a single 
scale quantity from a limited set of moments, we will not intend this here, but solely rely on 
the analytic calculation in $x$--space for all contributions.

In the large $x$ limit the first--order factorizable contributions of the irreducible diagrams 
have the structure indicated in Eq.~(\ref{eq:F1}).

In Mellin space the most singular term behaves like
%-------------------------------------------------------------------------------------------------
\begin{eqnarray}
\propto \frac{S_1^5(N)}{N}
\end{eqnarray}
%-------------------------------------------------------------------------------------------------
and all other terms also vanish as $N \rightarrow \infty$. The corresponding decrease with $N$ 
proceeds only slowly as is illustrated in Table~\ref{tab:decr}.
One obtains
%-------------------------------------------------------------------------------------------------
\begin{eqnarray}
a_{Qg}^{(3), x \rightarrow 1}(x) &=& 
        \frac{8}{3}  (\textcolor{blue}{C_A - C_F})^2 
 \ln ^5(1-x)
+ O(\ln^4(1-x))
\end{eqnarray}
%-------------------------------------------------------------------------------------------------
and
%-------------------------------------------------------------------------------------------------
\begin{eqnarray}
\Delta a_{Qg}^{(3), x \rightarrow 1}(x) &=& a_{Qg}^{(3), x \rightarrow 1}(x) 
\end{eqnarray}
%-------------------------------------------------------------------------------------------------
in the leading order. This equality holds numerically for $x \gsim 0.95$ at the level of up to 2.5~\%. 

Both the above large and small $x$ limits correspond to the first--order factorizable terms only 
and will receive additions from the diagrams which also receive contributions due to non--first--order 
factorizable master integrals in part, being dealt with in a forthcoming paper \cite{AQGFIN}.
%---------------------------------------------------------------------
\begin{center}
\begin{table}[H]\centering
\renewcommand*{\arraystretch}{1.4}
\begin{tabular}{|r|r|}
\hline
$N$  & $S_1^5(N)/N$
\\
\hline
$10^1$      &   21.556
\\
$10^2$      &   37.561
\\
$10^3$      &   23.502
\\
$10^4$      &    8.982
\\
$10^5$      &    2.583
\\
$10^6$      &    0.618
\\
$10^7$      &    0.130
\\
\hline
\end{tabular}
\caption[]{\sf Numerical illustration of the decrease of the most singular part of the 
first--order 
factorizable contributions in the large $x$ limit.
\label{tab:decr}}
\renewcommand*{\arraystretch}{1}
\end{table}
\end{center}
%-------------------------------------------------------------------------------------------------
%%%%%%%%%%%%%%%%%%%%%%%%%%%%%%%%%%%%%%%%%%%%%%%%%%%%%%%%%%%%%%%%%%%%%%%%%%%%%%%%%%%%%%%%%%%%%%%%%%%
\section{Numerical Results}
\label{sec:7}
%%%%%%%%%%%%%%%%%%%%%%%%%%%%%%%%%%%%%%%%%%%%%%%%%%%%%%%%%%%%%%%%%%%%%%%%%%%%%%%%%%%%%%%%%%%%%%%%%%%

\begin{figure}[H]
\centering
\includegraphics[width=0.7\textwidth]{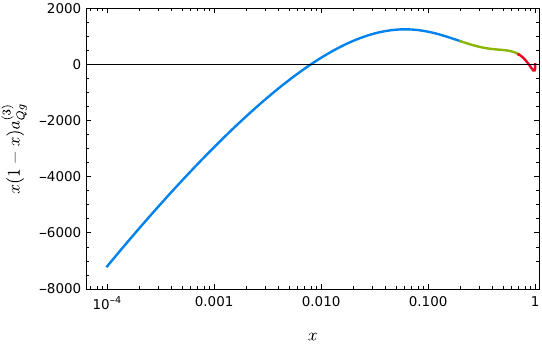}
\caption{\sf 
The contributions due to the first--order factorizable Feynman diagrams to $a_{Qg}^{(3)}(x)$  as a 
function of $x$ rescaled by the factor $x(1-x)$.
Full line (blue): expansion around $x=0$;
full line (green): expansion around $x=1/2$;
full line (red): expansion around $x=1$.}
\label{fig:1}
\end{figure}

\begin{figure}[H]
\centering
\includegraphics[width=0.7\textwidth]{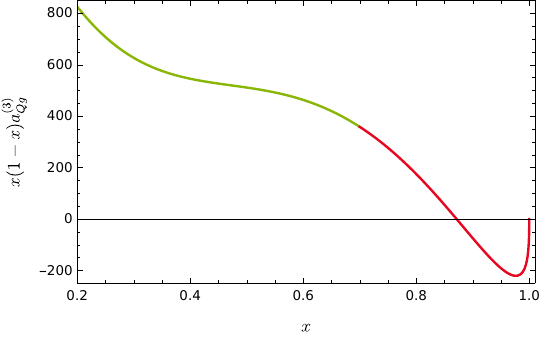}
\caption{\sf 
The contributions due to the first--order factorizable Feynman diagrams to $a_{Qg}^{(3)}(x)$  as a 
function of $x$ rescaled by the factor $x(1-x)$ for larger values of $x$.
Full line (green): expansion around $x=1/2$;
full line (red): expansion around $x=1$.}
\label{fig:2}
\end{figure}

\begin{figure}[H]
\centering
\includegraphics[width=0.7\textwidth]{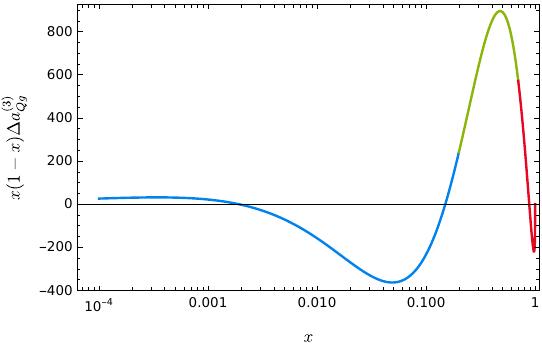}
\caption{\sf 
The contributions due to the first--order factorizable Feynman diagrams to $\Delta a_{Qg}^{(3)}(x)$  as a 
function of $x$ rescaled by the factor $x(1-x)$.
Full line (blue): expansion around $x=0$;
full line (green): expansion around $x=1/2$;
full line (red): expansion around $x=1$.}
\label{fig:3}
\end{figure}

\begin{figure}[H]
\centering
\includegraphics[width=0.7\textwidth]{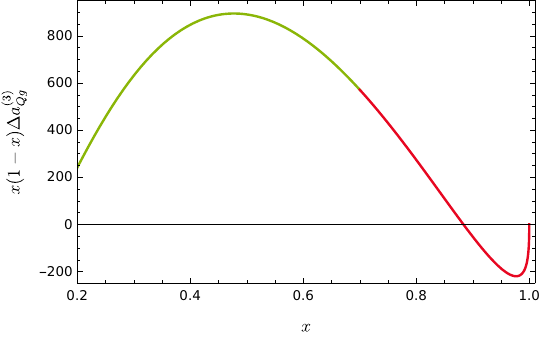}
\caption{\sf 
The contributions due to the first--order factorizable Feynman diagrams to $\Delta a_{Qg}^{(3)}(x)$  as a 
function of $x$ rescaled by the factor $x(1-x)$ for larger values of $x$.
Full line (green): expansion around $x=1/2$;
full line (red): expansion around $x=1$.}
\label{fig:4}
\end{figure}

\vspace*{1mm}
\noindent
In the following we illustrate numerically the analytic results obtained for the contributing irreducible Feynman 
diagrams which 
are first--order--factorizable. In Figures~\ref{fig:1} and \ref{fig:2} 
the sum of these contributions
to $a_{Qg}^{(3)}(x)$ are illustrated in the whole $x$ region and in the region of larger values of $x$
setting $\textcolor{blue}{N_F} = 3$. Here we use the expansions around $x = 0, 1/2$ and 1, 
which are matched in their respective overlap regions.
In Figures~\ref{fig:3} and \ref{fig:4} we show the corresponding results for the contributions to
$\Delta a_{Qg}^{(3)}(x)$.
These partial results of the analytic calculation are shown as quantitative illustrations,
but they cannot be used for phenomenological analyses yet.

Let us define
%-------------------------------------------------------------------------------------------------
\begin{eqnarray}
(\Delta) r(N) = \frac{\Mvec[(\Delta) a_{Qg}^{(3)}(x)](N)}{(\Delta){\rm MOM}(N)} - 1,
\end{eqnarray}
%-------------------------------------------------------------------------------------------------
%---------------------------------------------------------------------------------------------------------------------------$
%---------------------------------------------------------------------------------------------------------------------------
%---------------------------------------------------------------------------------------------------------------------------$
%---------------------------------------------------------------------------------------------------------------------------
\noindent
with MOM denoting the moments calculated analytically in Mellin space. We have compared the moments 
based on the analytic results with the 
sum of the Mellin moments of the 
first--order--factorizable irreducible Feynman diagrams for $N = 2, \ldots, 20$ in the unpolarized case
obtaining
%---------------------------------------------------------------------------------------------------------------------------$
%---------------------------------------------------------------------------------------------------------------------------
%---------------------------------------------------------------------------------------------------------------------------$
%---------------------------------------------------------------------------------------------------------------------------

%-------------------------------------------------------------------------------------------------
\begin{eqnarray}
r(N) &=& \{-1.74353 \cdot 10^{-8},
         4.73887 \cdot 10^{-10}, 
        -2.03360 \cdot 10^{-12}, 
        -1.05471 \cdot 10^{-14}, 
\nonumber\\ &&
        -5.99520 \cdot 10^{-15}, 
        -6.88338 \cdot 10^{-15}, 
        -6.77236 \cdot 10^{-15}, 
        -6.77236 \cdot 10^{-15}, 
\nonumber\\ &&
        -7.21645 \cdot 10^{-15}, 
        -7.43849 \cdot 10^{-15}
\}.
\end{eqnarray}
%-------------------------------------------------------------------------------------------------
Similarly, in the polarized case we obtain for $N = 3, \ldots, 21$ 
%-------------------------------------------------------------------------------------------------
\begin{eqnarray}
\Delta r(N) &=& 
\{
-5.44219 \cdot 10^{-9},
-2.52013 \cdot 10^{-11},
-1.39555 \cdot 10^{-13},
-6.99441 \cdot 10^{-15},
\nonumber\\ &&
-6.10623 \cdot 10^{-15},
-6.66134 \cdot 10^{-15},
-6.55032 \cdot 10^{-15},
-7.32747 \cdot 10^{-15},
\nonumber\\ &&
-7.43849 \cdot 10^{-15},
-7.32747 \cdot 10^{-15}
\}.
\end{eqnarray}
%-------------------------------------------------------------------------------------------------

\noindent
We have also solved the system of first order differential equations 
by calculating symbolic series expansions around different values of 
$t$ and numerically matching these at points where two neighboring
expansions converge, choosing $t = 0, 1/7, 1/4, 3/4, 4/3, 2, 4$ and $\infty$
More details regarding this method can be found 
in Refs.~\cite{Fael:2021kyg,Fael:2022miw}.
For the solution of large linear systems of equations, which we 
encounter here, we make use of finite field techniques implemented in 
\texttt{FireFly}~\cite{Klappert:2019emp,Klappert:2020aqs}.
A comparison to the analytic solution has been performed at the 
points  
%-------------------------------------------------------------------------------------------------
\begin{eqnarray}
x \in \{
10^{-4},
10^{-3},
10^{-2},
2 \cdot 10^{-2},
7 \cdot 10^{-2},
0.1,
0.2,
0.3,
0.4,
0.5,
0.6,
0.7,
0.8,
0.9,
0.95,
0.99\}
\end{eqnarray}
%-------------------------------------------------------------------------------------------------
for the quantity
%-------------------------------------------------------------------------------------------------
\begin{eqnarray}
(\Delta)r_{\rm num./an} = \frac{(\Delta) a_{Qg}^{(3), \rm irr.~numeric}}{(\Delta) a_{Qg}^{(3), \rm irr.~analytic}} - 1.
\end{eqnarray}
%-------------------------------------------------------------------------------------------------
One obtains
%-------------------------------------------------------------------------------------------------
\begin{eqnarray}
r_{\rm num./an} &=&  
\{
7.75 \cdot 10^{-16},
1.86 \cdot 10^{-15},  
1.96 \cdot 10^{-14},
4.27 \cdot 10^{-15},
2.57 \cdot 10^{-15},  
7.01 \cdot 10^{-15},
\nonumber\\ &&
5.49 \cdot 10^{-10},
8.08 \cdot 10^{-18},
8.92 \cdot 10^{-18},
1.04 \cdot 10^{-17},
1.29 \cdot 10^{-17},
2.39 \cdot 10^{-19},
\nonumber\\ &&
3.85 \cdot 10^{-19},
2.46 \cdot 10^{-19},
1.39 \cdot 10^{-19},
6.61 \cdot 10^{-20}
\}
\end{eqnarray}
%-------------------------------------------------------------------------------------------------
and
%-------------------------------------------------------------------------------------------------
\begin{eqnarray}
{\Delta r_{\rm num./an}} &=& \{
1.35 \cdot 10^{-15},
9.32 \cdot 10^{-15},
4.34 \cdot 10^{-15},
3.63 \cdot 10^{-15},
1.45 \cdot 10^{-14},
4.03 \cdot 10^{-14},
\nonumber \\ &&
9.13 \cdot 10^{-10},
5.22 \cdot 10^{-18},
5.42 \cdot 10^{-18},
6.22 \cdot 10^{-18},
7.62 \cdot 10^{-20},
9.15 \cdot 10^{-19},
\nonumber
\\ &&
1.78 \cdot 10^{-19},
3.86 \cdot 10^{-19},
1.51 \cdot 10^{-19},
6.80 \cdot 10^{-20}
\}.
\end{eqnarray}
%-------------------------------------------------------------------------------------------------
Both the above numerical checks confirm our analytic results. 

%%%%%%%%%%%%%%%%%%%%%%%%%%%%%%%%%%%%%%%%%%%%%%%%%%%%%%%%%%%%%%%%%%%%%%%%%%%%%%%%%%%%%%%%%%%%%%%%%%% 
\section{Conclusions} \label{sec:8} 
%%%%%%%%%%%%%%%%%%%%%%%%%%%%%%%%%%%%%%%%%%%%%%%%%%%%%%%%%%%%%%%%%%%%%%%%%%%%%%%%%%%%%%%%%%%%%%%%%%%

\vspace*{1mm}
\noindent
The massive OMEs $A_{Qg}^{(3)}$ and $\Delta A_{Qg}^{(3)}$ receive contributions from non--first--order 
and first--order factorizable terms. In the present paper we have calculated the latter 
contributions.
In Mellin $N$--space these are given by harmonic sums, generalized harmonic sums and nested (inverse)
binomial sums.  The Mellin inversion to $x$--space, in particular for the terms 
containing binomial 
contributions, is most efficiently done by using the $t$--space representation. This requires to 
study the five different regions $t \in (0,1/2], [1/2,1], [1,2], [2,4]$ and $t \in [4,\infty)$.
In transforming to $x$--space, only at $x=1$ an imaginary part is implied, which is not changed at the
pseudo--thresholds $x = 1/4, 1/2$ and the amplitudes vanish for $x > 1$. This implies continuity of
the amplitudes in $x \in (0,1]$. The result in $x$--space, either obtained by direct Mellin inversion
or by the $t$--space method, are given by G--functions over a {14 letter} alphabet and a large 
set of 
G--constants at special numbers in the main argument. The integral representations of these constants
can all be rationalized and cast into Kummer--Poincar\'e type Riemann--integrals of complex--valued 
letters after regularization, if needed. These constants are calculated to 100 digits numerically. The 
amplitudes
are finally expanded into logarithmic--modulated Taylor series up to 100 terms around $x = 0, x = 1/2$ and $x 
= 1$. 

We also discussed a series of results for the representations in Mellin $N$--space, including 
non--first--order factorizable contributions. We determined all recurrences which required 15000 moments in 
the unpolarized case and 11000 moments in the polarized case, not covering a smaller number of recurrences
requiring an even larger number of moments, such as the purely rational color--$\zeta$ terms $\propto 
\textcolor{blue}{T_F}$. This allowed us to compute all non--purely rational 
terms and non $\zeta_3$ terms, despite the fact that the contributing diagrams partly contain non--first--order 
factorizable contributions. The latter canceled in the sum when using the method of arbitrary high moments and 
one obtains
nested sum--product representations. For the purely rational terms and their associated terms $\propto 
\zeta_3$ we have observed that the corresponding recurrences are highly divergent in the limit $N 
\rightarrow \infty$, while the sum of these contributions vanishes in this limit. The rational terms
generate a factor of $\zeta_3$ dynamically for $N \rightarrow \infty$. One may calculate the 
fundamental systems of the asymptotic representations of the non--first--order factorizable recurrences, 
although this requires quite some effort for large recurrences. Furthermore, a series of 
representations in $N$--space diverge $\propto a^N,~~a = 2,4$. These divergences have to be 
arranged to cancel analytically.

We have also calculated the leading small and large $x$ contributions from the first--order 
factorizable terms to the singularities of $O(\ln(x)/x)$ in the unpolarized and of $O(\ln^5(x))$ in the 
polarized case, as well as of $O(\ln^5(1-x))$ for $x \rightarrow 1$. In the unpolarized case our results
agree with those for the $\zeta_2$ term given in the literature. In the polarized case we obtained the 
leading small $x$ contribution $\propto \textcolor{blue}{N_F}$. This term cannot be obtained by color 
rescaling from the pure singlet term.

The master integrals computed for the present part of the project form analytic base case integrals of 
the second part of the calculation of $(\Delta) a_{Qg}^{(3)}$, in which the non--first--order 
factorizable contributions are computed. The corresponding $_2F_1$ solutions have already been calculated 
in Ref.~\cite{Behring:2023rlq} to $O(\ep^0)$. Ancillary files to this paper contain larger formulae and a 
series of technical results.

\appendix
%%%%%%%%%%%%%%%%%%%%%%%%%%%%%%%%%%%%%%%%%%%%%%%%%%%%%%%%%%%%%%%%%%%%%%%%%%%%%%%%%%%%%%%%%%%%%%%%%%%
\section{The asymptotic expansions of the contributing generalized harmonic sums}
\label{sec:A}
%%%%%%%%%%%%%%%%%%%%%%%%%%%%%%%%%%%%%%%%%%%%%%%%%%%%%%%%%%%%%%%%%%%%%%%%%%%%%%%%%%%%%%%%%%%%%%%%%%%

\vspace*{1mm}
\noindent
In the following we list a series of examples of different depth of the asymptotic expansions of the 
generalized harmonic sums which contribute to the present calculations. We present the terms up to $O(1/N^{10})$.
The complete set of expressions is given in an ancillary file. The asymptotic representations
are required in $N$--space programs, cf.~\cite{BS}.
Some of the contributions are suppressed by a factor of $2^{-N}$ and a series of terms 
diverges like $\propto 2^N$. The latter behavior cancels in the amplitude, including contributions
from the terms containing finite binomial sums. The derivation of theses asymptotic expansions 
is not straightforward even using the package {\tt HarmonicSums}. In a series of cases the 
use of shuffle relations is required before these expansions are performed.
This is one reason to present 
the corresponding expressions in explicit form. For $|N| = 50$ the relative accuracy of the 
representations amounts to values between $2.48 \cdot 10^{-26}$ and $0.95 \cdot 10^{-6}$.

Examples for asymptotic expansions are given by
%-------------------------------------------------------------------------------------------------
\begin{eqnarray}
% LIN1
\lefteqn{S_1({\{-2\},N}) \approx} \nonumber\\ && 
-\ln(3)
+ (-2)^N \Biggl[
        \frac{2}{3 N}
        -\frac{2}{9 N^2}
        -\frac{2}{27 N^3}
        +\frac{2}{27 N^4}
        +\frac{10}{81 N^5}
\nonumber\\ &&   
        -\frac{14}{243 N^6}
        -\frac{98}{243 N^7}
     -\frac{106}{729 N^8}
        +\frac{4430}{2187 N^9}
        +\frac{2518}{729 N^{10}}
        -\frac{28402}{2187 N^{11}}
\Biggr], 
\\ 
% LIN11
\lefteqn{S_{1,2}\left({\left\{\frac{1}{2},-1\right\},N}\right) \approx} \nonumber\\ && 
2^{-N}
\Biggl[(-1)^N \Biggl[
\frac{1}{6 N^3}
-\frac{1}{2N^4}
+\frac{2}{3 N^5}
+\frac{1}{6 N^6}
-\frac{7}{3 N^7}
+\frac{11}{6 N^8}
+\frac{12}{N^9}
\nonumber\\ &&
-\frac{57}{2 N^{10}}
-\frac{229}{3 N^{11}}
\Biggr]
+\Biggl[
 \frac{1}{2 N}
-\frac{1}{N^2}
+\frac{3}{N^3}
-\frac{13}{N^4}
+\frac{75}{N^5}
-\frac{541}{N^6}
\nonumber\\ &&
+\frac{4683}{N^7}
-\frac{47293}{N^8}
+\frac{545835}{N^9}
-\frac{7087261}{N^{10}}
+\frac{102247563}{N^{11}}
\Biggr] 
\zeta_2
\Biggr]
\nonumber\\ &&
+\HA_{0,-2,-1}(1)
+\HA_{-2,0,-1}(1)
-\frac{1}{2} \zeta_2 \ln(3),
\\
% LIN21
\lefteqn{S_{1,1,2}\left({\left\{-2,\frac{1}{2},-1\right\},N}\right) \approx} \nonumber\\ && 
-\frac{1}{18 N^3}
+\frac{5}{24 N^4}
-\frac{79}{180 N^5}
+\frac{37}{72 N^6}
+\frac{1}{9 N^7}
-\frac{59}{36 N^8}
\nonumber\\ && 
+\frac{763}{540 N^9}
+\frac{941}{120 N^{10}}
+L \Biggl[
        -\HA_{0,-2,-1}(1)
        -\HA_{-2,0,-1}(1)
\nonumber\\ &&    
     -\HA_{-2,-1,0}(1)
\Biggr]
+\Biggl[
        \Biggl[
\frac{2}{3 N}
-\frac{2}{9 N^2}
-\frac{2}{27 N^3}
+\frac{2}{27 N^4}
+\frac{10}{81 N^5}
\nonumber\\ && 
-\frac{14}{243 N^6}
-\frac{98}{243 N^7}
-\frac{106}{729 N^8}
+\frac{4430}{2187 N^9}
+\frac{2518}{729 N^{10}}
\Biggr] [\HA_{0,-2,-1}(1) 
\nonumber\\ && 
+ \HA_{-2,0,-1}(1)]\Biggr] (-2)^N
+\Biggl[
-\frac{1}{2 N}
+\frac{1}{12 N^2}
-\frac{1}{120 N^4}
-\frac{1}{240 N^8}
\nonumber\\ && 
+\frac{1}{252 N^6}
+        \frac{1}{132 N^{10}}
\Biggr] 
[\HA_{0,-2,-1}(1)+\HA_{-2,0,-1}(1)+\HA_{-2,-1,0}(1)]
\nonumber\\ && 
-\frac{1}{2} \ln^2(2) \zeta_2
+\frac{1}{2} \Biggl[\HA_{-1,2}(1)
        +  \HA_{-2,1}(1)
\Biggr] \zeta_2
+\ln(3) \Biggl[
-\frac{1}{3 N}
+\frac{1}{9 N^2}
\nonumber\\ && 
+\frac{1}{27 N^3}
-\frac{1}{27 N^4}
-\frac{5}{81 N^5}
+\frac{7}{243 N^6}
+\frac{49}{243 N^7}
+\frac{53}{729 N^8}
\nonumber\\ && 
-\frac{2215}{2187 N^9}
-\frac{1259}{729 N^{10}}
\Biggr] (-2)^N 
\zeta_2
+\Biggl[
 \frac{1}{4 N^2}
-\frac{3}{4 N^3}
+\frac{9}{4 N^4}
-\frac{37}{4 N^5}
\nonumber\\ && 
+\frac{105}{2 N^6}
-\frac{1505}{4 N^7}
+\frac{25893}{8 N^8}
-\frac{130101}{4 N^9}
     +   \frac{748035}{2 N^{10}}
\Biggr] (-1)^N \zeta_2
\nonumber\\ && 
+\HA_{-2,1,-1,0}(1)
+\HA_{-2,-1,0,1}(1)
+\HA_{-2,-1,1,0}(1)
-\Biggl[
        \HA_{0,-2,-1}(1)
\nonumber\\ &&     
    +\HA_{-2,0,-1}(1)
\Biggr] \ln(3)
+\frac{1}{2} \zeta_2 \ln^2(3),
\\
%LIN38
\lefteqn{S_{1,1,1,2}\left(\left\{1,2,\frac{1}{2},-1\right\}\right)(N) \approx} \nonumber\\ && 
\ln(2) \Biggl[
        \Biggl[
 \frac{1}{2 N}
-\frac{1}{12 N^2}
+\frac{1}{120 N^4}
-\frac{1}{252 N^6}
+\frac{1}{240 N^8}
                -\frac{1}{132 N^{10}}
\Biggr] 
\nonumber\\ &&  \times
\HA_{0,-2,-1}(1)
        +\Biggl[
 \frac{1}{2 N}
-\frac{1}{12 N^2}
+\frac{1}{120 N^4}
-\frac{1}{252 N^6}
+\frac{1}{240 N^8}
\nonumber\\ && 
                -\frac{1}{132 N^{10}}
\Biggr] \HA_{-2,0,-1}(1)
        -\HA_{0,-1,-2,-1}(1)
        -\HA_{-1,0,-2,-1}(1)
\nonumber\\ &&    
    -\HA_{-1,-2,0,-1}(1)
        +\ln(3) \Biggl[
-\frac{1}{4 N}
+\frac{1}{24 N^2}
-\frac{1}{240 N^4}
+\frac{1}{504 N^6}
\nonumber\\ && 
-\frac{1}{480 N^8}
                +\frac{1}{264 N^{10}}
\Biggr] \zeta_2
        +\frac{1}{2} \HA_{-1,-2}(1) \zeta_2
\Biggr]
+L \Biggl[
        \ln(2) 
\nonumber\\ &&  \times
\Biggl[
                \HA_{0,-2,-1}(1)
                +\HA_{-2,0,-1}(1)
                -\frac{1}{2} \ln(3) \zeta_2
        \Biggr]
        -\HA_{0,-1,-2,-1}(1)
\nonumber\\ &&        
 -\HA_{-1,0,-2,-1}(1)
        -\HA_{-1,-2,0,-1}(1)
        +\frac{1}{2} \ln^2(2) \zeta_2
        +\frac{1}{2} \HA_{-1,-2}(1) 
\nonumber\\ &&  \times
\zeta_2
        +\frac{1}{4} \zeta_2^2
\Biggr]
+\ln^2(2) \Biggl[
         \frac{1}{2} \HA_{0,-2,-1}(1)
        +\frac{1}{2} \HA_{-2,0,-1}(1)
        +\Biggl[
 \frac{1}{4 N}
\nonumber\\ && 
-\frac{1}{24 N^2}
+\frac{1}{240 N^4}
-\frac{1}{504 N^6}
+\frac{1}{480 N^8}
                -\frac{1}{264 N^{10}}
\Biggr] \zeta_2
\nonumber\\ && 
        -\frac{1}{4} \zeta_2 \ln(3)
\Biggr]
+\Biggl[
\frac{1}{24 N^5}
-\frac{5}{16 N^6}
+\frac{53}{48 N^7}
-\frac{25}{16 N^8}
-\frac{359}{96 N^9}
\nonumber\\ && 
        +\frac{3763}{192 N^{10}}
\Biggr] (-1)^N
+\Biggl[
        \Biggr[
 \frac{4}{N^2}
+\frac{12}{N^3}
+\frac{60}{N^4}
+\frac{380}{N^5}
+\frac{2940}{N^6}
\nonumber\\ && 
+\frac{26908}{N^7}
+\frac{284508}{N^8}
+\frac{3413628}{N^9}
+\frac{45832380}{N^{10}}
\Biggr] 
\nonumber\\ && 
\times \HA_{0,-2,-1}(1)
        +\Biggl[
 \frac{4}{N^2}
+\frac{12}{N^3}
+\frac{60}{N^4}
+\frac{380}{N^5}
+\frac{2940}{N^6}
+\frac{26908}{N^7}
\nonumber\\ && 
+\frac{284508}{N^8}
+\frac{3413628}{N^9}
+\frac{45832380}{N^{10}}
\Biggr] \HA_{-2,0,-1}(1)
        +\ln(3) 
\nonumber\\ &&  \times
\Biggl[
-\frac{2}{N^2}
-\frac{6}{N^3}
-\frac{30}{N^4}
-\frac{190}{N^5}
-\frac{1470}{N^6}
-\frac{13454}{N^7}
-\frac{142254}{N^8}
\nonumber\\ && 
-\frac{1706814}{N^9}
-\frac{22916190}{N^{10}}
\Biggr] 
\zeta_2
\Biggr] 2^N
+\Biggl[
-\frac{1}{2 N}
+\frac{1}{12 N^2}
-\frac{1}{120 N^4}
\nonumber\\ && 
+\frac{1}{252 N^6}
-\frac{1}{240 N^8}
+        \frac{1}{132 N^{10}}
\Biggr] [\HA_{0,-1,-2,-1}(1) + \HA_{-1,0,-2,-1}(1) 
\nonumber\\ && 
+ \HA_{-1,-2,0,-1}(1)]
+\frac{1}{4} \ln^3(2) \zeta_2
+\Biggl[
         \frac{1}{2 N}
        -\frac{5}{8 N^2}
        +\frac{71}{72 N^3}
\nonumber\\ &&        
 -\frac{107}{48 N^4}
        +\frac{6757}{900 N^5}
        -\frac{8503}{240 N^6}
        +\frac{478466}{2205 N^7}
        -\frac{549383}{336 N^8}
\nonumber\\ &&     
    +\frac{10242761}{700 N^9}
        -\frac{36407611}{240 N^{10}}
        +\Biggl[
 \frac{1}{4 N}
-\frac{1}{24 N^2}
+\frac{1}{240 N^4}
\nonumber\\ && 
-\frac{1}{504 N^6}
+\frac{1}{480 N^8}
                -\frac{1}{264 N^{10}}
\Biggl] \HA_{-1,-2}(1)
        -\frac{1}{2} \HA_{-1,-1,-2}(1)
\nonumber\\ &&        
 -\frac{3}{8} \zeta_3
\Biggr] \zeta_2
+\Biggl[
 \frac{1}{8 N}
-\frac{1}{48 N^2}
+\frac{1}{480 N^4}
-\frac{1}{1008 N^6}
+\frac{1}{960 N^8}
\nonumber\\ && 
        -\frac{1}{528 N^{10}}
\Biggr] \zeta_2^2
+\HA_{0,-1,-1,-2,-1}(1)
+\HA_{-1,0,-1,-2,-1}(1)
\nonumber\\ && 
+\HA_{-1,-1,0,-2,-1}(1)
+\HA_{-1,-1,-2,0,-1}(1),
\\
%LIN39
\lefteqn{S_{1,1,1,2}\left(\left\{1,2,\frac{1}{2},1\right\}\right)(N) \approx} \nonumber\\ && 
\frac{67}{16} \zeta_5
+\frac{1}{4 N^2}
-\frac{11}{12 N^3}
+\frac{295}{96 N^4}
-\frac{953}{80 N^5}
+\frac{250711}{4320 N^6}
\nonumber\\ && 
-\frac{359743}{1008 N^7}
+\frac{162670639}{60480 N^8}
-\frac{36393227}{1512 N^9}
+\frac{15720627001}{63000 N^{10}}
\nonumber\\ && 
+\Biggl[
        -\frac{9}{16} \zeta_3
        -\frac{1}{N}
        +\frac{5}{4 N^2}
        -\frac{71}{36 N^3}
        +\frac{107}{24 N^4}
        -\frac{6757}{450 N^5}
\nonumber\\ &&        
 +\frac{8503}{120 N^6}
        -\frac{956932}{2205 N^7}
        +\frac{549383}{168 N^8}
        -\frac{10242761}{350 N^9}
        +\frac{36407611}{120 N^{10}}
\Biggr] \zeta_2
\nonumber\\ && 
+\Biggl[
        -\frac{17}{40} L
        -\frac{17}{80 N}
        +\frac{17}{480 N^2}
        -\frac{17}{4800 N^4}
        +\frac{17}{10080 N^6}
\nonumber\\ &&       
  -\frac{17}{9600 N^8}
        +\frac{17}{5280 N^{10}}
\Biggr] \zeta_2^2
+\Biggl[
 \frac{5}{2 N^2}
+\frac{15}{2 N^3}
+\frac{75}{2 N^4}
\nonumber\\ && 
+\frac{475}{2 N^5}
+\frac{3675}{2 N^6}
+\frac{33635}{2 N^7}
+\frac{355635}{2 N^8}
+\frac{4267035}{2 N^9}
\nonumber\\ &&   
     +\frac{57290475}{2 N^{10}}
\Biggr] 2^N \zeta_3,
\\
%LIN43
\lefteqn{S_{1,1,1,1,1}\left(\left\{\frac{1}{2},2,1,1,1\right\}\right)(N) \approx} \nonumber\\ && 
4 \Li_5\left(\frac{1}{2}\right)
+ \frac{1}{120} \ln^5(2)
-\frac{2}{N}
+\frac{1}{4 N^2}
+\frac{17}{324 N^3}
-\frac{3}{4 N^4}
\nonumber\\ &&
-\frac{29171}{45000 N^5}
+\frac{17837}{2160 N^6}
+\frac{13192844957}{62233920 N^7}
+\frac{3835123}{1120 N^8}
\nonumber\\ && 
+\frac{858280391021}{16329600 N^9}
+\frac{940261784593}{1134000 N^{10}}
+L^2 \Biggl[
-\frac{1}{N}
+\frac{25}{36 N^3}
+\frac{5}{2 N^4}
\nonumber\\ && 
+\frac{2993}{200 N^5}
+\frac{314}{3 N^6}
+\frac{3823133}{4410 N^7}
+\frac{500479}{60 N^8}
+\frac{1385608621}{15120 N^9}
\nonumber\\ &&      
   +\frac{1428138841}{1260 N^{10}}
\Biggr]
+L^3 \Biggl[
-\frac{1}{3 N}
-\frac{2}{9 N^3}
-\frac{2}{3 N^4}
-\frac{142}{45 N^5}
-\frac{58}{3 N^6}
\nonumber\\ && 
-\frac{9146}{63 N^7}
-\frac{1294}{N^8}
-\frac{601342}{45 N^9}
        -\frac{470906}{3 N^{10}}
\Biggr]
+L \Biggl[
        -\frac{2}{N}
        +\frac{1}{2 N^2}
\nonumber\\ &&     
    -\frac{19}{108 N^3}
        -\frac{1}{3 N^4}
        -\frac{190517}{18000 N^5}
        -\frac{18941}{180 N^6}
        -\frac{5946741}{5488 N^7}
\nonumber\\ &&    
     -\frac{3813722}{315 N^8}
        -\frac{134889193063}{907200 N^9}
        -\frac{25291803223}{12600 N^{10}}
        +\Biggl[
-\frac{1}{N}
-\frac{2}{3 N^3}
\nonumber\\ && 
-\frac{2}{N^4}
-\frac{58}{N^6}
-\frac{9146}{21 N^7}
-\frac{142}{15 N^5}
-\frac{3882}{N^8}
-\frac{601342}{15 N^9}
                -\frac{470906}{N^{10}}
\Biggr] 
\nonumber\\ &&  \times
\zeta_2
\Biggr]
+\ln(2) \Biggl[
        \Li_4\left(\frac{1}{2}\right)
        +\frac{12}{5} \zeta_2^2
\Biggr]
+\Biggl[
        \Li_4\left(\frac{1}{2}\right) \Biggl[ 
-\frac{1}{N}
+\frac{2}{N^2}
\nonumber\\ && 
-\frac{6}{N^3}
+\frac{26}{N^4}
-\frac{150}{N^5}
+\frac{1082}{N^6}
-\frac{9366}{N^7}
+\frac{94586}{N^8}
-\frac{1091670}{N^9}
\nonumber\\ && 
+\frac{14174522}{N^{10}}
\Biggr]
        +\ln^4(2) \Biggl[
-\frac{1}{24 N}
+\frac{1}{12 N^2}
-\frac{1}{4 N^3}
+\frac{13}{12 N^4}
\nonumber\\ && 
-\frac{25}{4 N^5}
+\frac{541}{12 N^6}
-\frac{1561}{4 N^7}
+\frac{47293}{12 N^8}
-\frac{181945}{4 N^9}
+\frac{7087261}{12 N^{10}}
\Biggr]
\nonumber\\ && 
        +\ln^2(2) \Biggl[
 \frac{1}{N}
-\frac{2}{N^2}
+\frac{6}{N^3}
-\frac{26}{N^4}
+\frac{150}{N^5}
-\frac{1082}{N^6}
+\frac{9366}{N^7}
\nonumber\\ && 
-\frac{94586}{N^8}
+\frac{1091670}{N^9}
-\frac{14174522}{N^{10}}
\Biggr] 
\zeta_2
        +\Biggl[
                \frac{4}{5 N}
-\frac{8}{5 N^2}
+\frac{24}{5 N^3}
\nonumber\\ && 
-\frac{104}{5 N^4}
+\frac{120}{N^5}
-\frac{4328}{5 N^6}
+\frac{37464}{5 N^7}
-\frac{378344}{5 N^8}
+\frac{873336}{N^9}
\nonumber\\ && 
-\frac{56698088}{5 N^{10}}
\Biggr] \zeta_2^2
\Biggr] 2^{-N}
+\frac{1}{3} \ln^3(2) \zeta_2
+\Biggl[
-\frac{1}{N}
+\frac{25}{36 N^3}
+\frac{5}{2 N^4}
\nonumber\\ && 
+\frac{2993}{200 N^5}
+\frac{314}{3 N^6}
+\frac{3823133}{4410 N^7}
+\frac{500479}{60 N^8}
+\frac{1385608621}{15120 N^9}
\nonumber\\ && 
+        \frac{1428138841}{1260 N^{10}}
\Biggr] \zeta_2
+\Biggl[
-\frac{2}{3 N}
-\frac{4}{9 N^3}
-\frac{4}{3 N^4}
-\frac{284}{45 N^5}
-\frac{116}{3 N^6}
\nonumber\\ && 
-\frac{18292}{63 N^7}
-\frac{2588}{N^8}
-\frac{1202684}{45 N^9}
        -\frac{941812}{3 N^{10}}
\Biggr] \zeta_3,
\end{eqnarray}
%------------------------------------------------------------------------------------------------- 
Some of the above constants can be expressed in terms of  polylogarithms. We present the relations up to 
depth three, since 
from depth 4 classical polylogarithms do not usually provide a good basis. One obtains
%------------------------------------------------------------------------------------------------- 
\begin{eqnarray} 
\HA_{-1,-2}(1)  &=& 
\frac{1}{2} \zeta_2
-\ln^2(2)
+\ln(2) \ln(3)
-\frac{1}{2} \ln^2(3)
- \Li_2\left[
        \frac{1}{3}\right],
%------------------------------------------------------------------------------------------------- 
\\ 
\HA_{-1,2}(1) &=& 
\ln ^2(2)
-\ln^2(3)
-2 \Li_2\left[
        \frac{1}{3}\right]
+\zeta_2,
%------------------------------------------------------------------------------------------------- 
\\
\HA_{-2,1}(1) &=& \Li_2\left[\frac{1}{3}\right],
\\
%------------------------------------------------------------------------------------------------- 
\HA_{0,4}(1) &=&
-2 \ln^2(2)
+2 \ln(2) \ln(3)
-\ln^2(3)
+\zeta_2
-2 \Li_2\left[
        \frac{1}{3}\right],
\\
%------------------------------------------------------------------------------------------------- 
\HA_{0,4,2}(1) &=& 
-\frac{3}{2} \zeta_2 \ln(2)
-\frac{5}{6} \ln^3(2)
+2 \ln^2(2) \ln(3)
-\ln(2) \ln^2(3)
+\frac{7}{2} \zeta_3
\nonumber\\ &&
-2 \ln(2) \Li_2\left[
        \frac{1}{3}\right]
+4 \Li_3\left[
        -\frac{1}{2}\right],
\\
%------------------------------------------------------------------------------------------------- 
\HA_{0,0,-2}(1) &=& - \Li_3\left[-\frac{1}{2}\right],
\\
%------------------------------------------------------------------------------------------------- 
\HA_{0,0,4}(1) &=& 
-2 \zeta_2 \ln(2)
+\frac{2}{3} \ln^3(2)
+\frac{7}{2} \zeta_3
+4 \Li_3\left[
        -\frac{1}{2}\right],
\\
%------------------------------------------------------------------------------------------------- 
\HA_{0,1,-2}(1) &=& 
-2 \zeta_2 (\ln(2)-\ln(3))
-\frac{1}{2} \ln^2(2) \ln(3)
+\ln(2) \ln^2(3)
-\frac{1}{2} \ln^3(3)
\nonumber\\ &&
+(\ln(2)-\ln(3)) \Li_2\left[
        \frac{1}{3}\right]
-\Li_3\left[
        \frac{1}{3}\right]
+\Li_3\left[
        \frac{2}{3}\right]
-\zeta_3,
%------------------------------------------------------------------------------------------------- 
\\
\HA_{0,-2,1}(1) &=& -\zeta_2 \ln\left(
        \frac{3}{2}\right)
-\frac{1}{6} \ln^3\left(
        \frac{3}{2}\right)
-\Li_3\left[
        -\frac{1}{2}\right]
+\Li_3\left[
        \frac{1}{3}\right],
%------------------------------------------------------------------------------------------------- 
\\
\HA_{0,-2,-1}(1) &=& 
-\frac{4}{3} \ln^3(2)
+3 \zeta_2 (\ln(2)-\ln(3))
+\ln^2(2) \ln(3)
-\frac{1}{2} \ln(2) \ln^2(3)
+\frac{1}{2} \ln^3(3)
\nonumber\\ && 
+\ln(2) \Li_2\left[
        \frac{1}{3}\right]
+\Li_3\left[
        -\frac{1}{3}\right]
-2 \Li_3\left[
        \frac{1}{3}\right]
-2 \Li_3\left[
        \frac{2}{3}\right]
+ \Li_3\left[
        \frac{3}{4}\right]
\nonumber\\ &&
+\frac{21}{8} \zeta_3,
%------------------------------------------------------------------------------------------------- 
\\ 
\HA_{-1,2,-1}(1) &=& 
\zeta_2 \ln(2)
+\ln^2(2) \ln(3)
-\ln(2) \ln^2(3)
+2 \Li_3\left[
        \frac{1}{3}\right]
-2 \Li_3\left[
        \frac{2}{3}\right],
%------------------------------------------------------------------------------------------------- 
\\
\HA_{-1,0,2}(1) &=& 
\frac{5}{6} \ln^3(2)
-\ln(2) \ln^2(3)
+\frac{1}{6} \ln^3(3)
+\frac{1}{2} \zeta_2 \Biggl[-\ln(2)+2 \ln(3)\Biggr]
\nonumber\\ && 
-2 \ln(2) \Li_2\left[
        \frac{1}{3}\right]
-\Li_3\left[
        -\frac{1}{3}\right]
-\Li_3\left[
        \frac{3}{4}\right]
+\frac{1}{4} \zeta_3,
%------------------------------------------------------------------------------------------------- 
\\
\HA_{-2,0,-1}(1) &=& 
\frac{4}{3} \ln^3(2)
-\ln^2(2) \ln(3)
+\frac{1}{2} \ln(2) \ln^2(3)
-\frac{1}{3} \ln^3(3)
+\frac{1}{2} \zeta_2 \Biggl[-6 \ln(2)
\nonumber\\ && 
+5 \ln(3)\Biggr]
-\ln(2) \Li_2\left[
        \frac{1}{3}\right]
+2 \Li_3\left[
        \frac{2}{3}\right]
-\Li_3\left[
        \frac{3}{4}\right]
-\zeta_3.
\end{eqnarray} 
%------------------------------------------------------------------------------------------------- 
Here we used the relation
%------------------------------------------------------------------------------------------------- 
\begin{eqnarray} 
\Li_2\left[\frac{1}{3}\right] 
 &=& - \frac{1}{2} \ln^2\left(\frac{2}{3}\right) -
\Li_2\left[-\frac{1}{2}\right] 
\end{eqnarray} 
%------------------------------------------------------------------------------------------------- 
and other ones for $\Li_2(x)$, cf.~\cite{LEWIN1,LEWIN2,Devoto:1983tc}.
The remaining generalized polylogarithms at argument $x=1$ can be calculated numerically, e.g. with
the program of Ref.~\cite{Vollinga:2004sn}, and we have listed their numerical values in ancillary
files at an accuracy of 100 digits for completeness.

%%%%%%%%%%%%%%%%%%%%%%%%%%%%%%%%%%%%%%%%%%%%%%%%%%%%%%%%%%%%%%%%%%%%%%%%%%%%%%%%%%%%%%%%%%%%%%%%%%%
\section{The asymptotic expansions of nested binomial sums}
\label{sec:B}
%%%%%%%%%%%%%%%%%%%%%%%%%%%%%%%%%%%%%%%%%%%%%%%%%%%%%%%%%%%%%%%%%%%%%%%%%%%%%%%%%%%%%%%%%%%%%%%%%%%

\vspace*{1mm}
\noindent
We will give only some examples for the asymptotic expansion of nested binomial sums in the 
following.\footnote{A series of 
representations has been given in Refs.~\cite{Ablinger:2014bra,Ablinger:2022wbb}.}
Also here aspects summarized in Ref.~\cite{Blumlein:2009ta} play a central role, see also 
\cite{Blumlein:2023vwi}. 
In $N$--space one first splits off lower--order factors and expands them individually asymptotically.
Then one considers the $t$--representations and splits off potential distribution--valued contributions, 
cf.~\cite{Behring:2023rlq}. The asymptotic expansion of these contributions is known
\cite{Blumlein:1998if}. After this the respective quantities $F(N)$ are viewed as Mellin transforms of  
functions $f(x)$, which are either analytic in the vicinity of $x=1$, or have to be first rewritten to obey 
this condition, \cite{Blumlein:2009ta}. These Mellin transforms can then be expanded into asymptotic series, 
because
they have representations in terms of factorial series \cite{FACT1,FACT2,FACT3},
%-------------------------------------------------------------------------------------------------
\begin{eqnarray}
\Omega(N) = \sum_{k=0}^\infty a_{k+1} \frac{k!}{N(N+1) \ldots (N+k)}.
\end{eqnarray}
%-------------------------------------------------------------------------------------------------
Terms $\propto S_1^k(N), k \geq 1, k \in \mathbb{N},$ are no factorial series, but their
asymptotic series are known.
Contributing $\Gamma$--functions of non--integer arguments may imply other powers of $N$, as 
e.g. factors of $\sqrt{N}$ and also factors $a^N$ with $a \in \mathbb{Z} \backslash \{0\}$.
A comprehensive treatment of these contributions is better given by the $t$--space representation,
being more systematic.
The examples for asymptotic representations given in the following were obtained by using the command 
{\tt BSExpansion}. Usually the resulting expressions have an involved form. One obtains 
the following asymptotic expansions
up to $O(1/N^{10})$
%-------------------------------------------------------------------------------------------------
\begin{eqnarray}
%--1
\tilde{F}_1(N) = \lefteqn{\sum_{\tau_1=1}^N 
\frac{\big(
        \tau_1!\big)^2}{\big(
        2 \tau_1\big)!} \approx}  
\nonumber\\ && \frac{1}{3}
+\frac{2 \pi}{9 \sqrt{3}} 
+\Biggl[
        -\frac{1}{3}
-\frac{19}{72 N}
+\frac{407}{3456 N^2}
-\frac{3587}{27648 N^3}
+\frac{612727}{2654208 N^4}
\nonumber\\ &&
-\frac{36974287}{63700992 N^5}
+\frac{640361849}{339738624 N^6}
-\frac{60994830787}{8153726976 N^7}
+\frac{54910715707991}{1565515579392 N^8}
\nonumber\\ && 
-\frac{2376532755785617}{12524124635136 N^9}
+\frac{699435110164273561}{601157982486528 N^{10}}
-\frac{115034280046636642783}{14427791579676672 N^{11}}
\Biggr]
\nonumber\\ && \times 
2^{-2 N} \sqrt{N} \sqrt{\pi },
\\
%-------------------------------------------------------------------------------------------------
%--2
\tilde{F}_2(N)= \lefteqn{\sum_{\tau_1=1}^N \frac{\big(
        2 \tau_1\big)!}{\big(
        \tau_1!\big)^2} \approx}
\nonumber\\
&&
\Biggl[
        \frac{4}{3}
+\frac{1}{18 N}
+\frac{59}{288 N^2}
+\frac{2425}{6912 N^3}
+\frac{576793}{663552 N^4}
+\frac{5000317}{1769472 N^5}
+\frac{953111599}{84934656 N^6}
\nonumber\\ && 
+\frac{107249721865}{2038431744 N^7}
+\frac{37133194953283}{130459631616 N^8}
+\frac{5464331904405803}{3131031158784 N^9}
\nonumber\\ && 
+\frac{1797410945424609151}{150289495621632 N^{10}}
\Biggr]
\frac{4^N}{\sqrt{N} \sqrt{\pi }},
%--------------------------------------------------------------------------------------
%--3
\\
\tilde{F}_3(N) = \lefteqn{\sum_{\tau_1=1}^N \frac{\big(
        \tau_1!\big)^2 
\sum_{\tau_2=1}^{\tau_1} \frac{(-1)^{\tau_2} \big(
        2 \tau_2\big)!}{\big(
        \tau_2!\big)^2 \tau_2^3}}{\big(
        2 \tau_1\big)!} \approx}
\nonumber\\ &&
-
\frac{16}{9}
-\Biggl[
        \frac{4}{9} \text{G}\Biggl[\Biggl\{
                \frac{\sqrt{1+4 \tau }}{\tau },\frac{1}{\tau }\Biggr\},1\Biggr]
        +\frac{2}{9} \text{G}\Biggl[\Biggl\{
                \frac{\sqrt{1+4 \tau }}{\tau },\frac{1}{\tau },\frac{1}{\tau 
}\Biggr\},1\Biggr]
\Biggr] \text{G}\Biggl[\Biggl\{
        \frac{\sqrt{1-\tau }}{4-\tau }\Biggr\},1\Biggr]
\nonumber\\ && 
-\Biggl[
        \frac{4}{9} \text{G}\Biggl[\Biggl\{
                \frac{\sqrt{1+4 \tau }}{\tau },\frac{1}{\tau }\Biggr\},1\Biggr]
        +\frac{2}{9} \text{G}\Biggl[\Biggl\{
                \frac{\sqrt{1+4 \tau }}{\tau },\frac{1}{\tau },\frac{1}{\tau 
}\Biggr\},1\Biggr]
\Biggr] \text{G}\Biggl[\Biggl\{
        \frac{\sqrt{1+4 \tau }}{1+\tau }\Biggr\},1\Biggr]
\nonumber\\ && 
+\frac{4}{9} 
        \text{G}\Biggl[\Biggl\{
                \frac{\sqrt{1+4 \tau }}{1+\tau },\frac{\sqrt{1+4 \tau }}{\tau 
},\frac{1}{\tau }\Biggr\},1\Biggr]
+\frac{2}{9} 
        \text{G}\Biggl[\Biggl\{
                \frac{\sqrt{1+4 \tau }}{1+\tau },\frac{\sqrt{1+4 \tau }}{\tau 
},\frac{1}{\tau },\frac{1}{\tau }\Biggr\},1\Biggr]
\nonumber\\ && 
+\Biggl[
\frac{2}{5 N^3}
-\frac{22}{25 N^4}
+\frac{24}{125 N^5}
+\frac{1188}{625 N^6}
-\frac{947}{3125 N^7}
-\frac{57253}{6250 N^8}
-\frac{28938}{15625 N^9}
\nonumber\\ && 
        +\frac{5225906}{78125 N^{10}} 
\Biggr] (-1)^N
+ \Biggl\{
        \Biggl[
-\frac{2}{3} \sqrt{N}
-\frac{19}{36} \frac{1}{\sqrt{N}}
+\frac{407 }{1728}	\frac{1}{N^{3/2}}
-\frac{3587 }{13824}\frac{1}{N^{5/2}}
\nonumber\\ && 
+\frac{612727 }{1327104}\frac{1}{N^{7/2}}
-\frac{36974287 }{31850496}\frac{1}{N^{9/2}}
+\frac{640361849 }{169869312}\frac{1}{N^{11/2}}
-\frac{60994830787 }{4076863488}\frac{1}{N^{13/2}}
\nonumber\\ && 
+\frac{54910715707991 }{782757789696}\frac{1}{N^{15/2}}
-\frac{2376532755785617 }{6262062317568}\frac{1}{N^{17/2}}
\nonumber\\ &&                
         +
                        \frac{699435110164273561}{300578991243264} \frac{1}{N^{19/2}}
                        -\frac{115034280046636642783 }{7213895789838336} \frac{1}{N^{21/2}}
\Biggr]
\nonumber\\ &&  \times 
\text{G}\Biggl[\Biggl\{
                        \frac{\sqrt{1+4 \tau }}{\tau },\frac{1}{\tau }\Biggr\},1\Biggr] 
\sqrt{\pi}
+                \Biggl[
-\frac{1}{3} \sqrt{N}
-\frac{19}{72} \frac{1}{\sqrt{N}}
+\frac{407}{3456} \frac{1}{N^{3/2}}
-\frac{3587}{27648} \frac{1}{N^{5/2}}
\nonumber\\ && 
+\frac{612727}{2654208} \frac{1}{N^{7/2}}
-\frac{36974287}{63700992} \frac{1}{N^{9/2}}
+\frac{640361849 }{339738624}\frac{1}{N^{11/2}}
-\frac{60994830787 }{8153726976}\frac{1}{N^{13/2}}
\nonumber\\ && 
+\frac{54910715707991 }{1565515579392}\frac{1}{N^{15/2}}
-\frac{2376532755785617 }{12524124635136}\frac{1}{N^{17/2}}
\nonumber\\ && 
+\frac{699435110164273561 }{601157982486528}\frac{1}{N^{19/2}}
                -\frac{115034280046636642783 }{14427791579676672}\frac{1}{N^{21/2}}
\Biggr] 
\nonumber\\ && 
\times
\text{G}\Biggl[\Biggl\{
                        \frac{\sqrt{1+4 \tau }}{\tau },\frac{1}{\tau },\frac{1}{\tau 
}\Biggr\},1\Biggr] \sqrt{\pi }
\Biggr\} 2^{-2 N}.
%-------------------------------------------------------------------------------------------------
\end{eqnarray}
%-------------------------------------------------------------------------------------------------
The different G--constants at $x=1$ have to be regulated in general and should  be 
rationalized, whenever possible. Then their letters should be partial fractioned, prior to 
calculating them numerically.

To perform the inverse Mellin transform, one can compute first the $t$--representation by using the command {\tt 
ComputeGeneratingFunction} 
%-------------------------------------------------------------------------------------------------
\begin{eqnarray}
F_1(t) &=&
t \Biggl\{
        -\frac{1}{2 (t-1)}
        -\frac{1}{(4-t)^{3/2} \sqrt{t}(t-1)}
\text{G}\left[ \left\{
                \sqrt{4-\tau } \sqrt{\tau }\right\},t\right]
\Biggr\},
\\
F_2(t) &=& \frac{1}{t-1} \Biggl[1 - \frac{1}{\sqrt{1-4t}}\Biggr],
\\
F_3(t) &=&
t \Biggl\{
        -\frac{5}{2 (t-1)}
        +\frac{\HA_{-1}(t)}{t-1}
        -\frac{\HA_{0,-1}(t)}{2 (t-1)}
        +\frac{(2+t) \HA_{0,0,-1}(t)}{2 (t-1) t}
+ \frac{1}{(4 - t)^{3/2} (t-1) \sqrt{t}} 
\nonumber\\ &&
\times \Biggl[
 -4  {\rm G}\Biggl[\Biggl\{\sqrt{(4 - \tau)\tau}\Biggr\}, t\Biggr]
+ 5  {\rm G}\Biggl[\Biggl\{\frac{\sqrt{4 - \tau}}{\sqrt{\tau(1+\tau)}}\Biggr\}, t\Biggr]
+ \frac{3}{2}
     {\rm G}\Biggl[\Biggl\{\sqrt{(4 - \tau)\tau}, \frac{1}{1+\tau}\Biggr\}, t\Biggr]
     \nonumber\\ && 
     - \frac{1}{2}
     {\rm G}\Biggl[\Biggl\{\sqrt{(4- \tau) \tau},\frac{1}{\tau},\frac{1}{1+\tau}\Biggr\}, 
t\Biggr]
+  {\rm G}\Biggl[\Biggl\{\sqrt{(4 - \tau) \tau}, \frac{1}{\tau}, \frac{1}{\tau}, \frac{1}{1+\tau}
\Biggr\}, t\Biggr]
\Biggr]
\Biggr\}.
\end{eqnarray}
%-------------------------------------------------------------------------------------------------
The binomial sums above can be represented by regularized Mellin transforms over an extended support
%-------------------------------------------------------------------------------------------------
\begin{eqnarray}
\tilde{F}_1(N) &=& \left[\frac{1}{3} + \frac{2 \pi}{9 \sqrt{3}}\right]
\left(1 - \frac{1}{4^N}\right) + \frac{1}{2 \cdot 4^N} \int_0^1 dy \frac{y^N-1}{(1-y)^{3/2}} \frac{y}{4-y},
\\
\tilde{F}_2(N) &=& \frac{1}{\pi} \int_0^4 dy \frac{y^N-1}{y-1} \frac{\sqrt{y}}{\sqrt{4-y}},
\\ 
\tilde{F}_3(N) &=& 
-\frac{16}{9}
+\Biggl(
        \frac{7}{9} 2^{1-2 N}
        +\frac{1}{9} \big(
                4^N-1\big) 4^{1-N} \text{G}\Biggl[\Biggl\{
                \frac{\sqrt{1-\tau }}{4-\tau }\Biggr\},1\Biggr]
\Biggr) \text{G}\Biggl[\Biggl\{
        \frac{1}{\tau },\frac{\sqrt{1+4 \tau }}{\tau }\Biggr\},1\Biggr]
\nonumber\\ && 
+\frac{4}{9} 
        \text{G}\Biggl[\Biggl\{
                \frac{1}{\tau },\frac{\sqrt{1+4 \tau }}{\tau },\frac{\sqrt{1+4 \tau }}{1+\tau 
}\Biggr\},1\Biggr]
-\frac{2}{9} 
        \text{G}\Biggl[\Biggl\{
                \frac{1}{\tau },\frac{1}{\tau },\frac{\sqrt{1+4 \tau }}{\tau },\frac{\sqrt{1+4 \tau 
}}{1+\tau }\Biggr\},1\Biggr]
\nonumber\\ &&  
-\frac{1}{9} \big(
        4^N-1
\big)
\text{G}\Biggl[\Biggl\{
                \frac{\sqrt{1-\tau }}{4-\tau }\Biggr\},1\Biggr] \text{G}\Biggl[\Biggl\{
                \frac{1}{\tau },\frac{1}{\tau },\frac{\sqrt{1+4 \tau }}{\tau }\Biggr\},1\Biggr] 
2^{1-2 N}
\nonumber\\ && 
-\frac{7}{9} 4^{-N} \text{G}\Biggl[\Biggl\{
        \frac{1}{\tau },\frac{1}{\tau },\frac{\sqrt{1+4 \tau }}{\tau }\Biggr\},1\Biggr]
+ f_{a3} 2^{-1-2N} \int_0^1 dx \frac{x(x^N-1)}{(1-x)^{3/2}(4-x)} 
\nonumber\\ && 
+ \int_0^1 dx \frac{x^{N+1} f_{b3}(x) 2 (-1)^N}{(1+x)(1+4x)^{3/2}},
\end{eqnarray}
%-------------------------------------------------------------------------------------------------
with
%-------------------------------------------------------------------------------------------------
\begin{eqnarray}
f_{a3} &=&
2 \text{G}\Biggl[\Biggl\{
        \frac{1}{\tau },\frac{\sqrt{1+4 \tau }}{\tau }\Biggr\},1\Biggr]
-\text{G}\Biggl[\Biggl\{
        \frac{1}{\tau },\frac{1}{\tau },\frac{\sqrt{1+4 \tau }}{\tau }\Biggr\},1\Biggr],
\\
f_{b3}(x) &=& 
2 \text{G}\Biggl[ \Biggl\{
        \frac{1}{\tau },\frac{\sqrt{1+4 \tau }}{\tau }\Biggr\},1\Biggr]
+2 \text{G}\Biggl[\Biggl\{
        \frac{\sqrt{1+4 \tau }}{\tau },\frac{1}{\tau }\Biggr\},x\Biggr]
\nonumber\\ &&
-\text{G}\Biggl[\Biggl\{
        \frac{1}{\tau },\frac{1}{\tau },\frac{\sqrt{1+4 \tau }}{\tau }\Biggr\},1\Biggr]
+\text{G}\Biggl[\Biggl\{
        \frac{\sqrt{1+4 \tau }}{\tau },\frac{1}{\tau },\frac{1}{\tau }\Biggr\},x\Biggr].
\end{eqnarray}
%-------------------------------------------------------------------------------------------------
While these functions of $N$ evaluate to rational numbers for $N \in \mathbb{N}$,
the $N$ dependence cannot be simply factored out as in the usual Mellin transforms.

The above constants have the following representation,
%-------------------------------------------------------------------------------------------------
\begin{eqnarray}
{\rm G} \left[\left\{\frac{\sqrt{1-\tau}}{4-\tau},\right\},1\right] &=& 2 - \frac{\pi}{\sqrt{3}},
\\
{\rm G} \left[\left\{ 
\frac{\sqrt{1+4\tau}}{\tau}, \frac{1}{\tau}\right\},1\right] &=& -
{\rm G} \left[\left\{\frac{1}{\tau}, 
\frac{\sqrt{1+4\tau}}{\tau}\right\},1\right], 
\\
{\rm G} \left[\left\{\frac{1}{\tau}, 
\frac{\sqrt{1+4\tau}}{\tau}\right\},1\right] &=&
-4
-4 \ln(2)
-\ln^2(2)
+4 \sqrt{5}
+4 \ln\big(
        \sqrt{5}-1\big)
\nonumber\\ && 
+2 \ln(2) \ln\big(
        \sqrt{5}-1\big)
-\ln^2\big(
        \sqrt{5}-1\big)
+2 \Li_2\left[
        \frac{1}{2}-\frac{\sqrt{5}}{2}\right],
\\
{\rm G} \left[\left\{\frac{1}{\tau},\frac{1}{\tau}, 
\frac{\sqrt{1+4\tau}}{\tau}\right\},1\right] &=&
-8
+\frac{2 \ln^3(2)}{3}
-i \ln^2(2) (-2 i+\pi )
+\ln(2) \big(
        -8
        +2 \zeta_2
\big)
\nonumber\\ && 
+8 \sqrt{5}
+2 \zeta_3
+\big(
        8
        -2 \ln^2(2)
        +2 \ln(2) (2+i \pi )
        -2 \zeta_2
\big) 
\nonumber\\ && \times 
\ln\big(
        \sqrt{5}-1\big)
+(-2+2 \ln(2)-i \pi ) \ln^2\big(
        \sqrt{5}-1\big)
\nonumber\\ && 
-\frac{2}{3} \ln^3\big(
        \sqrt{5}-1\big)
+4 \text{Li}_2\Biggl[
        \frac{1}{2} \big(
                1-\sqrt{5}\big)\Biggr]
+2 \text{Li}_3\Biggl[
        \frac{1}{2} \big(
                1-\sqrt{5}\big)\Biggr]
\nonumber\\ && 
-2 \text{Li}_3\Biggl[
        \frac{1}{2} \big(
                1+\sqrt{5}\big)\Biggr]
\\
\lefteqn{\hspace*{-4.5cm} 
{\rm G} \left[\left\{\frac{1}{\tau},\frac{1}{\tau}, 
\frac{\sqrt{1+4\tau}}{\tau},
\frac{\sqrt{1+4\tau}}{\tau}\right\}
,1\right] =}
\nonumber\\ &&
24
-8 \Li_4\left[\frac{1}{2}\right]
-
\frac{11}{3} \ln^4(2)
-\frac{8}{3} i \ln^3(2) (\pi - 2 i)
+\ln(2) \big(
        16
        +8 i \pi 
\nonumber\\ && 
        -16 \sqrt{5}
        -56 \zeta_2
        -16 i \pi  \zeta_2
        -7 \zeta_3
\big)
+\ln^2\big(
        \sqrt{5}-1
\big)
\Biggr[4
        +\ln^2(2)
\nonumber\\ && 
        -8 i \pi 
        -6 i \ln(2) \pi 
        +11 \zeta_2
        -2 \Li_2\left[\frac{1-\sqrt{5}}{2}\right]
        -2 \Li_2\left[\frac{1+\sqrt{5}}{2}\right]
\Biggr]
\nonumber\\ && 
+\ln\big(
        \sqrt{5}-1
\big)
\Biggl[-16
        +\frac{14}{3} \ln^3(2)
        +8 \ln^2(2) (1+i \pi )
        -8 i \pi
\nonumber\\ &&  
        +\ln(2) \Biggl[
                -8
                +8 i \pi 
                +2 \zeta_2
        \Biggr]
        +16 \sqrt{5}
        +48 \zeta_2
        +12 i \pi  \zeta_2
        +\frac{7}{2} \zeta_3
\nonumber\\ &&        
 +4 (2+\ln(2)) \Li_2\left[\frac{1-\sqrt{5}}{2}\right]
        +4 \ln(2) 
\Li_2\left[\frac{1+\sqrt{5}}{2}\right]
\nonumber\\ &&    
     +4 \Li_3\Biggl[
                \frac{1-\sqrt{5}}{2}\Biggr]
        -4 \Li_3\Biggl[
                \frac{1+\sqrt{5}}{2}\Biggr]
\Biggr]
+\ln^2(2) \big(
        4
        -14 \zeta_2
\big)
\nonumber\\ && 
-16 \sqrt{5}
-4 \HA_{-1,-1,-1,1}\big(
        \sqrt{5}\big)
-8 \zeta_2
-8 i \pi  \zeta_2
+\frac{68}{5} \zeta_2^2
\nonumber\\ && 
-\frac{2}{3} (4+5 \ln(2)
-i \pi ) \ln^3\big(
         \sqrt{5}-1\big)
+\ln^4\big(
        \sqrt{5}-1\big)
+\big(
        8 i \pi 
\nonumber\\ && 
        +4 i \ln(2) \pi 
        -12 \zeta_2
\big) \Li_2\Bigg[
        \frac{1-\sqrt{5}}{2}\Biggr]
+\big(
        8
        +4 \ln(2) (2+i \pi )
\nonumber\\ &&    
     +8 i \pi 
        -12 \zeta_2
\big) \Li_2\Biggl[
        \frac{1+\sqrt{5}}{2}\Biggr]
-4 (2+\ln(2)) \Li_3\Biggl[
        \frac{1-\sqrt{5}}{2} \Biggr]
\nonumber\\ && 
+4 \ln(2) \Li_3\Biggl[
        \frac{1+\sqrt{5}}{2} \Biggr]
+4 \Li_4\Biggl[
        \frac{1}{2}\Biggr]
-4 \Li_4\Biggl[
        \frac{1 - \sqrt{5}}{2}\Biggr],
\end{eqnarray}
%-------------------------------------------------------------------------------------------------
and
%-------------------------------------------------------------------------------------------------
\begin{eqnarray}
\HA_{-1,-1,-1,1}(\sqrt{5}) &=&
-\frac{2}{3} \ln^4(2)
+\Biggl[
        \frac{7}{6} \ln^3(2)
        +\frac{1}{2} \ln(2) \zeta_2
        +\frac{7}{8} \zeta_3
\Biggr] \ln\big(
        \sqrt{5}-1\big)
- \Biggl[
        \frac{3}{4} \ln^2(2)
\nonumber\\ && 
        +\frac{1}{4} \zeta_2
\Biggr] \ln^2\big(
        \sqrt{5}-1\big)
+\frac{1}{6} \ln(2) \ln^3\big(
        \sqrt{5}-1\big)
-\Li_4\left[
        \frac{1}{2}\right]
+\Li_4\left[
        \frac{1+\sqrt{5}}{2}\right]
\nonumber\\ && 
-\frac{7}{4} \ln(2) \zeta_3.
\end{eqnarray}
%-------------------------------------------------------------------------------------------------
The remaining constants are better calculated numerically, since the space of polylogarithms will
normally not suffice, but they have a representation in terms of Kummer--Poincar\'e iterated 
integrals, see Appendix~\ref{sec:C}.
These G--functions at $x=1$ are real and the imaginary contributions cancel those of  $\Li_k(y)$ for 
$y > 1$. If more than two different letters occur, the corresponding expressions become much more involved, but can 
still be rationalized if the root factor is the same. Then also letters containing general 
quadratic forms
in the denominators occur \cite{Ablinger:2021fnc}, which can be decomposed into (complex) Kummer--Poincar\'e 
type letters \cite{KUMMER1,KUMMER2,KUMMER3,POINCARE} by partial fractioning. This representation can be
obtained by applying the command {\tt GLToL} to the G--functions.

We also mention that the command {\tt SExpansion} may be used for a partial asymptotic expansion of binomial 
sums, mapping to other binomial sums. An example is given by
%-------------------------------------------------------------------------------------------------
\begin{eqnarray}
\sum_{\tau_1=1}^N \frac{\big(
        \tau_1!\big)^2 
\sum_{\tau_2=1}^{\tau_1} \frac{\big(
        2 \tau_2\big)! 
\sum_{\tau_3=1}^{\tau_2} \frac{1}{\tau_3}}{\big(
        \tau_2!\big)^2 \tau_2^2}}{\big(
        2 \tau_1\big)! \big(
        1+\tau_1\big)}
&\approx&
 \frac{1}{N^2}
+\frac{4}{9 N^3}
-\frac{5}{3 N^4}
+\frac{23}{15 N^5}
-\frac{67}{180 N^6}
-\frac{299}{630 N^7}
\nonumber\\ && 
+\frac{127}{2520 N^8}
+\frac{859}{1134 N^9}
+\frac{587}{6300 N^{10}}
+ \Biggl[
\frac{2}{N^2}
-\frac{8}{3 N^3}
+\frac{2}{N^4}
\nonumber\\ && 
-\frac{2}{3 N^5}
-\frac{1}{3 N^6}
+\frac{1}{3 N^7}
+\frac{1}{3 N^8}
-\frac{4}{9 N^9}
        -\frac{3}{5 N^{10}}
\Biggr] L
\nonumber\\ && 
-2 
\sum_{\tau_1=1}^N \frac{\big(
        \tau_1!\big)^2 
\sum_{\tau_2=1}^{\tau_1} \frac{\big(
        2 \tau_2\big)! 
\sum_{\tau_3=1}^{\tau_2} \frac{1}{\tau_3}}{\big(
        \tau_2!\big)^2 \tau_2^2}}{\big(
        2 \tau_1\big)! \tau_1^2}
+\Biggl[
        \frac{1}{\sqrt{N}}
-\frac{7}{8} \frac{1}{N^{3/2}}
\nonumber\\ && 
+\frac{113}{128} \frac{1}{N^{5/2}}  
-\frac{909}{1024} \frac{1}{N^{7/2}}
+\frac{29067}{32768} \frac{1}{N^{9/2}}
-\frac{232137}{262144} \frac{1}{N^{11/2}}
\nonumber\\ && 
+\frac{3715061}{4194304} \frac{1}{N^{13/2}}
-\frac{29759813}{33554432} \frac{1}{N^{15/2}}
+\frac{1904293555}{2147483648} \frac{1}{N^{17/2}}
\nonumber\\ && 
-\frac{15205631037}{17179869184} \frac{1}{N^{19/2}}
\Biggr] 2^{-2 N} 
\sqrt{\pi } 
\sum_{\tau_1=1}^N \frac{\big(
        2 \tau_1\big)! 
\sum_{\tau_2=1}^{\tau_1} \frac{1}{\tau_2}}{\big(
        \tau_1!\big)^2 \tau_1^2}
\nonumber\\ && 
+4 
\sum_{\tau_1=1}^N \frac{\big(
        \tau_1!\big)^2 
\sum_{\tau_2=1}^{\tau_1} \frac{\big(
        2 \tau_2\big)! 
\sum_{\tau_3=1}^{\tau_2} \frac{1}{\tau_3}}{\big(
        \tau_2!\big)^2 \tau_2^2}}{\big(
        2 \tau_1\big)! \tau_1}
-2 \zeta_2^2
-2 \zeta_2 \zeta_3
+6 \zeta_5~.
\nonumber\\
\end{eqnarray}
%-------------------------------------------------------------------------------------------------
In some cases, the contributing constants are related to solutions of algebraic equations of fourth order.

In the examples given in an ancillary file the representations contained up to four singularities in the 
region $x \in [0,1]$ which had to be regularized. Furthermore, one has to algebraically reduce the 
G--functions at $x=1$ to remove divergences, resulting from the symbol $G[\{1/(1-\tau)\},1]$. The contributing 
183 (root--valued) G--constants have a linear representation in terms of 7200 divergence free 
G--constants of the Kummer--Poincar\'e type. The asymptotic representations of the binomial sums given in an 
ancillary file to terms of $O(1/N^{10})$ have absolute  accuracies between $4.04 \cdot 10^{-7}$ and $2.85 \cdot 
10^{-45}$ at $|N| = 50$.
The explicit sum representation of the binomial sums can be obtained by using the command {\tt ToHarmonicSumsSum}.

For the individual asymptotic expansions of the binomial sums also G--constants containing two different 
root--factors contribute. Terms of this kind are absent in the physical amplitudes. These 
G--constants can also be rationalized as described in \cite{Raab:2021rwl}, see also \cite{Besier:2018jen}. 
We present a series of examples in Appendix~\ref{sec:C}.
%%%%%%%%%%%%%%%%%%%%%%%%%%%%%%%%%%%%%%%%%%%%%%%%%%%%%%%%%%%%%%%%%%%%%%%%%%%%%%%%%%%%%%%%%%%%%%%%%%%
\section{The calculation of special constants}
\label{sec:C}
%%%%%%%%%%%%%%%%%%%%%%%%%%%%%%%%%%%%%%%%%%%%%%%%%%%%%%%%%%%%%%%%%%%%%%%%%%%%%%%%%%%%%%%%%%%%%%%%%%%

\vspace*{1mm}
\noindent
In the following we describe the technical steps in the calculation of a series of G--constants at 
$x=1, 1/4$ determined by Kummer--Poincar\'e and  root--valued letters
%------------------------------------------------------------------------------------------------- 
\begin{eqnarray} 
&& \Biggl\{\frac{1}{2+x},\frac{1}{1+3x}, 
\frac{\sqrt{1-4x}}{x},
\frac{\sqrt{1-4x}}{1-x},
\frac{\sqrt{1-4x}}{1+x},
\frac{\sqrt{5-4x}-1}{1-x}, 
\frac{\sqrt{5-4x}}{1+x}
\frac{\sqrt{5-4x}-1}{2-x}, 
\nonumber\\ && 
\frac{\sqrt{5-4x}-1}{2+x} 
\Biggr\}. 
\end{eqnarray} 
%------------------------------------------------------------------------------------------------- 
As for the rationalizations to Kummer--Poincar\'e integrals, one transforms the square--roots
%------------------------------------------------------------------------------------------------- 
\begin{eqnarray} 
\sqrt{a + b x} \rightarrow \sqrt{1-x}.
\end{eqnarray} 
%------------------------------------------------------------------------------------------------- 
An example is
%------------------------------------------------------------------------------------------------- 
\begin{eqnarray} 
\lefteqn{\text{G}\Biggl[\Biggl\{\frac{\sqrt{5-4\tau}}{\tau}\Biggr\},1\Biggr]
\rightarrow} \nonumber\\ &&
{\tt GLToStandardForm[
  TransformGL[GL[\{Sqrt[5 - 4 VarGL]/(VarGL)\}, 5/(4 x)], x]] /. x \rightarrow 4/5}
\nonumber \\ &=&
\sqrt{5} \left[\text{G}\Biggl[\Biggl\{\frac{1}{\tau}\Biggr\},\frac{5}{4}\Biggr]+
\text{G}\Biggl[\Biggl\{\frac{\sqrt{1-\tau}}{\tau}\Biggr\},\frac{4}{5}\Biggr]
\right]~~{\tt~// SpecialGLToH}  
\nonumber \\ &=&
2
-2 \sqrt{5}
+2 \sqrt{5} \text{G}(\{-1\},1)
-\big(
        1+\sqrt{5}\big) \text{G}\left[
        \{-1\},\frac{1}{\sqrt{5}}\right]
+\text{G}\left[
        \{0\},\frac{4}{5}\right]
+\sqrt{5} \text{G}\left[
        \{0\},\frac{5}{4}\right]
\nonumber\\ &&
-\big(
        1-\sqrt{5}\big) \text{G}\left[
        \{1\},\frac{1}{\sqrt{5}}\right]
\nonumber \\ &=&
2
-2 \sqrt{5}
-2 \ln(2) \sqrt{5}
+\sqrt{5} \ln(5)
+2 \sqrt{5} \ln\big(
        \sqrt{5}-1\big),
\end{eqnarray} 
%------------------------------------------------------------------------------------------------- 
where we used the notation of \cite{Vollinga:2004sn} in the next to last line. 
Another example is
%------------------------------------------------------------------------------------------------- 
\begin{eqnarray} 
\lefteqn{
\text{G}\Biggl[\Biggl\{\frac{\sqrt{1-4\tau}}{\tau}, \frac{1}{1-\tau}\Biggr\},\frac{1}{4}\Biggr]
\rightarrow
\text{G}\Biggl[\Biggl\{\frac{\sqrt{1-\tau}}{\tau}, \frac{1}{1-4 \tau}\Biggr\},1\Biggr]} \nonumber\\ &=&
-4
+2 i \sqrt{3} \text{G}\left[
        \big\{-i \sqrt{3}\big\},1\right]
+2 \text{G}\left[\{-1\},1\right] \text{G}\left[
        \big\{-i \sqrt{3}\big\},1\right]
-2 i \sqrt{3} \text{G}\left[
        \big\{i \sqrt{3}\big\},1\right]
\nonumber\\ && 
+2 \text{G}\left[\{-1\},1\right] \text{G}\left[
        \big\{i \sqrt{3}\big\},1\right]
-2 \text{G}\left[
        \big\{-1,-i \sqrt{3}\big\},1\right]
-2 \text{G}\left[
        \big\{-1,i \sqrt{3}\big\},1\right]
\nonumber\\ && 
-\text{G}\left[
        \big\{-i \sqrt{3},-1\big\},1\right]
-\text{G}\left[
        \big\{-i \sqrt{3},1\big\},1\right]
-\text{G}\left[
        \big\{i \sqrt{3},-1\big\},1\right]
-\text{G}\left[
        \big\{i \sqrt{3},1\big\},1\right].
\nonumber\\
\end{eqnarray} 
%------------------------------------------------------------------------------------------------- 
Some of the emerging Kummer--Poincar\'e iterated integrals have to be regularized, for example.
%------------------------------------------------------------------------------------------------- 
\begin{eqnarray} 
\text{G}\left[\{a\},a\right] = - \ln(a) - S_1(\infty),~~\text{etc.}
\end{eqnarray} 
%------------------------------------------------------------------------------------------------- 
Furthermore, also expressions like
%------------------------------------------------------------------------------------------------- 
\begin{eqnarray} 
\text{G}\left[\{1,1\},1\right] &=& 	
\frac{1}{2} \text{G}^2\left[\{1\},1\right] = 	
 \frac{1}{2} S_1^2({\infty }),
\\
\text{G}\left[\{1,1,1\},1\right] &=& \frac{1}{6} \text{G}^3\left[\{1\},1\right] =
-\frac{1}{6} S_1^3({\infty }),~~\text{etc.}
\end{eqnarray} 
%------------------------------------------------------------------------------------------------- 
emerge, cf. also~\cite{Blumlein:2009cf}, where $S_1(\infty)$ drops out in the amplitude.

By using the relations
%------------------------------------------------------------------------------------------------- 
\begin{eqnarray} 
\label{eq:rep1}
\HA_{-2}(1) &=& \ln(3) - \ln(2),
\\
\HA_{0, -2}(1) &=&  - \Li_2\left(-\frac{1}{2}\right),
\\
\label{eq:rep2}
\HA_{-1, -3}(1) &=&  \frac{1}{2} \zeta_2 + \ln^2(2) -\ln(2) \ln(3)
+\Li_2\left(-\frac{1}{2}\right),
\end{eqnarray} 
%------------------------------------------------------------------------------------------------- 
one shows that the contribution of $O(\ln(x)/x)$ to $\Delta a_{Qg}^{(3)}$ vanishes.

In the $O(1/x)$--term also G--functions at argument $x=1/4,~1/2$ and 1 occur.
Their combination has to be shown to vanish too. Since these constants are of higher weight than in 
the case of the $O(\ln(x)/x)$--term, we will rationalize these constants to Kummer--Poincar\'e iterated integrals 
as has been outlined above. Of course, one can seek for basis representations of this type of integrals, 
cf.~\cite{Ablinger:2021fnc,Ablinger:2013cf}, but in the end also the specific relations of these numbers
have to be exploited, since the G--functions will occur at different main argument in any of these 
representations. The first constants are
%------------------------------------------------------------------------------------------------- 
\begin{eqnarray} 
\text{G}\Biggl[\Biggl\{
        \frac{\sqrt{5
        -4 \tau 
        }-1}{1-\tau}\Biggr\},1\Biggr] &=& -2
-2 \ln(2)
+2 \sqrt{5}
+2 \ln\big(
        \sqrt{5}-1\big),
\\
\text{G}\Biggl[\Biggl\{
        \frac{\sqrt{5
        -4 \tau 
        }}{1+\tau }\Biggr\},1\Biggr] &=& 2
-2 \sqrt{5}
+3 \ln(2)
-6 \ln\big(
        3-\sqrt{5}\big),
\\
\text{G}\Biggl[\Biggl\{
        \frac{\sqrt{1- 
        4 \tau }}{\tau}\Biggr\},\frac{1}{4}\Biggr] &=& -2,
\\
\text{G}\Biggl[\Biggl\{
        \frac{\sqrt{1- 
        4 \tau }}{2+\tau}\Biggr\},\frac{1}{4}\Biggr] &=& -2 + 3 \ln(2),
\\
\HA_{0,-1,-2}(1) &=&
-\frac{13}{8} \zeta_3
-\frac{1}{2} \zeta_2 \ln(2)
+\zeta_2 \ln(3)
-\frac{1}{6} \ln^3(3)
-\Li_3\Biggl(
        -\frac{1}{3}\Biggr)
\nonumber\\ &&
+2 \Li_3\Biggl(
        \frac{1}{3}\Biggr),
\\
\Li_2\left(-\frac{1}{3}\right) &=& 
-\zeta_2
-\ln^2(2)
+2 \ln(2) \ln(3)
-\frac{1}{2} \ln^2(3)
-2 \text{Li}_2\left(
        -\frac{1}{2}\right).
\end{eqnarray} 
%------------------------------------------------------------------------------------------------- 

By using the implementation of \cite{Vollinga:2004sn} one shall
in general rescale the main argument of the G--functions to $x=1$ for the numerical calculation.
The $O(1/x)$ term contains G--constants with root--valued letters and the above $\HA$--constants. 
It is given by
%------------------------------------------------------------------------------------------------- 
\begin{eqnarray} 
\label{eq:c1}
\lefteqn{\frac{c_1}{x} = 
(\textcolor{blue}{C_A - 2 C_F})^2 \textcolor{blue}{T_F}
\Biggl\{
        48 
                \text{G}\Biggl[\Biggl\{
                        \frac{1}{2-\tau },\frac{1}{1-\tau },\frac{-1+\sqrt{5
                        -4 \tau 
                        }}{1-\tau }\Biggr\},1\Biggr]
}
\nonumber\\ &&
        -32
                \text{G}\Biggl[\Biggl\{
                        \frac{-1+\sqrt{5
                        -4 \tau 
                        }}{1-\tau },\frac{1}{1-\tau }\Biggr\},1\Biggr]
        +48 
                \text{G}\Biggl[\Biggl\{
                        \frac{1}{2-\tau },\frac{-1+\sqrt{5
                        -4 \tau 
                        }}{1-\tau },\frac{1}{1-\tau }\Biggr\},1\Biggr\}
\nonumber\\ &&
        -32 
                \text{G}\Biggl[\Biggl\{
                        \frac{1}{\tau },\frac{1}{1-\tau },\frac{-1+\sqrt{5
                        -4 \tau 
                        }}{1-\tau }\Biggr\},1\Biggr]
        -32 
                \text{G}\Biggl[\Biggl\{
                        \frac{1}{\tau },\frac{-1+\sqrt{5
                        -4 \tau 
                        }}{1-\tau },\frac{1}{1-\tau }\Biggr\},1\Biggr]
\nonumber\\ &&
        +16 
                \text{G}\Biggl[\Biggl\{
                        \frac{-1+\sqrt{5
                        -4 \tau 
                        }}{1-\tau },\frac{1}{1-\tau },\frac{1}{1-\tau }\Biggr\},1\Biggr]
        -192 \ln(2) \left[2 - \frac{\pi}{\sqrt{3}}\right]
\nonumber\\ &&
        -96 \text{G}\Biggl[\Biggl\{
                \frac{\sqrt{1-4 \tau }}{1-\tau },\frac{1}{1-\tau }\Biggr\},\frac{1}{4}\Biggr]
        +96 \text{G}\Biggl[\Biggl\{
                \frac{\sqrt{5-4 \tau }}{2-\tau },\frac{1}{1-\tau }\Biggr\},1\Biggr]
      -96 \text{G}\Biggl[\Biggr\{
                \frac{1}{\tau },
                \frac{\sqrt{1-4 \tau }}{1-\tau }\Biggr\},\frac{1}{4}\Biggr]
\nonumber\\ &&   
        -64 \text{G}\Biggl[\Biggl\{
                \frac{\sqrt{1-4 \tau }}{\tau },\frac{1}{1-\tau }\Biggl\},\frac{1}{4}\Biggr]
        -\frac{160}{3} \text{G}\Biggl[\Biggl\{
                \frac{\sqrt{5-4 \tau }}{\tau },\frac{1}{1-\tau }\Biggr\},1\Biggr]
\nonumber\\ && 
        -160 \text{G}\Biggl[\Biggl\{
                \frac{\sqrt{1-4 \tau }}{1+\tau },\frac{1}{\tau }\Biggr\},\frac{1}{4}\Biggr]
        -48 \text{G}\Biggl[\Biggl\{
                \frac{\sqrt{1-4 \tau }}{\tau },\frac{1}{1-\tau },\frac{1}{1-\tau }\Biggr\},\frac{1}{4}\Biggr]
\nonumber\\ &&   
      +48 \text{G}\Biggl[\Biggl\{
                \frac{\sqrt{1-4 \tau }}{\tau },\frac{1}{1-\tau },\frac{1}{\tau }\Biggr\},\frac{1}{4}\Biggr]
        -32 \text{G}\Biggl[\Biggl\{
                \frac{\sqrt{1-4 \tau }}{\tau },\frac{1}{\tau },\frac{1}{1-\tau }\Biggr\},\frac{1}{4}\Biggr]
\nonumber\\ &&   
      +96 \text{G}\Biggl[\Biggl\{
                \frac{1}{1+\tau },\frac{1}{\tau },\frac{\sqrt{1-4 \tau }}{\tau }\Biggr\},\frac{1}{4}\Biggr]
        +96 \text{G}\Biggl[\Biggl\{
                \frac{1}{1+\tau },\frac{\sqrt{1-4 \tau }}{\tau },\frac{1}{\tau }\Biggr\},\frac{1}{4}\Biggr]
\nonumber\\ &&   
      +\frac{16}{3} \text{G}\Biggl[\Biggr\{
                \frac{\sqrt{5-4 \tau }}{1+\tau },\frac{1}{\tau },\frac{1}{1-\tau }\Biggr\},1\Biggr]
        +16 \text{G}\Biggl[\Biggl\{
                \frac{\sqrt{1-4 \tau }}{2+\tau },\frac{1}{1+\tau },\frac{1}{\tau },\Biggr\}\frac{1}{4}\Biggr]
        +\Biggl[
                -\frac{32}{3}
                -16 \ln(2)
\nonumber\\ &&                
 +16 \sqrt{5}
                +48 \ln(3)
                -16 \ln\big(
                        3-\sqrt{5}\big)
                +48 \ln\big(\sqrt{5}-1\big)
        \Biggr] \zeta_2
        -120 \zeta_3
        +384 \ln^2(2)
        -8 \ln^3(3)
\nonumber\\ && 
        -96 \ln^2(5)
        -192 \text{Li}_2\left[
                \frac{1}{5}\right]
        -12 \text{Li}_3\left[
                \frac{1}{9}\right]
        +144 \text{Li}_3\left[
                \frac{1}{3}\right]
\Biggr\}  \expm 0.
\end{eqnarray} 
%------------------------------------------------------------------------------------------------- 

\noindent
One may perform the algebraic reduction \cite{Blumlein:2003gb}. After the rationalization there are still  
quadratic forms in the denominators. The main argument will generally be different from 1.
The decomposition into Kummer--Poincar\'e numbers leads to
197 terms, which are rescaled to  main argument $x=1$. These numbers were calculated to 1000 digits. 
In this way we showed that the $O(1/x)$ term vanishes at an accuracy of $10^{-998}$. This accuracy can be 
further improved by calculating the G--functions to an even higher precision. 
We also compared all these constants with the result obtained by using {\tt NumericalValues}, which delivers
6--7 digits only. The numerical values of the constants occurring in (\ref{eq:c1}) are given in an ancillary file.
  
Accordingly, the expansion of the first--order factorizable terms of the amplitude around $x = 0, 1/2$ and 1 to 
100 terms lead to a much larger number of G--constants. Their calculation proceeds in the same way.

In the asymptotic expansion of the finite binomial sums also G--functions with two different root--factors 
out of the set 
%------------------------------------------------------------------------------------------------- 
\begin{eqnarray} 
\left\{ \sqrt{x(x+1)},~~\sqrt{x(1-x)},~~\sqrt{1-x^2}\right\}
\end{eqnarray} 
%------------------------------------------------------------------------------------------------- 
emerge. Structures of this kind can be rationalized by the transformation, cf.~\cite{Raab:2021rwl}, 
%------------------------------------------------------------------------------------------------- 
\begin{eqnarray} 
x = \frac{2 y^2}{y^4 + 1},~~~~y = \frac{1}{\sqrt{2x}}[\sqrt{1+x} - \sqrt{1-x}].
\end{eqnarray} 
%------------------------------------------------------------------------------------------------- 
After rationalization, one expects letters of cyclotomy 4,~\cite{Ablinger:2011te}, i.e. the emergence of $\pi$ 
and the Catalan number {\sf C} \cite{CATALAN}, as well as of other cyclotomic constants.
For the simplest integrals one obtains
%------------------------------------------------------------------------------------------------- 
\begin{eqnarray} 
\int_0^1 dx \sqrt{x(1+x)} &=& 4 \sqrt{2} \int_0^1 dy \frac{y^2 (1+y^2)(1-y^4)}{(1+y^4)^3} = 
\frac{3}{2\sqrt{2}} - \frac{1}{4} \ln(1+\sqrt{2})
\nonumber\\
\\
\int_0^1 dx \sqrt{x(1-x)} &=& 4 \sqrt{2} \int_0^1 dy \frac{y^2 (1-y^2) (1-y^4)}{(1+y^4)^3} =
\frac{\pi}{8} 
\\
\int_0^1 dx \sqrt{1-x^2} &=& 4 \int_0^1 dy \frac{y(1-y^4)^2}{(1+y^4)^3} = \frac{\pi}{4}
\\
\int_0^1 dx \sqrt{x(1+x)} \int_0^x dy  \sqrt{y(1-y)}
&=& 32 \int_0^1 dx \int_0^x dy
\frac{x^2 (1-x^2) (1+x^2)^2 y^2 (1+y^2) (1-y^2)^2}
{\big(1+x^4\big)^3 \big(1+y^4\big)^3}
\nonumber\\ &=&
-\frac{17}{96}+\frac{\sf C}{16}+\pi  \Biggl[
        \frac{3}{16} \frac{1}{\sqrt{2}}
        -\frac{1}{32} \ln\big(
                1
                +\sqrt{2}
        \big)
\Biggr].
\end{eqnarray}
\begin{eqnarray}
\lefteqn{\int_0^1 dx \sqrt{x(1+x)} \int_0^x dy  \sqrt{1-y^2} = \int_0^1 dx \int_0^x dy 
\frac{16 \sqrt{2} x^2 (1-x^2)  \big(
        1+x^2\big)^2 y (1-y^4)^2 }{\big(
        1+x^4\big)^3 \big(
        1+y^4\big)^3}}
\nonumber\\ 
&=&
\pi \Biggl[-\frac{27 }{256}
+\frac{3}{8}   \frac{1}{\sqrt{2}} \Biggr]
+\frac{i}{16}  \Biggl\{\text{Li}_2\left[
        1-\sqrt[4]{-1}\right]
-  \text{Li}_2\left[
        1+(-1)^{3/4}\right]
\nonumber\\ && 
+  \text{Li}_2\left[
        \frac{1}{2} \big(
                (1-i)-\sqrt{2}\big)\right]
-  \text{Li}_2\left[
        \frac{1}{2} \big(
                (1+i)-\sqrt{2}\big)\right]
+  \text{Li}_2\left[
        \frac{1}{2} \big(
                (1+i)-i \sqrt{2}\big)\right]
\nonumber\\ && 
-  \text{Li}_2\left[
        \frac{1}{2} i \big(
                (-1-i)+\sqrt{2}\big)\right]
+  \text{Li}_2\left[
        \frac{2}{(1-i)+\sqrt{2}}\right]
-  \text{Li}_2\left[
        \frac{2}{(1+i)+\sqrt{2}}\right]
\nonumber\\ && 
+ 2 \text{Li}_2\left[
                \frac{1}{4}(1+i)
\big((1-i)+\sqrt{2}\big)\right]
- 2  i \text{Li}_2\left[
                \frac{1}{4}(1-i)
\big((1+i)+\sqrt{2}\big)\right]\Biggr\}.
\label{eq:C24}
\end{eqnarray} 
%------------------------------------------------------------------------------------------------- 
One may easily write Eq.~(\ref{eq:C24}) in terms of several generalized hypergeometric functions. 
However, the representation in terms of polylogarithms and cyclotomic functions is obtained only 
by rationalizing the integrand. 
%%%%%%%%%%%%%%%%%%%%%%%%%%%%%%%%%%%%%%%%%%%%%%%%%%%%%%%%%%%%%%%%%%%%%%%%%%%%%%%%%%%%%%%%%%%%%%%%%%%
\section{\boldmath Analytic continuation to $N$-space}
\label{sec:D}
%%%%%%%%%%%%%%%%%%%%%%%%%%%%%%%%%%%%%%%%%%%%%%%%%%%%%%%%%%%%%%%%%%%%%%%%%%%%%%%%%%%%%%%%%%%%%%%%%%%

\vspace*{1mm}
\noindent
One may use the representations of Eq.~(\ref{eq:F0}--\ref{eq:F1}) to perform the analytic 
continuation to $N$--space in the analyticity region in  $N \in \mathbb{C}$. The following integrals
contribute
%------------------------------------------------------------------------------------------------- 
\begin{eqnarray} 
I_1(N,k,l;a) &=& \int_0^a dx x^{N-1+k} \ln^l(x),~~~
k \geq 0, l \leq 5, k,l \in \mathbb{N}, a \in [0,1],
\\
%----
I_2(N,k;b,a) &=& \int_a^b dx x^{N-1} \left(\frac{1}{2} - x \right)^k,~~~ b > \frac{1}{2} > a, a,b \in [0,1],
\\
I_3(N,m;b) &=& \int_b^1 dx x^{N-1} \ln^m(1-x),~~~ 
m \leq 5, m \in \mathbb{N}, b \in [0,1].
\end{eqnarray} 
%------------------------------------------------------------------------------------------------- 
These integrals are related to incomplete Beta- and $\Gamma$--functions \cite{NIST}. However, we 
will use different representations in the following, based on generalized harmonic sums, 
cf.~\cite{Ablinger:2013cf}, to allow for simpler representations for $N \in \mathbb{C}$. 

One obtains
%------------------------------------------------------------------------------------------------- 
\begin{eqnarray} 
I_1(N,k,l;a) = a^{N+k} \sum_{m=0}^l \frac{l!}{(l-m)!} \frac{(-1)^m \ln^{l-m}(a)}{(N+k)^{m+1}}
\end{eqnarray} 
%------------------------------------------------------------------------------------------------- 
and $I_1$ may be expressed in terms of generalized harmonic sums
%------------------------------------------------------------------------------------------------- 
\begin{eqnarray} 
\frac{a^{N+k}}{(N+k)^\nu} =  S_\nu(a)(N+k) - S_\nu(a)(N+k-1),~~ \nu \leq 6, \nu \in \mathbb{N}, a \in [0,1],
\end{eqnarray} 
%------------------------------------------------------------------------------------------------- 
through which its first--order recurrence in $N$ and the asymptotic representations are provided. The function $I_2$ 
is written as 
%------------------------------------------------------------------------------------------------- 
\begin{eqnarray} 
I_2(N,k;a,b) = \bar{I}_2(N,k;b) - \bar{I}_2(N,k;a);~~~~\bar{I}_2(N,k;a) = I_2(N,k;a,0).
\end{eqnarray} 
%------------------------------------------------------------------------------------------------- 
$\bar{I}_2$ obeys
%------------------------------------------------------------------------------------------------- 
\begin{eqnarray} 
\bar{I}_2(N,k;a) &=& \frac{a^N}{N} \left(\frac{1}{2} - a\right)^k 
+ \frac{k}{N} \bar{I}_2(N+1,k-1;a),
\\
\label{inco1}
\bar{I}_2(N,0;a)
&=& \frac{a^N}{N}.
\end{eqnarray} 
%------------------------------------------------------------------------------------------------- 
One may start with the initial condition (\ref{inco1}) and needs no asymptotic expansion for the representation
in $N \in \mathbb{C}$ in this case.

We write $I_3$ by
%------------------------------------------------------------------------------------------------- 
\begin{eqnarray} 
I_3(N,m;b) = \frac{(-1)^m m!}{N} S_{
\underbrace{{\scriptstyle 1, \ldots,1}}_{m}}
- \bar{I}_3(N,m;b)
\\
\bar{I}_3(N,m;b) = b^N \Mvec[\ln^m(1-x b)](N),
\end{eqnarray} 
%------------------------------------------------------------------------------------------------- 
where
%------------------------------------------------------------------------------------------------- 
\begin{eqnarray} 
\bar{I}_3(N,0;b) &=& \frac{b^N}{N},
\\
\bar{I}_3(N,1;b) &=& - \frac{1}{N} \Biggl[
S_1(\{b\})
+\big(
        1-b^N\big) \ln(1-b)\Biggr],
\\
%-----------------
\bar{I}_3(N,2;b) &=&
\frac{1}{N} \Biggl[
        2 \big(
                S_1 S_1(\{b\})
                +S_2(\{b\})
                -S_{1,1}(\{b,1\})
        \big)
        +2 \big(
                S_1
                -S_1(\{b\})
        \big) \ln(1-b)
\nonumber\\ &&
        -\big(
                1-b^N\big) \ln^2(1-b)
\Biggr].
\end{eqnarray} 
%------------------------------------------------------------------------------------------------- 
The higher terms are given in an ancillary file. The recursions for the generalized harmonic sums 
are given in Eq.~(\ref{eq:rec1}).
The asymptotic expansions of the generalized sums for $|N| \rightarrow \infty$ read
%------------------------------------------------------------------------------------------------- 
\begin{eqnarray} 
S_1(\{b\},N) &=&
-\ln(1-b)
+\Biggl[
        -\frac{1}{(1-b) N}
        +\frac{1}{(1-b)^2 N^2}
        -\frac{1+b}{(1-b)^3 N^3}
        +\frac{1+4 b+b^2}{(1-b)^4 N^4}
\nonumber\\ && -        \frac{(1+b) \big(
                1+10 b+b^2\big)}{(1-b)^5 N^5}
\Biggr] b^{N+1}
+ O\left(\frac{1}{N^6}\right),
\\
%-------------------------------------
S_2(\{b\},N) &=&
\text{Li}_2(b)
+ \Biggl[
        -\frac{1}{(1-b) N^2}
        +\frac{2}{(1-b)^2 N^3}
        -\frac{3 (1+b)}{(1-b)^3 N^4}
        + \frac{4 \big(
                1+4 b+b^2\big)}{(1-b)^4 N^5}
\Biggr] b^{N+1}
\nonumber\\ &&
+ O\left(\frac{1}{N^6}\right),
\\
%-------------------------------------
S_{1, 1}(\{b, 1\}, N) &=&
 \frac{1}{2} \ln^2(1-b)
+\text{Li}_2(b)
+\Biggl[
        -\frac{3-b}{2 (1-b)^2 N^2}
        +\frac{31+4 b+b^2}{12 (1-b)^3 N^3}
        -\frac{43+82 b+7 b^2}{12 (1-b)^4 N^4}
\nonumber\\ && 
+       -\frac{-549-3414 b-1964 b^2-74 b^3+b^4}{120 (1-b)^5 N^5}
        +L \Biggl[
                -\frac{1}{(1-b) N}
                +\frac{1}{(1-b)^2 N^2}
\nonumber\\ &&                
 -\frac{1+b}{(1-b)^3 N^3}
                +\frac{1+4 b+b^2}{(1-b)^4 N^4}
               - \frac{(1+b) \big(
                        1+10 b+b^2\big)}{(1-b)^5 N^5}
        \Biggr]
\Biggr] b^{N+1}
\nonumber\\ && 
+ O\left(\frac{1}{N^6}\right),~~\text{etc.}
\end{eqnarray} 
%------------------------------------------------------------------------------------------------- 
The representations given in this appendix will 
also apply to the complete representations for $a_{Qg}^{(3)(N)}$ and $\Delta a_{Qg}^{(3)(N)}$.

\vspace*{5mm}
\noindent
{\bf Acknowledgments}.\\
We would like to thank S.~Klein for calculating a series of Mellin moments in the polarized case 
\cite{SKLEIN} by using {\tt MATAD}, which we have used for comparison and thank P.~Marquard for discussions. 
This work was supported by the Austrian Science Fund (FWF) grants P33530 and P34501N.

%\newpage
%-----------------------------------------------------------------------------------------------------

%-----------------------------------------------------------------------------------
\end{document}

%% file: a3unp
a_{Qg}^{(3)} &=& \frac{1}{2}[1+(-1)^N] 
\nonumber\\ && \times \Biggl\{\textcolor{blue}{C_F} \Biggl\{
        \textcolor{blue}{C_A T_F}  \Biggl[
                \frac{8 S_1^2 \zeta_2 P_{13}}{3 (N-1) N^2 (1+N)^2 (2+N)^2}
                -\frac{288 \zeta_2^2 P_{14}}{5 (N-1) N^2 (1+N)^2 (2+N)^2}
\nonumber\\ &&             
    +\frac{\zeta_2 P_{25}}{18 (N-1) N^3 (1+N)^3 (2+N)^3}
                -\frac{4 S_1 \zeta_2 P_{29}}{9 (N-1) N^3 (1+N)^3 (2+N)^3}
\nonumber\\ &&                
 + {\sf B_4} \Biggl[
                        \frac{32 P_{14}}{(N-1) N^2 (1+N)^2 (2+N)^2}
                        +32 \pqgn S_1
                \Biggr]
                +\pqgn \Biggl[
                        -\frac{12 S_2 \zeta_2 P_1}{(N-1) N (1+N) (2+N)}
\nonumber\\ &&                  
       +\Biggl[
                                -\frac{8 \big(
                                        1+3 N+3 N^2\big) \zeta_2}{N (1+N)}
                                +16 S_1 \zeta_2
                        \Biggr] S_{-2}
                        +(32 S_1^3 
                        -8 S_3 
                        -8 S_{-3} 
                        +16 S_{-2,1}) \zeta_2
\nonumber\\ &&                        
 -\frac{288}{5} S_1 \zeta_2^2
                \Biggr]
        \Biggr]
        +\textcolor{blue}{T_F^2} \Biggl[
                \frac{2 \zeta_2 P_{34}}{9 (N-1) N^4 (1+N)^4 (2+N)^3}
                +\textcolor{blue}{N_F} \Biggl[
                        -\frac{32 S_1^2 P_7}{81 N^2 (1+N)^2 (2+N)}
\nonumber\\ &&                
         -\frac{16 S_3 P_{24}}{81 (N-1) N^3 (1+N)^3 (2+N)^2}
                        +\frac{2 (N-2) \zeta_2 P_{30}}{9 (N-1) N^4 (1+N)^4 (2+N)^3}
\nonumber\\ &&                     
    +\frac{8 S_2 P_{33}}{27 (N-1) N^4 (1+N)^4 (2+N)^3}
                        +\frac{P_{37}}{243 (N-1) N^6 (1+N)^6 (2+N)^5}
\nonumber\\ &&                      
   +\pqgn
                         \Biggl[
                                -
                                \frac{56 \zeta_3 P_{15}}{9 (N-1) N^2 (1+N)^2 (2+N)}
                                +\Biggl[
                                        -\frac{256}{27} S_3
                                        -\frac{128}{3} S_{2,1}
                                        +\frac{224 \zeta_3}{9}
                                \Biggr] S_1
                                +\Biggl[
                                        -\frac{64}{9} S_2
\nonumber\\ &&                
                         -\frac{16 \zeta_2}{3}
                                \Biggr] S_1^2
                                -\frac{32}{27} S_1^4
                                -\frac{128}{9} S_2^2
                                +\frac{256}{9} S_4
                                -\frac{128}{3} S_{3,1}
                                +\frac{256}{3} S_{2,1,1}
                        \Biggr]
\nonumber\\ &&                
         +\Biggl[
                                -\frac{16 P_{12}}{243 N^2 (1+N)^3 (2+N)}
                                +\frac{32 \big(
                                        24+83 N+49 N^2+10 N^3\big) S_2}{27 N^2 (1+N) (2+N)}
\nonumber\\ &&                
                 +\frac{16 \big(
                                        12+28 N+11 N^2+5 N^3\big) \zeta_2}{9 N^2 (1+N) (2+N)}
                        \Biggr] S_1
                        +\frac{32 \big(
                                24+83 N+49 N^2+10 N^3\big) S_1^3}{81 N^2 (1+N) (2+N)}
\nonumber\\ &&                
         -\frac{128 \big(
                                -2-3 N+N^2\big) S_{2,1}}{3 N^2 (1+N) (2+N)}
                \Biggr]
                +\frac{80 \big(
                        6+11 N+4 N^2+N^3\big) S_1 \zeta_2}{9 N^2 (1+N) (2+N)}
                +\pqgn \Biggl[
                        -\frac{40}{3} S_1^2 \zeta_2
\nonumber\\ &&                
         +8 S_2 \zeta_2
                \Biggr]
        \Biggr]
\Biggr\}
+\textcolor{blue}{C_A^2 T_F} \Biggl\{
        \frac{144 (N-2) (3+N) \zeta_2^2 P_2}{5 (N-1) N^2 (1+N)^2 (2+N)^2}
\nonumber\\ &&   
      -\frac{4 S_1^2 \zeta_2 P_{18}}{3 (N-1) N^2 (1+N)^2 (2+N)^2}
        +
        \frac{4 S_1 \zeta_2 P_{31}}{9 (N-1)^2 N^3 (1+N)^3 (2+N)^3}
\nonumber\\ &&        
 +\frac{2 \zeta_2 P_{35}}{9 (N-1)^2 N^4 (1+N)^4 (2+N)^4}
        +{\sf B_4} \Biggl[
                -\frac{4 P_{17}}{(N-1) N^2 (1+N)^2 (2+N)^2}
                -16 \pqgn S_1
        \Biggr]
\nonumber\\ && 
        +\pqgn \Biggl[
                -\frac{4 S_2 \zeta_2 P_8}{3 (N-1) N (1+N) (2+N)}
                +\Biggl[
                        -32 S_2 \zeta_2
                        +\frac{288 \zeta_2^2}{5}
                \Biggr] S_1
\nonumber\\ &&                
 +\Biggl[
                        -\frac{8 \zeta_2 P_9}{3 (N-1) N (1+N) (2+N)}
                        -48 S_1 \zeta_2
                \Biggr] S_{-2}
                +(-16 S_1^3
                -8 S_3
                -8 S_{-3} 
\nonumber\\ &&                
 +16 S_{-2,1}) \zeta_2
        \Biggr]
\Biggr\}
+\textcolor{blue}{C_A T_F^2} \Biggl[
        -\frac{4 \zeta_2 P_{28}}{9 (N-1) N^3 (1+N)^3 (2+N)^3}
\nonumber\\ && 
        +\textcolor{blue}{N_F} \Biggl[
 -\frac{64 S_{2,1} P_4}{27 N (1+N)^2 (2+N)^2}
                -\frac{160 S_{-3} P_5}{27 N (1+N)^2 (2+N)^2}
                -\frac{16 S_1^3 P_{10}}{81 N (1+N)^2 (2+N)^2}
\nonumber\\ &&                
 +\frac{64 S_{-2,1} P_{16}}{9 (N-1) N^2 (1+N)^2 (2+N)^2}
                +\frac{8 S_1^2 P_{20}}{81 N (1+N)^3 (2+N)^3}
\nonumber\\ &&                
 -\frac{32 S_3 P_{22}}{81 (N-1) N^2 (1+N)^2 (2+N)^2}
                +\frac{8 S_2 P_{26}}{81 (N-1) N^3 (1+N)^3 (2+N)^3}
\nonumber\\ &&                
 -
                \frac{4 \zeta_2 P_{27}}{9 (N-1) N^3 (1+N)^3 (2+N)^3}
                -\frac{8 P_{36}}{243 (N-1) N^5 (1+N)^5 (2+N)^5}
\nonumber\\ &&                
 +\pqgn \Biggl[
                        \Biggl[
                                \frac{1888}{27} S_3
                                +\frac{224}{9} S_{2,1}
                                -64 S_{-2,1}
                                -\frac{224}{9} \zeta_3
                        \Biggr] S_1
                        +\frac{160}{3} S_{-3} S_1
                        +\Biggl[
                                \frac{176}{9} S_2
                                +\frac{16 \zeta_2}{3}
                        \Biggr] S_1^2
\nonumber\\ &&                
         +\frac{32}{27} S_1^4
                        +\frac{80}{9} S_2^2
                        +\frac{640}{9} S_4
                        +\Biggl[
                                -\frac{64 (-1+2 N) S_1}{(N-1) N}
                                +32 S_1^2
                                +\frac{160}{3} S_2
                                +\frac{32}{3} \zeta_2
                        \Biggr] S_{-2}
\nonumber\\ &&                
         +\frac{352}{9} S_{-4}
                        -\frac{32}{9} S_{3,1}
                        +\frac{64 (-1+2 N) S_{-2,1}}{(N-1) N}
                        -\frac{128}{3} S_{-2,2}
                        -\frac{160}{3} S_{-3,1}
                        -\frac{416}{9} S_{2,1,1}
\nonumber\\ &&                
         +64 S_{-2,1,1}
                        +\frac{16}{3} S_2 \zeta_2
                        +\frac{448 \big(
                                1+N+N^2\big) \zeta_3}{9 (N-1) N (1+N) (2+N)}
                \Biggr]
                +\Biggl[
                        -\frac{16 \zeta_2 P_6}{9 N (1+N)^2 (2+N)^2}
\nonumber\\ &&                
         -\frac{16 S_2 P_{11}}{27 N (1+N)^2 (2+N)^2}
                        +\frac{16 P_{32}}{243 (N-1) N^2 (1+N)^4 (2+N)^4}
                \Biggr] S_1
\nonumber\\ &&                
 +\Biggl[
                        -\frac{64 S_1({N}
                        ) P_{16}}{9 (N-1) N^2 (1+N)^2 (2+N)^2}
                        +\frac{32 P_{21}}{81 N (1+N)^3 (2+N)^3}
                \Biggr] S_{-2}
        \Biggr]
\nonumber\\ && 
        +\pqgn \Biggl[
                \frac{40}{3} S_1^2 \zeta_2
                +\frac{40}{3} S_2 \zeta_2
                +\frac{80}{3} S_{-2} \zeta_2
        \Biggr]
        +\frac{160 \big(
                4-N+N^2+4 N^3+N^4\big) S_1 \zeta_2}{9 N (1+N)^2 (2+N)^2}
\Biggr]
\nonumber\\ && 
+\textcolor{blue}{C_F^2 T_F} \Biggl[
        -\frac{16 {\sf B_4} (N-1) \big(
                -2+3 N+3 N^2\big)}{N^2 (1+N)^2}
        +\frac{4 S_1^2 \zeta_2 P_3}{N^2 (1+N)^2 (2+N)}
\nonumber\\ &&      
  +\frac{8 S_1 \zeta_2 P_{19}}{N^3 (1+N)^3 (2+N)}
        +\frac{\zeta_2 P_{23}}{2 N^4 (1+N)^4 (2+N)}
        +\pqgn \Biggl[
                \Biggl[
                        -\frac{16 \zeta_2}{N (1+N)}
                        +32 S_1 \zeta_2
                \Biggr] S_{-2}
\nonumber\\ &&           
     -16 S_1^3 \zeta_2
                -\frac{8 \big(
                        2+3 N+3 N^2\big) S_2 \zeta_2}{N (1+N)}
                +(32 S_1 S_2 
                +16 S_3 
                +16 S_{-3} 
                -32 S_{-2,1}) \zeta_2
        \Biggr]
\nonumber\\ && 
        +\frac{144 (N-1) \big(
                -2+3 N+3 N^2\big) \zeta_2^2}{5 N^2 (1+N)^2}
\Biggr]
-\frac{64}{9} \pqgn \textcolor{blue}{T_F^3} \zeta_3\Biggr\} + O(\text{rat}) + O(\zeta_3),
\end{eqnarray}
%---------------------------------------------------------------------------------------------------
with the polynomials $P_i$
%---------------------------------------------------------------------------------------------------
\begin{eqnarray}
P_1&=&N^4+2 N^3-3 N^2-4 N-4,
\\ 
P_2&=&3 N^4+6 N^3+7 N^2+4 N+4,
\\ 
P_3&=&3 N^4+14 N^3+43 N^2+48 N+20,
\\ 
P_4&=&5 N^4+11 N^3+50 N^2+85 N+20,
\\ 
P_5&=&5 N^4+14 N^3+53 N^2+82 N+20,
\\ 
P_6&=&5 N^4+20 N^3+59 N^2+76 N+20,
\\ 
P_7&=&10 N^4+185 N^3+789 N^2+521 N+141,
\\ 
P_8&=&11 N^4+22 N^3-59 N^2-70 N-48,
\\ 
P_9&=&11 N^4+22 N^3-47 N^2-58 N-36,
\\ 
P_{10}&=&20 N^4+107 N^3+344 N^2+439 N+134,
\\ 
P_{11}&=&40 N^4+151 N^3+544 N^2+779 N+214,
\\ 
P_{12}&=&230 N^5-924 N^4-5165 N^3-7454 N^2-10217 N-2670,
\\ 
P_{13}&=&N^6-9 N^5-120 N^4-137 N^3+29 N^2+56 N+36,
\\ 
P_{14}&=&3 N^6+9 N^5-5 N^4-25 N^3-14 N^2-16,
\\ 
P_{15}&=&3 N^6+9 N^5-N^4-17 N^3-38 N^2-28 N-24,
\\ 
P_{16}&=&5 N^6-9 N^5-24 N^4-61 N^3-143 N^2-20 N+36,
\\ 
P_{17}&=&9 N^6+27 N^5-15 N^4-75 N^3-62 N^2-20 N-56,
\\ 
P_{18}&=&11 N^6+33 N^5-87 N^4-85 N^3+4 N^2-116 N-48,
\\ 
P_{19}&=&13 N^6+36 N^5+39 N^4+8 N^3-21 N^2-29 N-10,
\\ 
P_{20}&=&22 N^6+271 N^5+2355 N^4+6430 N^3+6816 N^2+3172 N+1256,
\\ 
P_{21}&=&47 N^6+278 N^5+1257 N^4+2552 N^3+1794 N^2+284 N+448,
\\ 
P_{22}&=&65 N^6+429 N^5+1155 N^4+725 N^3+370 N^2+496 N+648,
\\ 
P_{23}&=&-153 N^8-612 N^7-1196 N^6-1374 N^5-947 N^4-222 N^3+176 N^2+184 N
\nonumber\\ &&
+48,
\\ 
P_{24}&=&29 N^8-976 N^7-2382 N^6+1736 N^5+9129 N^4+7472 N^3+6832 N^2+5376 N
\nonumber\\ &&
+3888,
\\ 
P_{25}&=&-261 N^9-1566 N^8+2506 N^7+27752 N^6+65115 N^5+88078 N^4+76456 N^3
\nonumber\\ &&
+46032 N^2
+17200 N+3552,
\\ 
P_{26}&=&4 N^9-117 N^8+806 N^7+6901 N^6+15770 N^5+7720 N^4-6644 N^3-3128 N^2
\nonumber\\ &&
-4032 N
-1728,
\\ 
P_{27}&=&15 N^9+90 N^8+146 N^7-32 N^6-501 N^5-610 N^4-244 N^3-48 N^2
\nonumber\\ &&
+224 N+96,
\\ 
P_{28}&=&69 N^9+414 N^8+1148 N^7+2134 N^6+3171 N^5+5180 N^4+6500 N^3+5352 N^2
\nonumber\\ &&
+3008 N
+672,
\\ 
P_{29}&=&337 N^9+2088 N^8+4868 N^7+5338 N^6+1523 N^5-3602 N^4-5768 N^3-3392 N^2
\nonumber\\ &&
+48 N
+288,
\\ 
P_{30}&=&45 N^{10}+405 N^9+1606 N^8+3842 N^7+6717 N^6+9325 N^5+10888 N^4+9804 N^3
\nonumber\\ &&
+6232 N^2
+3264 N+864,
\\ 
P_{31}&=&103 N^{10}+515 N^9+1124 N^8+1298 N^7-809 N^6-601 N^5+3078 N^4+2820 N^3
\nonumber\\ &&
+680 N^2
-2160 N-864,
\\ 
P_{32}&=&491 N^{10}+5292 N^9+23603 N^8+61598 N^7+87216 N^6+29418 N^5-73982 N^4
\nonumber\\ &&
-73764 N^3
-11408 N^2-2672 N+864,
\\ 
P_{33}&=&57 N^{11}+547 N^{10}+1416 N^9-398 N^8-6819 N^7-13965 N^6-19422 N^5
\nonumber\\ &&
-31384 N^4
-39024 N^3-27808 N^2-16992 N-5184,
\\ 
P_{34}&=&333 N^{11}+2331 N^{10}+6142 N^9+6930 N^8+617 N^7-6377 N^6-4900 N^5
\nonumber\\ &&
+4444 N^4
+11680 N^3+8944 N^2+4992 N+1728,
\\ 
P_{35}&=&69 N^{13}+552 N^{12}+1190 N^{11}-997 N^{10}-8937 N^9-17658 N^8-16952 N^7
\nonumber\\ &&
-19177 N^6
-36634 N^5-54000 N^4
-57200 N^3-39184 N^2-17184 N-3456,
\\ 
P_{36}&=&3597 N^{15}+44514 N^{14}+237011 N^{13}+692290 N^{12}+1139033 N^{11}+849246 N^{10}
\nonumber\\ &&
-377441 N^9-1484940 
N^8-1459136 N^7-806374 N^6-465872 N^5-281016 N^4
\nonumber\\ &&
-22912 N^3+33504 N^2+18432 N+3456,
\\ 
P_{37}&=&15777 N^{17}+186525 N^{16}+879391 N^{15}+1874085 N^{14}+575913 N^{13}-5568833 N^{12}
\nonumber\\ &&
-10465411 N^{11}
-2970289 N^{10}+11884298 N^9+12640320 N^8-10343664 N^7
\nonumber\\ &&
-40750480 N^6-55711424 N^5-53947712 N^4-42534912 N^3
-23256576 N^2
\nonumber\\ &&
-7865856 N -1244160

%% file: a3pol
\begin{eqnarray}
\label{eq:apol}
\lefteqn{\Delta a_{Qg}^{(3)} = \frac{1}{2}[1 - (-1)^N] \Biggl\{\textcolor{blue}{C_F} \Biggl[
         \textcolor{blue}{T_F^2} \Biggl[
                 \textcolor{blue}{N_F} \Biggl[
                         \frac{8 S_2 Q_{18}}{27 N^4 (1+N)^4 (2+N)}
                         -
                         \frac{16 S_3 Q_{12}}{81 N^3 (1+N)^3 (2+N)}}
\nonumber\\ &&
                         +\frac{Q_{20}}{243 N^6 (1+N)^6 (2+N)}
+\Biggl(
                                 \frac{32 \big(10 N^3 + 49 N^2 + 19 N -24\big) S_2}{27 N^2 (1+N) (2+N)}
                                 -\frac{16 Q_5}{243 N^2 (1+N)^2 (2+N)}
\nonumber\\ &&                                 
-\frac{256}{27} {\Dpgqn} S_3
                                 -\frac{128}{3} {\Dpgqn} S_{2,1}
                         \Biggr) S_1
                         -\Biggl(
                                 \frac{32 \big(
                                         651+442 N+175 N^2+10 N^3\big)}{81 N^2 (1+N) (2+N)}
                                 +\frac{64}{9} {\Dpgqn} S_2
                         \Biggr) S_1^2
\nonumber\\ &&        
                  +\frac{32 \big(
                                 -24+19 N+49 N^2+10 N^3\big) S_1^3}{81 N^2 (1+N) (2+N)}
                        + {\Dpgqn} \Biggl(-\frac{32}{27}  S_1^4
                         -\frac{128}{9}  S_2^2
                         -\frac{128}{3}  S_{3,1}
                         +\frac{256}{3}  S_{2,1,1}
\nonumber\\ &&                
          +\frac{256}{9}  S_4 \Biggr)
                         -\frac{128 S_{2,1}}{3 N^2}
                         + {\Dpgqn} \Biggl(
                                 \frac{2 Q_{13}}{9 N^3 (1+N)^3}
                                 +\frac{16  (6+5 N) S_1}{9 N}
                                 -\frac{16}{3}  S_1^2
                         \Biggr) \zeta_2
\nonumber\\ &&                
          +   {\Dpgqn} \Biggl(
                                 -\frac{56  Q_1}{9 N^2 (1+N)^2}
                                 +\frac{224  S_1}{9}
                         \Biggr) \zeta_3
                 \Biggr]
                 +
                         {\Dpgqn} \Biggl(\frac{2  Q_{15}}{9 N^3 (1+N)^3}
                         +\frac{80 
                          (3+N) S_1}{9 N}
\nonumber\\ &&                
          -
                         \frac{40}{3}  S_1^2
                         +8  S_2
                 \Biggr) \zeta_2
         \Biggr]
%------------------------
         +\textcolor{blue}{C_A T_F} \Biggl[
                 {\sf B_4} {\Dpgqn} \Biggl(
                         \frac{32  \big(
                                 -5+3 N+3 N^2\big)}{N (1+N)}
                         +32  S_1
                 \Biggr)
  +\Biggl(
                         -\frac{4 S_1 Q_9}{9 N^3 (1+N)^3}
\nonumber\\ &&                
                         +\frac{Q_{16}}{18 N^4 (1+N)^4}
                         +\frac{8 \big(
                                 27-40 N-12 N^2+N^3\big) S_1^2}{3 N^2 (1+N)^2}
                         +  32{\Dpgqn}  S_1^3
                         -12 {\Dpgqn}^2 (2+N) S_2
\nonumber\\ &&                
          -8 {\Dpgqn} S_3
                         + {\Dpgqn} \Biggl(
                                 -\frac{8  \big(
                                         1+3 N+3 N^2\big)}{N (1+N)}
                                 +16  S_1
                         \Biggr) S_{-2}
                         -8 {\Dpgqn} S_{-3}
                         +16 {\Dpgqn} S_{-2,1}
                 \Biggr) \zeta_2
\nonumber\\ &&                
  +\Biggl(
                         -\frac{288 {\Dpgqn} \big(
                                 -5+3 N+3 N^2\big)}{5 N (1+N)}
                         -\frac{288}{5} {\Dpgqn} S_1
                 \Biggr) \zeta_2^2
         \Biggr]
 \Biggr]
\nonumber\\ && 
%--------------
 +\textcolor{blue}{C_A T_F^2}  \Biggl[
         \textcolor{blue}{N_F} \Biggl[
                 -\frac{32 S_3 Q_4}{81 N^2 (1+N)^2 (2+N)}
                 +\frac{8 S_2 Q_{11}}{81 N^3 (1+N)^3 (2+N)}
                 -\frac{8 Q_{19}}{243 N^5 (1+N)^5 (2+N)}
\nonumber\\ &&                
  +\Biggl(
                         \frac{16 Q_{10}}{243 N (1+N)^4 (2+N)}
                         -\frac{16 \big(
                                 139+38 N+71 N^2+40 N^3\big) S_2}{27 N (1+N)^2 (2+N)}
                         +\frac{1888}{27} {\Dpgqn} S_3
\nonumber\\ &&                
          +\frac{224}{9} {\Dpgqn}
                          S_{2,1}
                         -64 {\Dpgqn} S_{-2,1}
                 \Biggr) S_1
                 +\Biggl(
                         \frac{8 Q_3}{81 N (1+N)^3 (2+N)}
                         +\frac{176}{9} {\Dpgqn} S_2
                 \Biggr) S_1^2
\nonumber\\ &&               
  -\frac{16 \big(
                         83+82 N+67 N^2+20 N^3\big) S_1^3}{81 N (1+N)^2 (2+N)}
                 +  {\Dpgqn} \Biggl(\frac{32}{27}  S_1^4
                 +\frac{80}{9}  S_2^2
                 +\frac{640}{9}  S_4 \Biggr)
 +\Biggl(
                         32 {\Dpgqn} S_1^2
\nonumber\\ &&                
          +\frac{160}{3} {\Dpgqn} S_2
                         +\frac{32 \big(
                                 -296+49 N-40 N^2+47 N^3\big)}{81 N (1+N)^3}
                         -\frac{64 \big(
                                 7-6 N+5 N^2\big) S_1}{9 N (1+N)^2}
                 \Biggr) S_{-2}
\nonumber\\ &&                
  +\Biggl(
                         -\frac{160 \big(
                                 7-6 N+5 N^2\big)}{27 N (1+N)^2}
                         +\frac{160}{3} {\Dpgqn} S_1
                 \Biggr) S_{-3}
                 +\frac{352}{9} {\Dpgqn} S_{-4}
                 -\frac{64 \big(
                         7-9 N+5 N^2\big) S_{2,1}}{27 N (1+N)^2}
\nonumber\\ &&                
  -\frac{32}{9} {\Dpgqn} S_{3,1}
                 +\frac{64 \big(
                         7-6 N+5 N^2\big) S_{-2,1}}{9 N (1+N)^2}
                 + {\Dpgqn} \Biggl(-\frac{128}{3}  S_{-2,2}
                 -\frac{160}{3}  S_{-3,1}
                 -\frac{416}{9}  S_{2,1,1}
\nonumber\\ &&                
  +64  S_{-2,1,1}\Biggr)
                 +\Biggl(
                         -\frac{4 Q_6}{9 N^3 (1+N)^3}
                         -\frac{16 \big(
                                 7+5 N^2\big) S_1}{9 N (1+N)^2}
                         + {\Dpgqn} (\frac{16}{3}  S_1^2
                         +\frac{16}{3}  S_2
                         +\frac{32}{3} 
                          S_{-2})
                 \Biggr) \zeta_2
\nonumber\\ &&                
  + {\Dpgqn} \Biggl(
                         \frac{448 }{9 N (1+N)}
                         -\frac{224 S_1}{9}
                 \Biggr) \zeta_3
         \Biggr]
         +\Biggl(
                 -\frac{4 Q_7}{9 N^3 (1+N)^3}
                 +\frac{160 (-2+N) (2+N) S_1}{9 N (1+N)^2}
\nonumber\\ &&                
                 + {\Dpgqn} \Biggl(\frac{40}{3}  S_1^2
  +\frac{40}{3}  S_2
                 +\frac{80}{3}  S_{-2}\Biggr)
         \Biggr) \zeta_2
 \Biggr]
 +\textcolor{blue}{C_F^2 T_F} \Biggl[
         -\frac{16 {\sf B_4} {\Dpgqn} \big(
                 -2+3 N+3 N^2\big)}{N (1+N)}
\nonumber\\ && 
%--------------------------------------
         + {\Dpgqn} \Biggl(
                 -\frac{Q_{14}}{2 N^3 (1+N)^3}
                 +\Biggl(
                         \frac{8  Q_2}{N^2 (1+N)^2}
                         +32  S_2
                 \Biggr) S_1
                 +\frac{4  (2+N) (5+3 N) S_1^2}{N (1+N)}
                 -16  S_1^3
\nonumber\\ &&                
  -\frac{8  \big(
                         2+3 N+3 N^2\big) S_2}{N (1+N)}
                 +16  S_3
                 +\Biggl(
                         -\frac{16}{N (1+N)}
                         +32  S_1
                 \Biggr) S_{-2}
                 +16  S_{-3}
                 -32  S_{-2,1}
         \Biggr) \zeta_2
\nonumber\\ && 
         +\frac{144 {\Dpgqn} \big(
                 -2+3 N+3 N^2\big)}{5 N (1+N)} \zeta_2^2
 \Biggr]
%-----
 + \textcolor{blue}{C_A^2 T_F} \Biggl[
         {\sf B_4} \Biggl(
                 -36 {\Dpgqn}^2 (2+N)
                 -16 {\Dpgqn} S_1
         \Biggr)
\nonumber\\ &&  
        + \Biggl(
                 \frac{2 Q_{17}}{9 N^4 (1+N)^4}
                 +\Biggl(
                         \frac{4 Q_8}{9 N^3 (1+N)^3}
                         -32 {\Dpgqn} S_2
                 \Biggr) S_1
                 -\frac{4 \big(
                         24-83 N+11 N^3\big) S_1^2}{3 N^2 (1+N)^2}
\nonumber\\ &&                
  -16 {\Dpgqn} S_1^3
                 -\frac{4 {\Dpgqn} \big(
                         -48+11 N+11 N^2\big) S_2}{3 N (1+N)}
                 -8 {\Dpgqn} S_3
                 -8 {\Dpgqn} S_{-3}
\nonumber\\ && 
                 +\Biggl(
                         -\frac{8 {\Dpgqn} \big(
                                 -36+11 N+11 N^2\big)}{3 N (1+N)}
                         -48 {\Dpgqn} S_1
                 \Biggr) S_{-2}
  +16 {\Dpgqn} S_{-2,1}
         \Biggr) \zeta_2
\nonumber\\ &&        
  + {\Dpgqn} \Biggl(
                 \frac{144  \big(
                         -8+3 N+3 N^2\big)}{5 N (1+N)}
                 +\frac{288}{5}  S_1
         \Biggr) \zeta_2^2
 \Biggr]
 -\frac{64}{9} {\Dpgqn} \textcolor{blue}{T_F^3} \zeta_3\Biggr\} + 
O(\text{rat}) + O(\zeta_3),
\end{eqnarray}
with the polynimials
\begin{eqnarray}
  Q_1&=&3 N^4+6 N^3-N^2-4 N+12,\\                           
  Q_2&=&13 N^4+23 N^3+4 N^2-14 N-5,\\
  Q_3&=&22 N^4+183 N^3+1027 N^2+2022 N+580,\\
  Q_4&=&65 N^4+364 N^3+883 N^2+614 N-648,\\
  Q_5&=&230 N^4-1154 N^3-2405 N^2-709 N-66,\\
  Q_6&=&15 N^5+15 N^4-103 N^3+33 N^2-20 N-36,\\
  Q_7&=&69 N^5+69 N^4-55 N^3+51 N^2-338 N-36,\\
  Q_8&=&103 N^5+103 N^4-79 N^3+317 N^2-612 N-144,\\
  Q_9&=&337 N^5+403 N^4-541 N^3-583 N^2-300 N+108,\\
  Q_{10}&=&491 N^5+2837 N^4+6440 N^3+10244 N^2+10934 N+1328,\\
  Q_{11}&=&4 N^6-201 N^5-143 N^4+246 N^3-1328 N^2+1368 N+1296,\\
  Q_{12}&=&29 N^6-1005 N^5-3859 N^4-5139 N^3-4486 N^2-2172 N+1944,\\
  Q_{13}&=&45 N^6+135 N^5+211 N^4+101 N^3-68 N^2+384 N+216,\\
  Q_{14}&=&153 N^6+459 N^5+527 N^4+217 N^3-4 N^2-48 N+8,\\
  Q_{15}&=&333 N^6+999 N^5+805 N^4-7 N^3-14 N^2-300 N-216,\\
  Q_{16}&=&-261 N^7-522 N^6+3712 N^5+5362 N^4-5623 N^3-7144 N^2-276 N
\nonumber\\ &&
+144,\\
  Q_{17}&=&69 N^7+138 N^6-667 N^5-541 N^4+952 N^3-1277 N^2+990 N+432,\\
  Q_{18}&=&57 N^8+376 N^7+1488 N^6+1958 N^5-461 N^4-510 N^3+1676 N^2-1656 N
\nonumber\\ &&
-1296,\\
  Q_{19}&=&3597 N^{10}+19335 N^9+36218 N^8+27506 N^7-1294 N^6-13534 N^5
\nonumber\\ &&
+12977 
N^4-7 N^3-4122 N^2-8388 N-3240,\\
  Q_{20}&=&15777 N^{12}+76086 N^{11}+111457 N^{10}-96922 N^9-540757 N^8
-841318 N^7
\nonumber\\ &&
-810709 N^6-26710 N^5+826216 N^4+92256 N^3-345888 N^2-289440 N
\nonumber\\ &&
-77760
\end{eqnarray}